\documentclass[fleqn,usenatbib]{mnras}

\usepackage{newtxtext,newtxmath}

\usepackage{ulem}

\usepackage[T1]{fontenc}

\DeclareRobustCommand{\VAN}[3]{#2}
\let\VANthebibliography\thebibliography
\def\thebibliography{\DeclareRobustCommand{\VAN}[3]{##3}\VANthebibliography}

\usepackage{graphicx}	
\usepackage{amsmath}	
\usepackage{xcolor}

\title[Neutron star mergers as the r-process site]{Neutron star mergers as the astrophysical site of the r-process in the Milky Way and its satellite galaxies}

\author[S. Wanajo et al.]{
Shinya Wanajo$^{1,2}$\thanks{E-mail: shinya.wanajo@aei.mpg.de},
Yutaka Hirai$^{3,4}$\thanks{JSPS Research Fellow}
and
Nikos Prantzos$^{5}$
\\
$^{1}$Max Planck Institute for Gravitational Physics (Albert Einstein Institute), Am M\"uhlenberg 1, Potsdam-Golm, D-14476, Germany\\
$^{2}$iTHEMS Research Group, RIKEN, Wako, Saitama 351-0198, Japan\\
$^{3}$Astronomical Institute, Tohoku University, 6-3 Aramaki, Aoba-ku, Sendai, Miyagi 980-8578, Japan\\
$^{4}$RIKEN Center for Computational Science, 7-1-26 Minatojima-minami-machi, Chuo-ku, Kobe, Hyogo 650-0047, Japan\\
$^{5}$Institut d'Astrophysique de Paris, UMR7095 CNRS, Sorbonne Université, 98bis Bd. Arago, F-75104 Paris, France
}

\date{Accepted XXX. Received YYY; in original form ZZZ}

\pubyear{2020}

\begin{document}
\label{firstpage}
\pagerange{\pageref{firstpage}--\pageref{lastpage}}
\maketitle

\begin{abstract}
Recent progress of nucleosynthesis work as well as the discovery of a kilonova associated with the gravitational-wave source GW170817 indicates that neutron star mergers (NSM) can be a site of the r-process. Several studies of galactic chemical evolution, however, have pointed out inconsistencies between this idea and the observed stellar abundance signatures in the Milky Way: (a) the presence of Eu at low  (halo) metallicity and (b) the descending trend of Eu/Fe at high  (disc) metallicity. In this study, we explore the galactic chemical evolution of the Milky Way's halo,  disc and satellite dwarf galaxies. Particular attention is payed to the forms of delay-time distributions for both type Ia supernovae (SN Ia) and NSMs. The Galactic halo is modeled as an ensemble of  independently evolving building-block galaxies with different masses. The single building blocks  as well as the disc and satellite dwarfs are treated as well-mixed one-zone systems. Our results indicate that the aforementioned inconsistencies can be resolved and thus NSMs can be the unique r-process site in the Milky Way, provided that the delay-time distributions satisfy the following conditions: (i) a long delay ($\sim 1$ Gyr) for the appearance of the first SN Ia (or a slow early increase of its number) and (ii) an additional early  component providing $\gtrsim 50\%$ of all NSMs with a delay of  $\sim 0.1$ Gyr. In our model, r-process-enhanced and r-process-deficient stars in the halo appear to have originated from ultra-faint dwarf-sized and massive building blocks, respectively. 
Our results also imply that the natal kicks of binary neutron stars have a little impact on the evolution of Eu in the disc.
\end{abstract}

\begin{keywords}
stars: abundances -- stars: neutron -- Galaxy: abundances -- Galaxy: evolution.
\end{keywords}



\section{Introduction}

The origin of the elements produced by the rapid neutron-capture process (r-process) is still uncertain \cite[see][for a recent comprehensive review]{Cowan2021}. Although the observation of a radioactively powered electromagnetic emission \citep[kilonova,][]{Li1998,Metzger2010} associated with the neutron star merger (NSM) GW170817 \citep{Abbott2017} implies the production of neutron-capture elements such as Sr \citep{Watson2019,Domoto2021} and lanthanides \citep{Arcavi2017,Chornock2017,Nicholl2017,Tanaka2017}, no trace of heavy r-process elements such as Eu and Th has been found in the kilonova ejecta. This can be due to difficulties in identification of elements in highly Doppler-shifted spectra as well as a lack of the relevant atomic data; however, it is suggested that the overall observational behaviour of this kilonova may be explained with little production of such heavy r-process elements \citep{Wanajo2018b}. 

From a nucleosynthesis point of view, NSMs are widely regarded as a promising r-process site \citep{Lattimer1974,Symbalisty1982,Eichler1989,Meyer1989,Freiburghaus1999,Goriely2011,Korobkin2012,Bauswein2013}. Recent nucleosynthesis studies based on the numerical models of NSMs indicate that all (i.e., light and heavy) r-process nuclei are produced in the early dynamical ejecta \citep{Wanajo2014,Sekiguchi2015,Sekiguchi2016,Goriely2015,Radice2018}. In addition, the post-merger ejecta from the subsequently formed accretion discs are predicted to be enriched by either  light \citep{Just2015,Lippuner2017,Shibata2017,Fujibayashi2018,Fujibayashi2020,Fujibayashi2020b,Fernandez2020} or all r-process nuclei \citep{Wu2016,Siegel2017,Fernandez2019}. Conversely, core-collapse supernovae (CCSN) appear to be excluded from the candidates for the r-process site \citep{Wanajo2013,Wanajo2011,Wanajo2018}. However, subsets of CCSNe   are suggested to be viable sources of the r-process elements, like, e.g., collapsars \citep{Siegel2019} or magneto-rotational supernovae \citep[MRSN,][]{Winteler2012,Nishimura2015,Reichert2020}; but see  \citealt{Fujibayashi2020c} and \citealt{Moesta2018} for implications of the former and latter case, respectively.

The study of galactic chemical evolution (GCE) has been a powerful tool to disentangle the different astrophysical sources (e.g., low and intermediate-mass stars, CCSNe and SNe Ia) of various elements (e.g., $\alpha$ and iron-group elements) from an increasing number of measured stellar abundances in the Milky Way \citep[MW, e.g.,][]{Timmes1995,Chiappini1999,Goswami2000,Prantzos2018,Kobayashi2020}. In a similar manner, GCE models have also been applied to account for the observational trend of measured Eu as representative of the r-process elements \citep[owing to its 95\% of pure r-process origin in the solar system, e.g.,][]{Goriely1999,Prantzos2020}.
However, various simplifications and shortcomings of  phenomenological GCE models, in particular for the MW halo, made it difficult to interpret the observational trend of Eu. 

Early work on the GCE of Eu in the MW \citep{Mathews1990,Mathews1992} has concluded that low-mass CCSNe with short (but non-negligible) time delay  are suitable sources of the r-process elements, being able to account for the appearance of stars with measured Eu at [Fe/H]\footnote{The logarithmic abundance defined by [A/B] $= \log\, (N_\mathrm{A}/N_\mathrm{B}) - \log\, (N_\mathrm{A}/N_\mathrm{B})_\odot$ for the elements A and B with numbers $N_\mathrm{A}$ and $N_\mathrm{B}$.} $\sim -3$ \citep[see also][]{Pagel1995}. NSMs were disfavored, because their binary lifetimes appeared too long to allow for a substantial contribution at low metallicity. However, since the discovery of large star-to-star scatter in [Eu/Fe] at low metallicity \citep[more than two orders of magnitude,][]{McWilliam1995,Ryan1996}, it becomes clear that commonly used one-zone models with instantaneous mixing (producing single evolutionary tracks) have difficulty in interpreting such a distinct observational trend of Eu.

The first attempt to reconcile GCE with such star-to-star scatter in [Eu/Fe] has been made by introducing some degree of inhomogeneity in the models, by assuming the chemical compositions of stars to be a mixture of supernova ejecta and ambient gas swept up by their blast waves \citep{Ishimaru1999,Tsujimoto1999,Argast2004}. However, these models still treated  the MW halo as a single system, in contrast to the  paradigm of hierarchically merging sub-haloes \citep[e.g.,][]{Hirschmann2012}. For this reason, the favoured r-process site (low-mass CCSNe) was unchanged from previous studies. Otherwise, an extremely short binary lifetime such as 0.001--0.01 Gyr, being appreciably shorter than those estimated for the observed binaries \citep[$\ge 0.05$ Gyr,][]{Stovall2018}, had to be invoked for NSMs to be the major sources of the r-process elements \citep{Argast2004,DeDonder2004,Komiya2014,Matteucci2014,Tsujimoto2014,Cescutti2015,Wehmeyer2015,Cote2017}.

An alternative approach for the MW halo has been proposed by \citet{Ishimaru2015,Ojima2018} based on the scenario of hierarchical sub-halo merging \citep{Prantzos2006,Prantzos2008b}, in which the halo is assumed to be composed of dwarf-like building-block galaxies with different stellar masses \citep[see also][]{Komiya2016}. They showed that NSMs with binary lifetimes of 0.1 Gyr could be the predominant sources of the r-process elements in the halo, assuming a smaller star formation efficiency for a less-massive building-block galaxy \citep[see also][for a similar conclusion in the chemodynamical simulations of dwarf galaxies]{Hirai2015,Hirai2017}. These models also indicated that the r-process-enhanced stars in the halo originated from the building-blocks with masses as small as ultra-faint dwarf (UFD) galaxies. This is consistent with the recent discoveries of r-process-enriched UFDs Reticulum II \citep{Ji2016,Roederer2016}, Tucana III \citep{Hansen2017} and Grus II \citep{Hansen2020}.

Recently, cosmological zoom-in simulations of MW-analogous galaxies have become feasible, in which the processes such as sub-halo merging and metal mixing can be more self-consistently incorporated \citep{Shen2015,Voort2015,Naiman2018,Haynes2019,Voort2020}. The results appear to be, however, highly dependent on the treatment of metal mixing \citep{Hirai2017b,Naiman2018}. Moreover, the spatial and mass-scale resolutions are still insufficient to explore the GCE in UFD-sized structures. For this reason, isolated \citep{Hirai2015,Hirai2017} or zoom-in \citep{Safarzadeh2017,Tarumi2020} simulations for satellite dwarf galaxies were also performed. The results are controversial; some studies favour a subset of CCSNe \citep[e.g., MRSNe,][]{Haynes2019,Voort2020}, while others support NSMs as the main sources of r-process elements. In any case, the limitations in resolution as well as the uncertainties in metal mixing make it difficult to draw a firm conclusion.

Another problem has recently been pointed out on the evolution of Eu in the disc. The values of [Eu/Fe] predicted by the models with NSMs being the origin of Eu do not decrease at high metallicity ([Fe/H] $> -1$) as opposed to its observational trend, when the commonly used delay-time distributions of $\propto t^{-1}$ are adopted for both SNe Ia and NSMs \citep{Komiya2016,Cote2017,Hotokezaka2018,Molero2021}. 
It should be noted, however, that the delay-time distributions, in particular shortly after the binary formation ($< 1$ Gyr), cannot be well constrained from observation for either of SNe Ia \citep{Maoz2014,Strolger2020} or NSMs \citep{Beniamini2019,Galaudage2021}.

The purpose of this study is to explore if NSMs can be the unique r-process site across the different components of the MW (halo and disc) and the satellite dwarf galaxies by simultaneously resolving the above problems, i.e., the observational behaviours of Eu at low and high metallicity. Particular attention is given to the uncertainties in the delay-time distributions for both SNe Ia and NSMs. The GCE model in \citet{Ishimaru2015} is adopted, in which we assume the MW halo as an ensemble of building-block (dwarf) galaxies over a wide range of stellar masses. The single building blocks, the satellite dwarfs and the disc are treated as well-mixed one-zone systems (\S~\ref{sec:method}). The results are presented in \S~\ref{sec:result}, in which a possible effect of the natal kicks of binary neutron stars is also discussed for the disc. We then discuss several relevant issues based on our results across different components of the MW (\S~\ref{sec:discussion}). Finally, the conclusions of this study are presented in \S~\ref{sec:conclusion}.

\section{Methods}
\label{sec:method}

In this study, we use the code of GCE, \texttt{iGCE}, which is based on the work in \citet{Ishimaru1999,Ishimaru2004,Ishimaru2015}. The MW halo is modeled as an ensemble of non-interacting dwarf  galaxies (``building blocks''). The GCE of each galaxy is computed following a standard recipe \citep[e.g.,][]{Prantzos2008b,Pagel2009,Matteucci2012}, in which a one-zone, homogeneous gaseous system with stars loses its mass via gas outflow as described in \S~\ref{subsec:block}. This single one-zone model is also applied for the GCE of satellite dwarfs. The MW disc is modelled as a one-zone system but with the infall of pristine gas (\S~\ref{subsec:infall}). 

Note that we explore the GCE of the MW halo more extensively than those in \citet{Ishimaru2015} and \citet{Ojima2018} by adding the contribution of SNe Ia with a delay-time distribution (\S~\ref{subsec:dtdsnia}). Moreover, the fixed delays of NSMs for the rapid (0.001 Gyr) and long (0.1 Gyr) channels in their work are replaced by a single delay-time distribution (\S~\ref{subsec:dtdnsm}). The star-to-star scatter of [Eu/Fe] resulting from the rarity of NSMs \citep[not considered in][]{Ishimaru2015} is represented by an ensemble of probabilistic evolutionary tracks for building-blocks (\S~\ref{subsubsec:eu_halo}) without using the Monte Carlo method  in \citet{Ojima2018}.

Throughout this paper, mass and time are given in units of $M_\odot$ and Gyr, respectively.

\subsection{GCE of a building-block (or dwarf) galaxy}
\label{subsec:block}

We consider the temporal evolution of the gas mass fraction, normalized to the initial mass of a single building block (or dwarf),
\begin{equation}
    f_\mathrm{gas}(t) = \frac{M_\mathrm{gas}(t)}{M_0},
	\label{eq:fgas}
\end{equation}
where $M_\mathrm{gas}(t)$ is the gas mass at a given time $t$ from the beginning and $M_0 = M_\mathrm{gas}(0)$. The evolution of $f_\mathrm{gas}(t)$
is given by the differential equation
\begin{equation}
    \frac{d f_\mathrm{gas}(t)}{dt} = -\psi(t) -\varphi(t) +[s(t) +b(t)],
	\label{eq:gas}
\end{equation}
where $\psi(t)$ is the star formation rate, $\varphi(t)$ the outflow rate, $s(t)$ and $b(t)$ the mass ejection rates of single and binary stars, respectively, and
$f_\mathrm{gas}(0) = 1$. 
 Similarly, the evolution of the  stellar mass fraction,
\begin{equation}
    f_\mathrm{star}(t) = \frac{M_\mathrm{star}(t)}{M_0},
	\label{eq:fstar}
\end{equation}
can be computed such as
\begin{equation}
    \frac{d f_\mathrm{star}(t)}{dt} = \psi(t) -[s(t) +b(t)]
	\label{eq:star}
\end{equation}
with $f_\mathrm{star}(0) = 0$.

The star formation rate and the gas outflow rate normalized by $M_0$ are defined by
\begin{equation}
    \psi(t) = k_\mathrm{SF}\, f_\mathrm{gas}(t)
	\label{eq:psi}
\end{equation}
and
\begin{equation}
    \varphi(t) = k_\mathrm{OF}\, f_\mathrm{gas}(t),
	\label{eq:varphi}
\end{equation}
respectively, where the coefficients $k_\mathrm{SF}$ and $k_\mathrm{OF}$ (star formation efficiency, SFE, and outflow efficiency, OFE) will be determined in \S~\ref{subsec:mfunc}.

The function $b(t)$ in equation~(\ref{eq:gas}) describes 
the mass ejections from binary stars and here indicates SNe Ia and NSMs; we consider the contribution of only single stars for other components. 

At a given time, the mass fraction of the ejecta from the dying single stars between the progenitor masses $m$ and $m+dm$ at birth is represented by $[m-m_\mathrm{rem}(m)]\, \phi(m)\, dm$, 
where
\begin{equation}
    \phi(m) \propto m^{-\alpha}
	\label{eq:phi}
\end{equation}
is the initial mass function (IMF) with the broken power of $\alpha = 0.3$, 1.3, 2.3 and 2.7 for $m < 0.08$, $0.08 \le m < 0.5$, $0.5 \le m < 1$ and $m \ge 1$ \citep{Kroupa2002} and $m_\mathrm{rem}(m)$ the remnant mass of the star with $m$. The stars of $m < 9$ are assumed to leave white dwarfs with $m_\mathrm{rem}(m) = 0.446 + 0.106\, m$ \citep{Iben1984} and those of $m \ge 9$ massive white dwarfs or neutron stars (or black holes) with $m_\mathrm{rem}(m) = 1.4$. For the latter, the same mass is adopted for black holes, which does not affect the result because of their diminishing weight near the high-mass end of IMF. Equation~(\ref{eq:phi}) is normalized as $\int m\, \phi(m)\, dm = 1$ between $m_\mathrm{low} = 0.05$ and $m_\mathrm{up} = 60$. The gas ejection rate per baryon mass from dying single stars is, therefore, given by
\begin{equation}
    s(t) = \int_{m_t}^{m_\mathrm{up}}\, \psi[t-\tau(m)]\, [m-m_\mathrm{rem}(m)]\, \phi(m)\, dm,
	\label{eq:gt}
\end{equation}
where $\tau(m)$ is the lifetime of the star with $m$ adopted from \citet{Schaller1992} and $m_t$ the mass of the star with $\tau(m) = t$. In this study, $b(t) = 0$ is taken because of its negligible contribution to the total gas mass in the galaxy.

Similarly, the evolution of element $i$ in the form of gas, $f_{\mathrm{gas},i}(t)$, is given by
\begin{equation}
    \frac{d f_{\mathrm{gas},i}(t)}{dt} = -\frac{f_{\mathrm{gas},i}(t)}{f_\mathrm{gas}(t)}\, [\psi(t) +\varphi(t)] +[s_i(t) +b_i(t)].
	\label{eq:gasi}
\end{equation}
In the right-hand side, the first term in the second bracket is the mass ejection rate of element $i$ from dying single stars given by
\begin{equation}
    s_i(t) = \int_{m_t}^{m_\mathrm{up}}\, \psi[t-\tau(m)]\, Y_i[m, Z(t-\tau(m))]\, \phi(m)\, dm,
	\label{eq:sit}
\end{equation}
where $Y_i[m, Z(t)]$ is the yield of element $i$ (in $M_\odot$) from the star with $m$ at birth and metallicity $Z(t)$. The contribution of binaries (SNe Ia and NSMs) is expressed as
\begin{equation}
    b_i(t) = C\, Y_i \int_0^t D(t') \, \psi(t-t')\, dt',
	\label{eq:bit}
\end{equation}
where the star formation rate is convolved with the delay-time distribution, $D(t)$, and $C \equiv B\int \phi(m)\, dm$ is a constant integrated between the mass range of progenitors ($B$ is the binary fraction) described in \S~\ref{subsec:yield}. Each contribution of SNe Ia or NSMs is separately added such as $b_i(t) = b_{\mathrm{SNIa}, i}(t) + b_{\mathrm{NSM}, i}(t)$. In this study, the yield $Y_i$ is taken to be a constant for each type of events, being independent of progenitor mass, mass ratio and metallicity.

\subsection{Stellar yields}
\label{subsec:yield}

\begin{table*}
	\centering
	\caption{Adopted parameters for the four considered cases of delay-time distributions of SNe Ia and NSMs in equations~(\ref{eq:dtdia}) and (\ref{eq:dtdnsm}), respectively. The minimum delays for SNe Ia (second column) and NSMs (third column) as well as the median of the log-normal component $\exp \mu$ (fourth column) in equation~(\ref{eq:dtdnsm}) are given in Gyr. The fifth column presents the fraction of the power-law component $A$ (fifth column) in equation~(\ref{eq:dtdnsm}). The effective SFE, $K$, in the sixth column is the resultant coefficient in the right-hand side of equation~(\ref{eq:sfof}) for each case.}
	\label{tab:dtd}
	\begin{tabular}{lcccccc} 
		\hline
		        & $t_\mathrm{min}$ (SN Ia) & $t_\mathrm{min}$ (NSM) & $\exp \mu$ & $A$ & $K$  & Type of NSM delay-time distribution                   \\
		\hline
		case 1  & 1.0                         & 0.020                      & 0.10             & 0.5 & 0.045 & standard (power-law + log-normal)   \\
		case 2  & 1.0                         & 0.020                      & ---              & 1.0 & 0.045 & power-law                           \\
		case 3  & 1.0                         & 0.005                      & 0.01             & 0.0 & 0.045 & log-normal, short delay (mimicking subsets of CCSNe) \\
		case 4  & 0.1                         & 0.020                      & 0.10             & 0.5 & 0.017 & standard (but short delay of SNe Ia)    \\
		\hline
	\end{tabular}
\end{table*}

In this study, the evolution of Mg and Eu with respect to Fe is explored, which are assumed to be supplied from CCSNe, SNe Ia and NSMs. As our main focus is placed on the enrichment of r-process elements, the contributions of s-process elements are not considered. A decrease of H (or an increase of He) is not computed, although the gas return to the system from dying stars over all the mass range is included in equation~(\ref{eq:gt}). For this reason, the logarithmic abundance of a given element $i$ with respect to H is derived such as [$i$/H] = $\log\, (N_i/N_\mathrm{H}) - \log\, (N_i/N_\mathrm{H})_\odot$ = $\log\, (N_i/N_{i,\odot}) - \log\, (N_\mathrm{H}/N_\mathrm{H,\odot})$ $\approx \log\, (X_i/X_{i,\odot})$, where $X_i(t) \equiv f_{\mathrm{gas},i}(t)/f_\mathrm{gas}(t)$ is the mass fraction of element $i$. Here, we assume $\log\, (X_\mathrm{H}/X_\mathrm{H,\odot}) \approx 0$.

For Mg and Fe, we adopt the metallicity-dependent yields of CCSNe, which cover the progenitor mass range of 13--$40\, M_\odot$ \citep{Kobayashi2006,Nomoto2006}. In this study, the progenitor mass range of CCSNe is taken to be 10--$60\, M_\odot$. The yields at $40\, M_\odot$ also are applied for 40--$60\, M_\odot$. Otherwise, the yields for a given mass and a metallicity are obtained by a linear interpolation of tabulated values.

Fe is also produced by SNe Ia, for which the yield of the W7 model, $0.74\, M_\odot$ \citep{Iwamoto1999}, is adopted. The coefficient in equation~(\ref{eq:bit}) is set to $C = 1.4\times 10^{-3}$ such that the IMF-folded fraction of SNe Ia with respect to that of CCSNe becomes the present value of 0.23 for MW-analogous galaxies estimated from  nearby supernova observations \citep{Li2011}. This corresponds to the binary fraction of $B = 0.052$ when the progenitor mass range of SNe Ia is taken to be 3--$8\, M_\odot$.

Eu is assumed to be exclusively produced by NSMs. The Eu yield is taken to be $2.5\times 10^{-5}\, M_\odot$, a value similar to that in the dynamical ejecta of a NSM in \citet{Wanajo2014}. The coefficient in equation~(\ref{eq:bit}) is set to $C = 1.2\times 10^{-5}$ such that the resulting [Eu/Fe] at [Fe/H] $\sim -1.5$ approximately matches the measured values in the MW halo (for case 1; see \S~\ref{subsubsec:eu_halo}). This corresponds to the binary fraction of $B = 0.002$ when the progenitor mass range is assume to be 10--$60\, M_\odot$. Provided that the rate of CCSNe is $2\times 10^{-2}$ yr$^{-1}$ \cite[e.g.,][]{Diehl2013}, the present-day NSM rate in the MW becomes $4\times 10^{-5}$ yr$^{-1}$ (see also \S~\ref{subsubsec:kick}), a value that resides in the range inferred from various observations \citep[e.g., of short gamma-ray bursts and GW170817 with a number density of MW-analogous galaxies $\approx 0.01$ Mpc$^{-3}$,][]{Hotokezaka2018}.

\subsection{Delay-time distribution for SNe Ia}
\label{subsec:dtdsnia}

\begin{figure}
	\includegraphics[width=0.86\columnwidth]{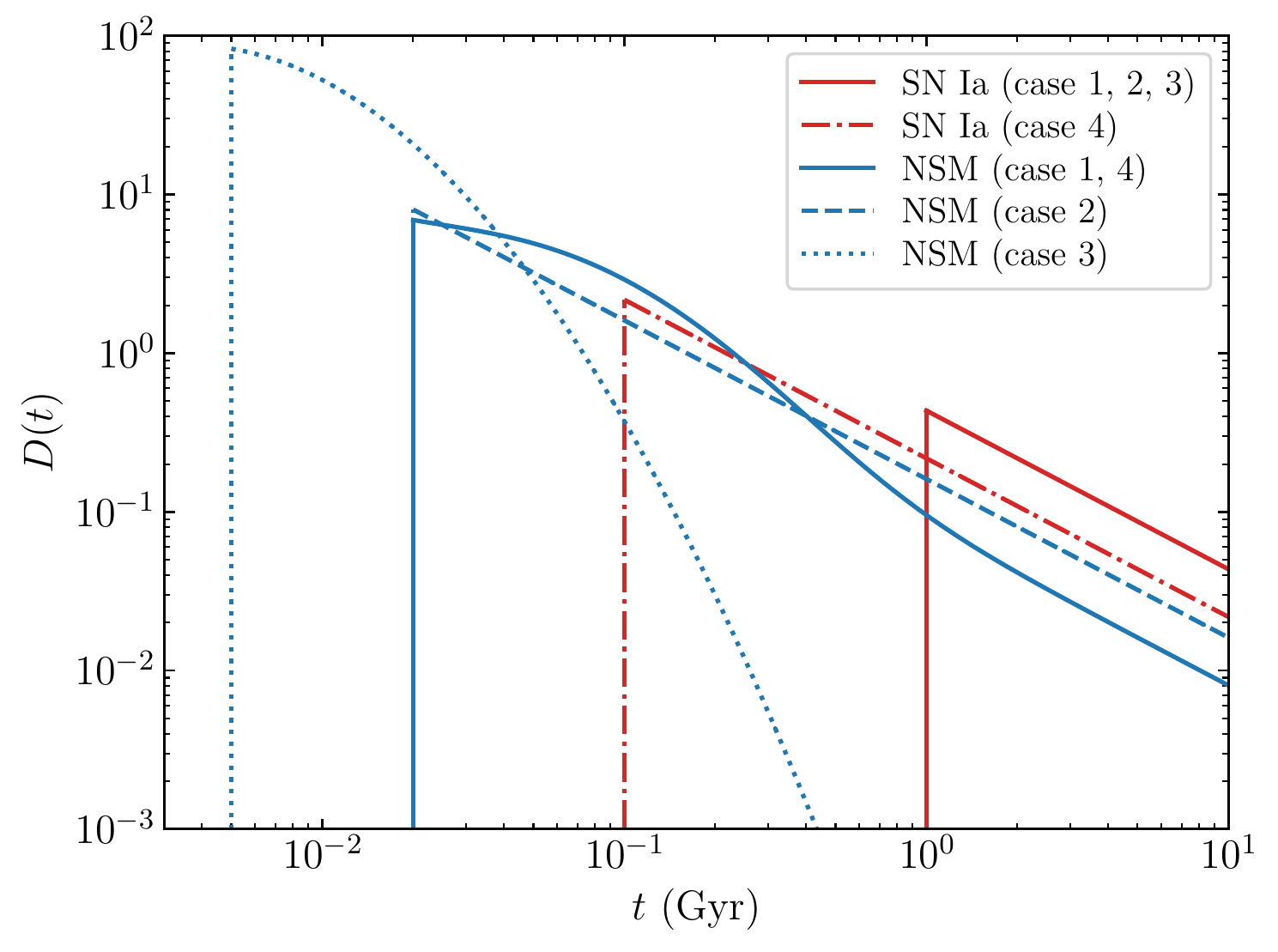}
    \caption{Adopted delay-time distributions for SNe Ia (red) and NSMs (blue). The distributions are normalised to $\int{D(t)\, dt} = 1$. The line styles indicate different types of distributions (cases 1-4 in Table~\ref{tab:dtd}).}
    \label{fig:dtd}
\end{figure}

For SNe Ia, the observationally inferred, empirical delay-time distribution \citep{Maoz2014},
\begin{equation}
D(t) = \frac{1}{\ln\, (t_\mathrm{max}/t_\mathrm{min})}\, t^{-1}
	\label{eq:dtdia}
\end{equation}
(for $t_\mathrm{min} < t < t_\mathrm{max}$; $D(t) = 0$ otherwise) is taken, where $t_\mathrm{min}$ and $t_\mathrm{max}$ (= 10 Gyr) are the minimum and maximum delays, respectively. While this empirical form appears robust for $> 1$ Gyr, the observed rate of SNe Ia is found to be substantially smaller than the prediction of equation~(\ref{eq:dtdia}) for $< 1$ Gyr \citep{Strolger2020}. Moreover, nearly constant stellar ratios of [Mg/Fe] in the MW halo imply a subdominant role of SNe Ia during the first $\sim 1$ Gyr. We adopt, therefore, $t_\mathrm{min} = 1.0$ Gyr (cases 1--3 in Table~\ref{tab:dtd}; red solid line in Fig.~1), although a case of short delay, $t_\mathrm{min} = 0.1$ Gyr (case 4; dash-dotted line), is also considered for comparison purposes (see additional tests in APPENDIX \ref{sec:appendex}). 



\subsection{Delay-time distribution for NSMs}
\label{subsec:dtdnsm}

For NSMs, we assume the functional form of delay-time distribution in \citet{Beniamini2019},
\begin{equation}
    D(t) = \frac{A}{\ln\, (t_\mathrm{max}/t_\mathrm{min})}\, t^{-1} + \frac{1-A}{\sqrt{2\pi \sigma^2}\, t} \exp \left[-\frac{(\ln t - \mu)^2}{2\sigma^2} \right]
	\label{eq:dtdnsm}
\end{equation}
(for $t_\mathrm{min} < t < t_\mathrm{max}$; $D(t) = 0$ otherwise) with power-law and additional log-normal terms, where $t_\mathrm{max}$ = 10 Gyr and $\sigma =1.0$. Equation~(\ref{eq:dtdnsm}) well reproduces the estimated delay-time distribution for currently observed binary neutron stars in the MW, which exhibits an apparent excess in the early population ($< 1$ Gyr) compared to the classical power-law distribution. For our fiducial model (cases 1 and 4), we assume an equal weight for the power-law and log-normal components, i.e., $A = 0.5$. The median of the log-normal component is taken to be $\exp \mu = 0.1$ Gyr \citep[cases 1 and 4 in Table~\ref{tab:dtd}; blue solid curve in Fig.~\ref{fig:dtd}; instead of 0.3 Gyr in][]{Beniamini2019} in agreement with \citet[][a constant delay of 0.1 Gyr]{Ishimaru2015}. A case without the log-normal term, i.e., $A = 1.0$ (case 2; blue dashed line), is also considered for comparison purposes. The minimum delay is assume to be $t_\mathrm{min}$ = 0.020 Gyr \citep[instead of 0.035 Gyr in][]{Beniamini2019}, a value consistent with the observation of short gamma-ray bursts \citep{Wanderman2015}. We also consider a distribution with little delay, mimicking that for subsets of CCSNe such as collapsars or MRSNe by taking $t_\mathrm{min}$ = 0.005 Gyr, $\exp \mu = 0.01$ Gyr and $A = 0$ (case 3 in Table~\ref{tab:dtd}; blue dotted curve in Fig.~\ref{fig:dtd}; see additional tests in APPENDIX \ref{sec:appendex}).

\subsection{Star formation rate and outflow rate}
\label{subsec:sfof}

We assume that the building blocks of the MW halo obey the mass-metallicity relation of dwarf galaxies observed in the Local Group, that is, the mean metallicity being scaled as $10^\mathrm{[Fe/H]} \propto (M_*/10^6)^{0.30}$ \citep[][]{Kirby2013},
where $M_*$ is the present-day stellar mass of a given galaxy (see a caution in \S~\ref{sec:discussion}). 
In \cite{Prantzos2008}, it is shown that the metallicity distribution function of the dwarf satellites of the MW can be explained in the framework of simple, analytical, one-zone GCE models by assuming that the ratio SFE/OFE is proportional to some power of $M_*$. We thus adopt \citep{Ishimaru2015}
\begin{equation}
\frac{k_\mathrm{SF}}{k_\mathrm{OF}} = K\left(\frac{M_*}{10^6}\right)^{0.3}.
	\label{eq:sfof}
\end{equation}
However, neither $k_\mathrm{SF}$ or $k_\mathrm{OF}$ is constrained for a given building-block galaxy. In this study, therefore, we set $k_\mathrm{OF} = 1.0$ (Gyr$^{-1}$) for all galaxies, while $k_\mathrm{SF}$ is determined from equation~(\ref{eq:sfof}) for a given mass galaxy (see an additional test in APPENDIX \ref{sec:appendex}). \citet[][]{Ishimaru2015} showed that this choice (their Case 1) is in  qualitative agreement with the evolution of Eu in the MW halo \citep[with the assumption of its source being NSMs, see also][]{Ojima2018}. The coefficient in equation~(\ref{eq:sfof}), $K$ (hereafter effective SFE) in Table~\ref{tab:dtd} (sixth column), will be determined in \S~\ref{subsec:halo}.

\subsection{Mass function of the building-block galaxies in the MW halo}
\label{subsec:mfunc}

\begin{figure}
	\includegraphics[width=0.86\columnwidth]{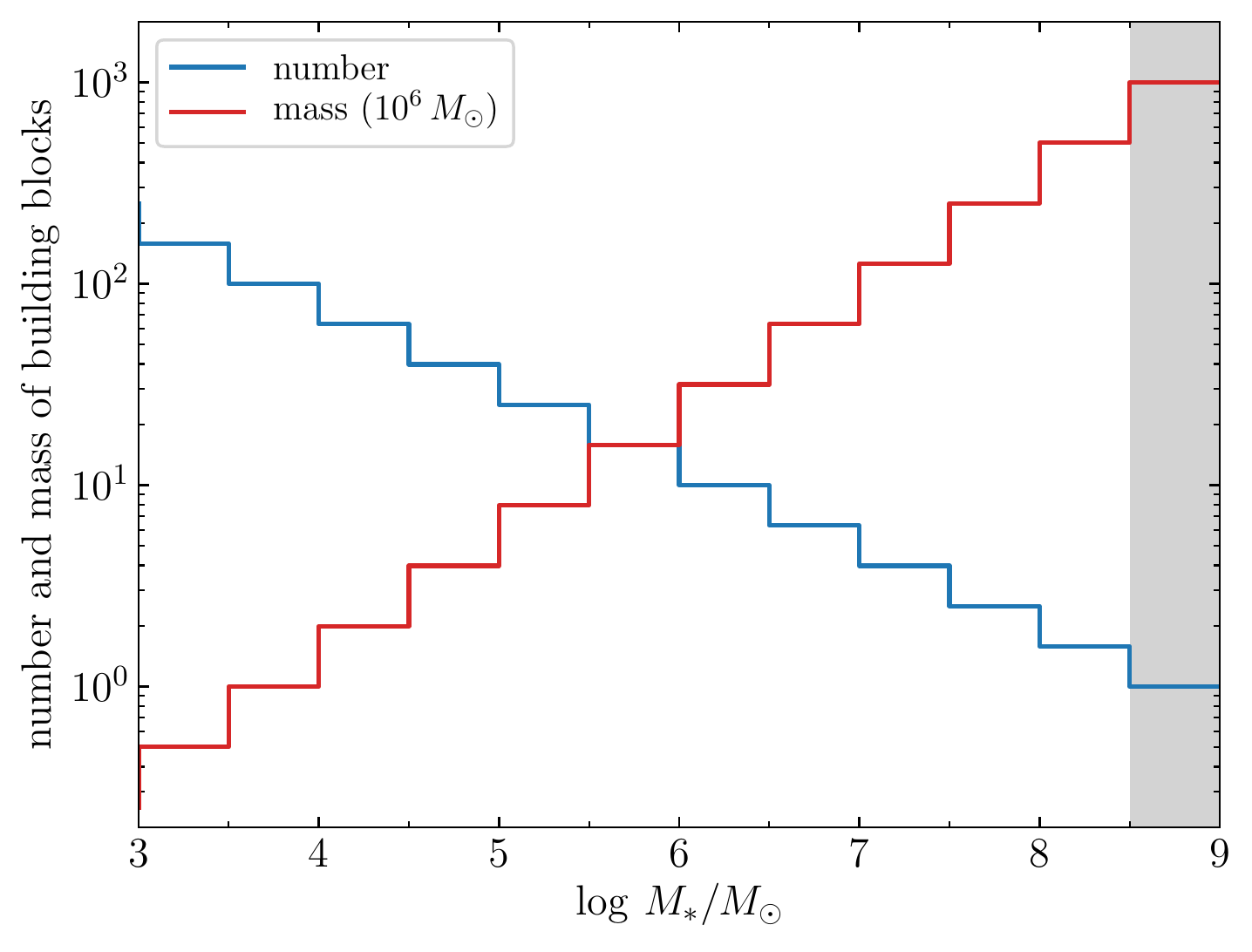}
    \caption{Number (blue) and mass (red; in $10^6\, M_\odot$) of building-block galaxies per $\Delta \log\, (M_*/M_\odot) = 0.5$, when those of the last bin at $10^{8.5}$--$10^9\, M_\odot$ are assumed to be 1 and $1\times 10^9\, M_\odot$, respectively. The nominal values are set at the high-mass end of each interval. The most massive building block can be regarded as the in situ stellar halo (indicated by the gray shaded region) and the remainder the accreted stellar haloes.}
    \label{fig:msub}
\end{figure}

As we consider the MW halo being an ensemble of building-block galaxies, their mass function should be defined. To this end, we begin with the relation between the number ($N_\mathrm{DH}$) and mass ($M_\mathrm{DH}$) of dark-matter sub-haloes, 
\begin{equation}
\frac{dN_\mathrm{DH}}{dM_\mathrm{DH}} \propto {M_\mathrm{DH}}^{-2},
	\label{eq:dm}
\end{equation}
which is indicated by a cosmological simulation of the MW halo \citep{Diemand2007,Griffen2016}. This appears to be valid at early times, being back to the redshift of $z = 5$ \citep{Salvadori2007}. Thus, we use this relation to estimate the stellar-mass function of the building-block galaxies as in \cite{Prantzos2008}.

From equation~(\ref{eq:sfof}), we have 
\begin{equation}
\frac{M_*}{M_\mathrm{out}} \propto {M_*}^{0.3}
	\label{eq:sout}
\end{equation}
for a given building block with $M_*$ at the end of evolution, where $M_\mathrm{out}$ is the gas mass lost from the system. Due to the fact that $M_\mathrm{out} \gg M_*$ at the end of evolution, we also have 
\begin{equation}
M_\mathrm{out} \approx M_\mathrm{out} + M_* = M_0 \propto M_\mathrm{DH},
	\label{eq:mout}
\end{equation}
where the cosmic baryon-to-dark matter ratio is assumed to be preserved.

Finally, we obtain the stellar-mass function of building-block galaxies (at the end of evolution), $\Phi(M_*)$, by combining equations~(\ref{eq:dm})--(\ref{eq:mout}) as
\begin{equation}
\Phi(M_*) = \frac{dN_\mathrm{BB}}{dM_*} = \frac{0.6\, M_\mathrm{halo}}{{M_\mathrm{up}}^{0.6}-{M_\mathrm{low}}^{0.6}}\, {M_*}^{-1.4},
	\label{eq:mfunc}
\end{equation}
where $N_\mathrm{BB}$ is the number of building blocks. As we find from equations~(\ref{eq:dm}) and (\ref{eq:mfunc}), the stellar mass function is flatter than that of dark-matter sub-halos. This is due to the fact that an initially more massive building-block galaxy locks up more baryonic matter into stars owing to its greater SFE or smaller OFE (the former is the case in this study; \S~\ref{subsec:sfof}) as can be seen in equations~(\ref{eq:sfof}) and (\ref{eq:sout}). 
Equation~(\ref{eq:mfunc}) is normalized such as
$\int_{M_\mathrm{low}}^{M_\mathrm{up}} M_* \Phi(M_*)\, dM_* = M_\mathrm{halo}$,
where $M_\mathrm{low} = 1\times 10^3$ and $M_\mathrm{up} = 10^9$ are, respectively, the minimum and maximum stellar masses of building-block galaxies adopted in this study. These choices are dictated by (a) the estimated masses of UFDs \citep{Kirby2013} and (b) the stellar mass of the MW halo observed today, $M_\mathrm{halo} = 1.5 \times 10^9$ \citep[][]{Deason2019}.

From equation~(\ref{eq:mfunc}), we have $dN_\mathrm{BB}/d\log\, M_* \propto {M_*}^{-0.4}$ and $dM_\mathrm{*,tot}/d\log\, M_* \propto {M_*}^{0.6}$, where $M_\mathrm{*,tot}$ is the total stellar mass. These indicate that the stars from low-mass building-block galaxies are subdominant in the MW halo, even though the number of such galaxies is large, as illustrated in Fig.~\ref{fig:msub}. In other words, the stars in the MW halo originate predominantly from a small number of massive building-block galaxies, being in agreement with the indications from observation \citep[e.g.,][]{Helmi2018,DiMatteo2019,Helmi2020} and cosmological simulations \citep[e.g.,][]{Bignone2019,Mackereth2019,Fattahi2020,Santistevan2020}. 
The cosmological simulations by \citet{Monachesi2019,Font2020} suggest that the mass fraction of accreted stellar haloes (out to $\sim 20$ kpc from the center) accounts for about 50\% of the total halo mass. In our model, this is expressed  as
\begin{equation}
\int_{M_\mathrm{low}}^{M_\mathrm{up, acc}} M_* \Phi(M_*)\, dM_* = 0.5\, M_\mathrm{halo},
	\label{eq:macc}
\end{equation}
where $M_\mathrm{up, acc}$ is the maximum mass of the accreted building blocks. Equation~(\ref{eq:macc}) gives $M_\mathrm{up, acc} = 3.2 \times 10^8$ and thus $M_\mathrm{halo} - M_\mathrm{up, acc} \sim M_\mathrm{up}$. Therefore, our model is consistent with these simulations if we interpret the most massive building block of $M_\mathrm{up}$ and the remainder of $10^3$--$10^{8.5}\, M_\odot$ being the in situ and accreted stellar haloes, respectively (Fig.~\ref{fig:msub}).



\subsection{GCE of the MW disc}
\label{subsec:infall}

We model the MW disc as a simple one-zone region, progressively formed by gaseous infall \citep{Prantzos2008b,Pagel2009,Matteucci2012}. This type of  standard infall model describes well the GCE of the thin disc but cannot account for that of the thick disc, e.g., the observationally identified dichotomy of $\alpha$-element distribution \citep{Adibekyan2013,
Bensby2014,Hayden2015,Queiroz2020}. 
Such a bimodal abundance distribution may also concern the r-process elements (\citealt{Griffith2019}; but see \citealt{Guiglion2018}).

In addition to a standard infall model, therefore, we also examine a two-infall model \citep{Chiappini1997}, in which the first and second infall episodes are assumed to correspond to those of the thin and thick discs, respectively \citep{Spitoni2019,Tsujimoto2019,Palla2020}. Note that the two-infall model assumes that the formation of the thick and thin discs are sequential, in which the infalling material in the second episode is assumed to be a mixture of pristine (or low-metallicity) gas and that of the thick disc. In fact, such distinct phases of the evolution can be found in some cosmological zoom-in simulations, being separated by cessation of gas accretion over a certain period of time \citep{Noguchi2018,Mackereth2019,Buck2020,Khoperskov2021}. Conversely, \citet{Agertz2021} and \citet{Renaud2021,Renaud2021b} have demonstrated that the thick and thin discs are in part coeval, with distinct gas flows being responsible for the formation of these two components. 
Considering the situation with the thick disc formation being unclear at present, 
we adopt here a two-infall model because of its simplicity.

\begin{table}
	\centering
	\caption{Parameters for the standard (one-) and two-infall models adopted in equations~(\ref{eq:infall}) and (\ref{eq:inf12}). The second to last columns present the coefficient $A$, the infall timescales (in Gyr) for the thick and thin discs, the delay time for the thin disc (in Gyr) and the coefficients of SFE for the thick and thin discs (in Gyr$^{-1}$).}
	\label{tab:infall}
	\begin{tabular}{lcccccc} 
		\hline
		Type       & $A$  & $\tau_1 $ & $\tau_2 $ & $t_2 $ & $k_\mathrm{SF}$ (thick) & $k_\mathrm{SF}$ (thin) \\
		\hline
		one infall & 1.0  & --- & 7.0 & 0.0 & --- & 0.8\\
		two infall & 0.75 & 1.5 & 3.0 & 5.0 & 2.0 & 1.0\\
		\hline
	\end{tabular}
\end{table}

The gas infall rate ($-\varphi(t)$ in equation~(\ref{eq:gas})) of primordial gas with respect to the total mass $M_0$ in the  the disc is defined by
\begin{equation}
    -\varphi(t) = (1-A) f_1(t) + A f_2(t),
	\label{eq:infall}
\end{equation}
where
\begin{equation}
    f_j(t) = \frac{1}{\tau_j [1-\exp(-(T-t_j)/\tau_j)]}\, \exp \left(-\frac{t-t_j}{\tau_j}\right)
	\label{eq:inf12}
\end{equation}
is either of the thick ($j = 1$) or thin ($j = 2$) component of the MW disc with the infall timescale of $\tau_j$. $A$ sets the fraction of mass accreted onto each disc component at $T = 12$ Gyr, the assumed age of the MW disc. The range of time is $t_j < t < T$, where $t_j$ is the delay time ($t_1 = 0$). A choice of $A = 1$ and $t_2 = 0$ reduces to the standard infall model with a single accretion episode. Note $M_\mathrm{gas}(0) = 0$ and $M_0 = M_\mathrm{gas}(T) + M_\mathrm{star}(T)$ in equation~(\ref{eq:fgas}) for the disc. 


The star formation rate is assume to be proportional to ${\Sigma_\mathrm{gas}}^{1.5}$ \citep{Schmidt1959,Kennicutt1998}, where $\Sigma_\mathrm{gas}$ is the local gas surface density in the disc. This reduces to
\begin{equation}
    \psi(t) = k_\mathrm{SF}\, {f_\mathrm{gas}(t)}^{1.5}
	\label{eq:psi_}
\end{equation}
because of $\Sigma_\mathrm{gas} \propto M_\mathrm{gas}(t) = M_0 f_\mathrm{gas}(t)$.

The adopted parameters for the standard (one-) and two-infall models are presented in Table~\ref{tab:infall}. For both, $k_\mathrm{SF}$ is adjusted such that the present-day gas fraction  ($= f_\mathrm{gas}(T)/[f_\mathrm{gas}(T) + f_\mathrm{star}(T)]$) becomes $\approx 0.2$ \citep{Kubryk2015}. 
The locally observed mass ratio of the thin/thick disc determines the fraction of gas accreted onto the thin disc to be $A \approx 0.75$ \citep{Prantzos2008b}. The other parameters, $\tau_1$, $\tau_2$ and $t_2$, are determined as to reproduce reasonably well the evolution of [Mg/Fe] in the disc.

\section{Results}
\label{sec:result}

In this section, we present the results of GCE for the MW halo (\S~\ref{subsec:halo}), the satellite dwarf galaxies (\S~\ref{subsec:dwarf}) and the MW disc (\S~\ref{subsec:disc}). For the disc, we also explore a possible effect of natal kicks imparted to binary neutron stars on the GCE of Eu (\S~\ref{subsubsec:kick}).

\subsection{MW halo}
\label{subsec:halo}

We compute the GCE of building block galaxies with masses from $M_\mathrm{low}$ to $M_\mathrm{up}$ with a step of $\Delta \log M_* = 0.05$ (121 different mass galaxies). Each building block is evolved up to 2 Gyr as in \citet{Ishimaru2015,Ojima2018}, which appears to be a reasonable timescale for the assembly of building blocks estimated from the ages of globular clusters \citep[see fig. 9 in][]{Kruijssen2020}. The evolution of the halo as a whole is then constructed according to the mass function of building blocks (equation~(\ref{eq:mfunc})).

\subsubsection{Metallicity distribution}
\label{subsubsec:dmd}

\begin{figure*}
	\includegraphics[width=0.86\columnwidth]{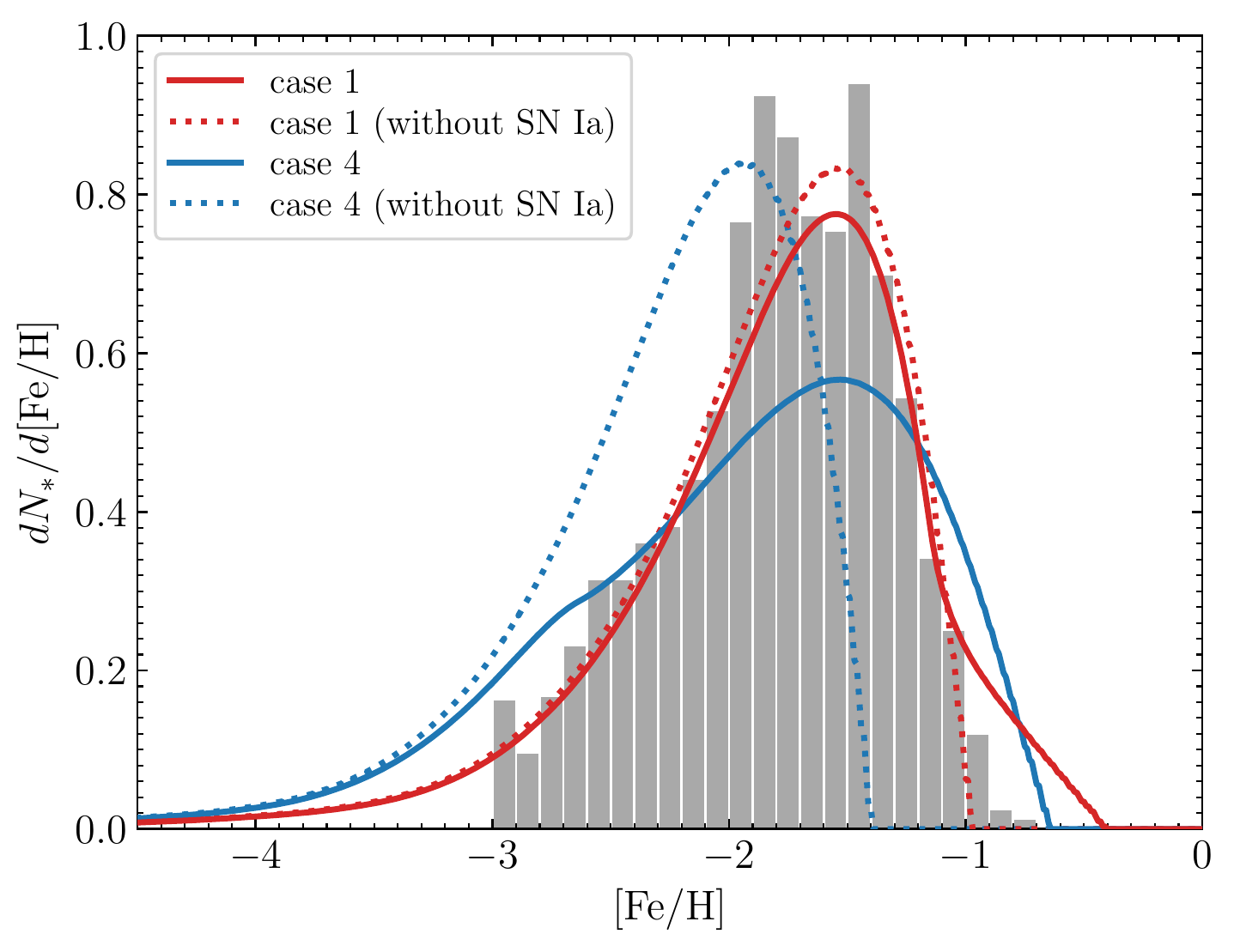}
	\includegraphics[width=0.86\columnwidth]{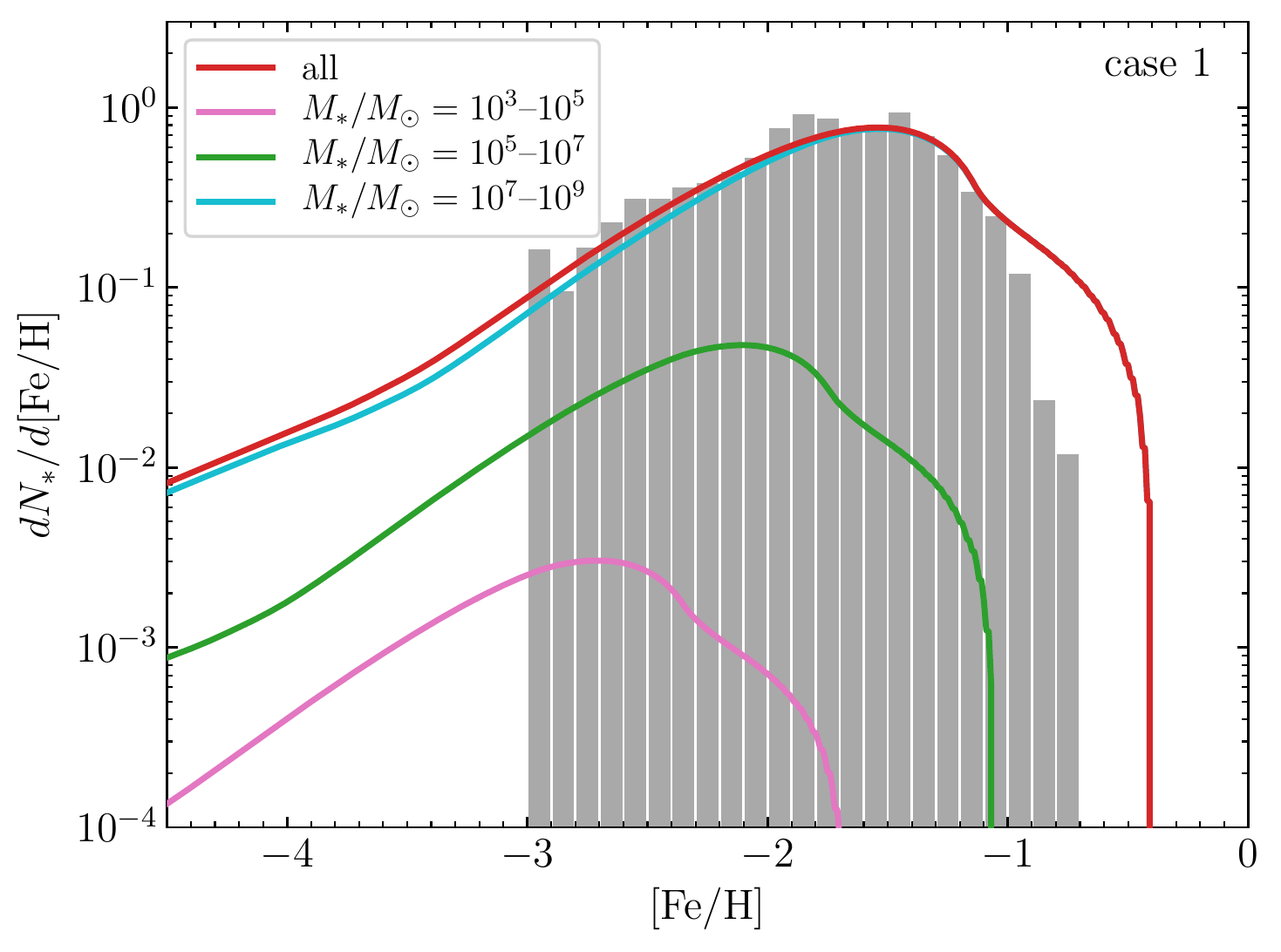}
    \caption{Normalized metallicity distribution $dN_*/d$[Fe/H] ($N_*$ is the number of stars today) of the MW halo, which is compared to the observationally inferred distribution \citep[gray histogram,][]{An2013}. \textit{Left}: the metallicity distributions for case 1 (red) and case 4 (blue) with (solid curves) and without (dotted curves) contributions of SNe Ia. $K$ in equation~(\ref{eq:sfof}) is determined such that the peak (with SNe Ia) of the distribution is obtained at [Fe/H] $\approx -1.55$, being similar to that of observation. Note that cases 2 and 3 give the same distribution to that of case 1. \textit{Right}: the contributions of the building-block galaxies (in logarithmic scale) with $M_*$ (in $M_\odot$) in the ranges $10^3$--$10^5$ (magenta), $10^5$--$10^7$ (green) and $10^7$--$10^9$ (cyan) to the total (red).}
    \label{fig:dmd_halo}
\end{figure*}

First of all, the effective SFE, $K$, in equation~(\ref{eq:sfof}) should be specified, which controls the evolution of building blocks as well as satellite dwarf galaxies. We attempt to fit the computed metallicity distribution $dN_*/d$[Fe/H] ($N_*$ is the number of stars today) of the halo to that inferred by observation \citep{An2013}. In fact, the observational metallicity distribution of the halo is somewhat uncertain, in particular at the low and high-metallicity ends \citep{Youakim2020}. Nevertheless, the metallicity at the peak of distribution appears relatively robust, being [Fe/H] $\approx -1.8$--$-1.3$. Therefore, we set the value of $K$ as to obtain the peak of distribution at [Fe/H]$_\mathrm{peak} \approx -1.55$. Note that the distributions for cases 2 and 3 are the same as that for case 1. 

As can be seen in the left panel of Fig.~\ref{fig:dmd_halo}, the (normalized) metallicity distribution for case 1 (red solid curve) is in good agreement with that of \citet[][gray histogram]{An2013}. We find that, for case 1, the contribution of SNe Ia starting from 1 Gyr is unimportant except for that at the high-metallicity end ([Fe/H] $\gtrsim -1$), when compared to the distribution without SNe Ia (red dotted curve). By contrast, SNe Ia play an important role to the metallicity distribution for case 4, for which the contribution starts from 0.1 Gyr, as can be seen in Fig.~\ref{fig:dmd_halo} (left, blue solid and dotted curves). However, the early contribution of SNe Ia leads to too many stars at low metallicity to reconcile our model with the observed distribution of \citet{An2013}. Either a long delay ($\sim$1 Gyr) for the appearance of the first SN Ia is required, or a  slow early increase of its number, instead of the abrupt one displayed in Fig. \ref{fig:dtd} (see APPENDIX \ref{sec:appendex}).

The right panel of Fig.~\ref{fig:dmd_halo} depicts the contributions of the building block galaxies in the ranges of $10^3 \leq M_* < 10^5$ (magenta), $10^5 \leq M_* < 10^7$ (green) and $10^7 \leq M_* < 10^9$ (cyan) to the metallicity distribution (red) for case 1. We find that the distribution peaks at lower metallicity for a less-massive building block as anticipated from its smaller $k_\mathrm{SF}$ defined by equation~(\ref{eq:sfof}). As described in \S~\ref{subsec:mfunc}, the contribution to the total number of stars progressively increases for more massive building blocks despite their smaller number. For instance, UFD-sized building blocks ($M_* < 10^5$) account for only a few percent of the stars at [Fe/H] $\lesssim -3$. 


\subsubsection{Evolution of Mg}
\label{subsubsec:mg_halo}

\begin{figure*}
	\includegraphics[width=0.86\columnwidth]{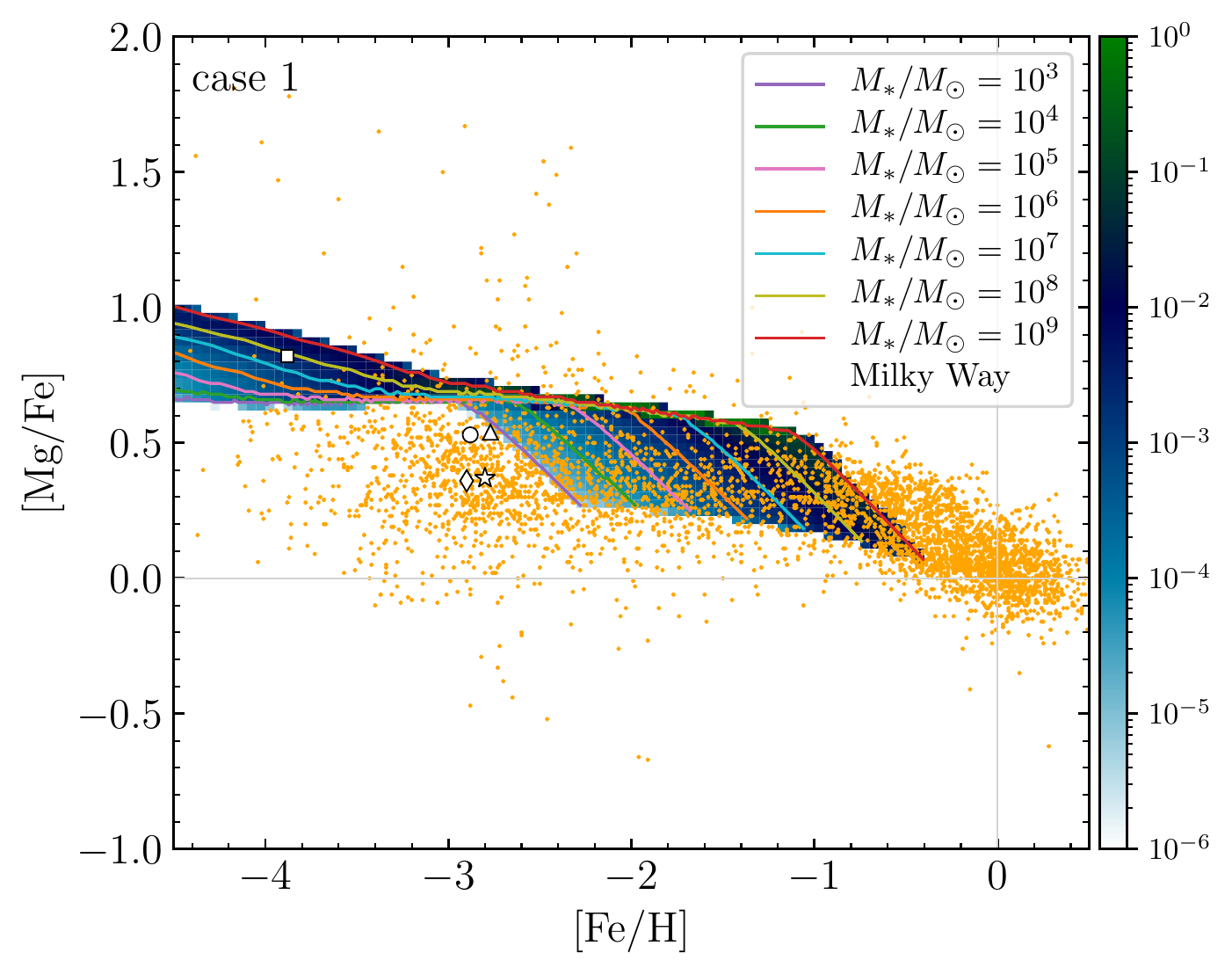}
	\includegraphics[width=0.86\columnwidth]{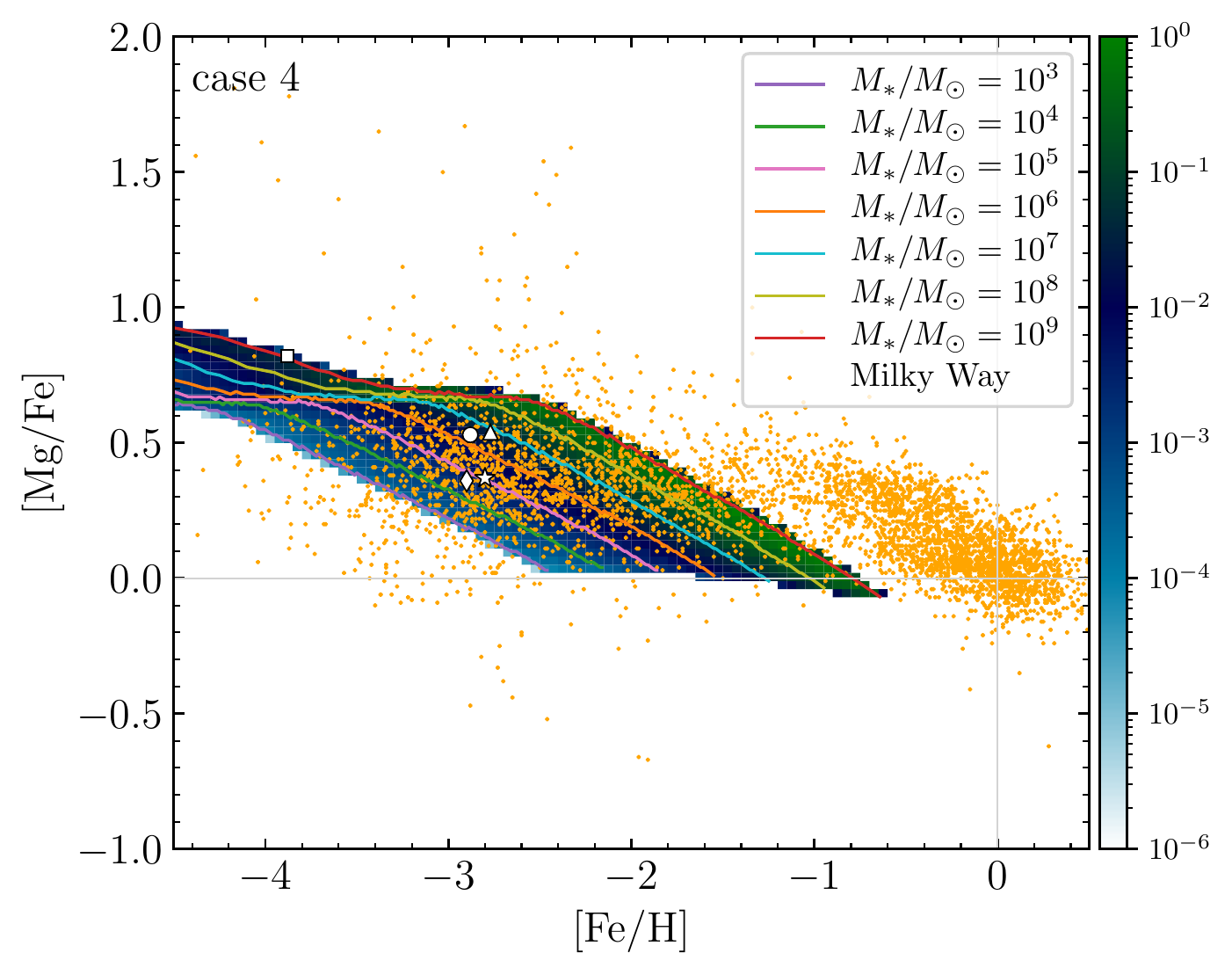}
    \caption{[Mg/Fe] as a function of [Fe/H] in the MW halo for case 1 (left) and case 4 (right). The results for cases 2 and 3 are the same as that for case 1. The colour scale indicates the relative number of stars with respect to the maximum value in each panel. The curves with different colours show the evolution of building-block galaxies with stellar masses specified in the legend. The orange dots are the measured values in the MW taken from the SAGA data base \citep{Suda2008,Suda2011}. See the caption of Fig.~\ref{fig:eufe_halo} for the description of symbols.}
    \label{fig:mgfe_halo}
\end{figure*}

Measured stellar values of $\alpha$-elements such as Mg in the MW constitute another constraint to our model of the halo, in addition to the metallicity distribution (\S~\ref{subsubsec:dmd}). The stellar values of Mg are characterized by nearly constant ratios of [Mg/Fe] $\sim 0.4$ for [Fe/H] $< -1$ with small star-to-star scatter \citep[orange dots in Fig.~\ref{fig:mgfe_halo},][]{Suda2008,Suda2011}. This small scatter in [Mg/Fe] is particularly a salient feature when a well-selected homogeneous sample is taken \citep[within 0.2 dex,][]{Arnone2005}.

Fig.~\ref{fig:mgfe_halo} displays the evolution of [Mg/Fe] in the MW halo as a function of [Fe/H] for case 1 (left) and case 4 (right), where the number density of stars are colour-coded. Note that the results for cases 2 and 3 are the same as that for case 1. The evolutionary tracks for individual building-block galaxies are indicated by the curves with different colours (see the legend). For case 1, our model of the halo reproduces relatively a flat trend of [Mg/Fe] with a modest scatter, being qualitatively consistent with observation (see also Fig.~\ref{fig:appendix_mg}). This is due to the late contribution of SNe Ia starting from 1 Gyr. As a result, a knee (at which [Mg/Fe] starts decreasing) appears in each evolutionary track of a building block, e.g., at [Fe/H] $\sim -3$ and $\sim -1$ for $M_* = 10^3$ and $10^9$, respectively. The knees at different metallicities lead to 
a scatter of about 0.4 dex, while the different values of [Mg/Fe] from CCSNe make a scatter of about 0.3 dex for [Fe/H] $< -3$. Note that, however, the global trend of [Mg/Fe] is governed by the stars from relatively massive building blocks as anticipated from the right panel of Fig.~\ref{fig:dmd_halo}. Therefore, the scatter of [Mg/Fe] among the bulk of stars is expected to be smaller (see the red area in Fig.~\ref{fig:appendix_mg}). It should be noted that the predicted [Mg/Fe] is systematically higher than the measured values by about 0.3 dex, which might be attributed to uncertainties in the adopted  nucleosynthetic yields or to an IMF  with the most massive stars ending as black holes with no explosion. 

We find in the right panel of Fig.~\ref{fig:mgfe_halo} that case 4 appears to be inconsistent with observation, because [Mg/Fe] starts declining at too low [Fe/H] to be compatible with the observational trend (see also the blue area in Fig.~\ref{fig:appendix_mg}). This is due to the fact that the short delay ($t_\mathrm{min} = 0.1$ Gyr) leads to substantial number of SNe Ia coming into play in the GCE of building blocks. We consider, therefore, that the delay-time distribution of equation~(\ref{eq:dtdia}) with $t_\mathrm{min} \sim 1$ Gyr for SNe Ia is preferable because of the observational constraints on the metallicity distribution (\S~\ref{subsubsec:dmd}) and the evolution of Mg in the halo, at least in the framework of this model (see additional tests in APPENDIX \ref{sec:appendex}). 

\subsubsection{Evolution of Eu}
\label{subsubsec:eu_halo}

\begin{figure*}
	\includegraphics[width=0.86\columnwidth]{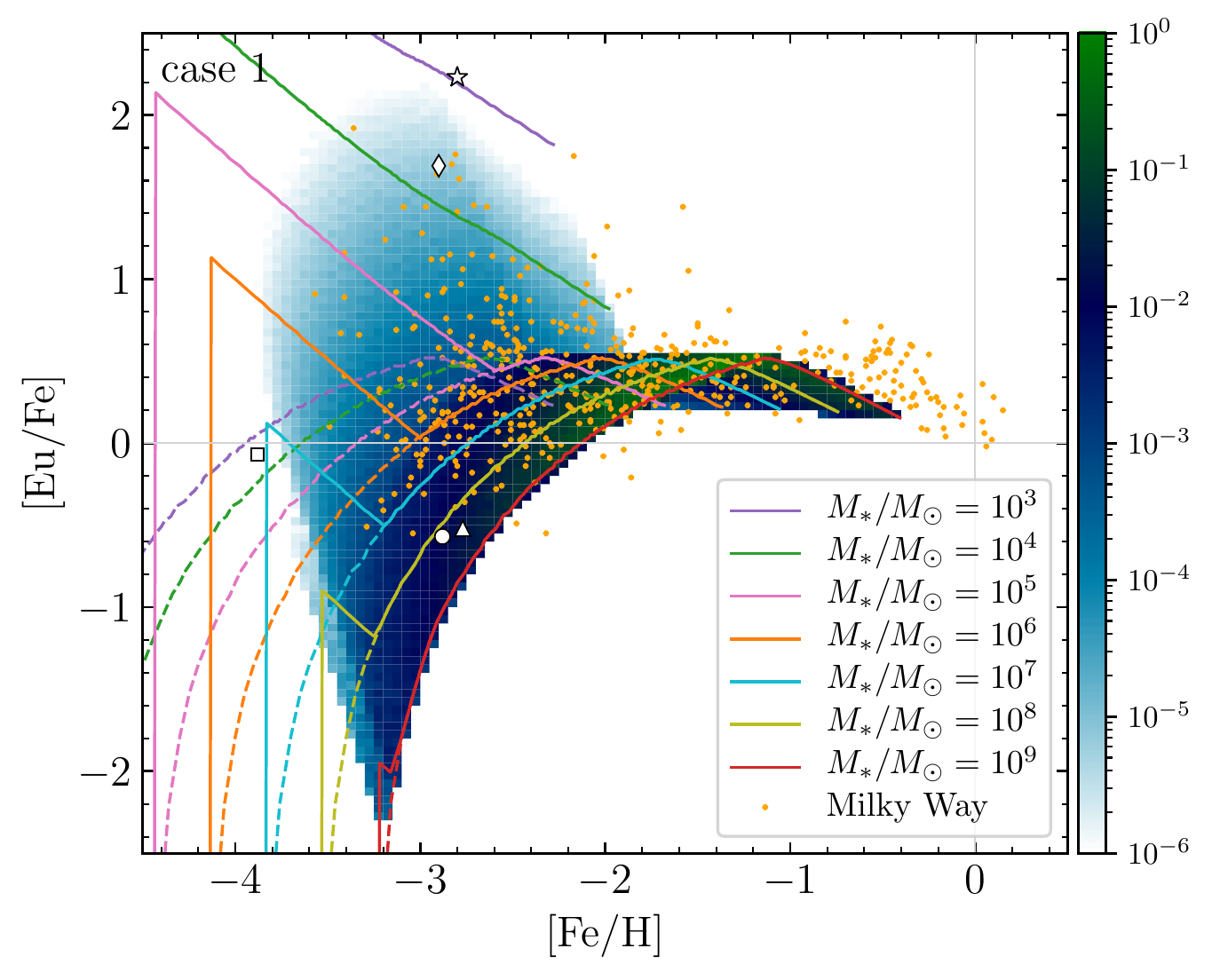}
	\includegraphics[width=0.86\columnwidth]{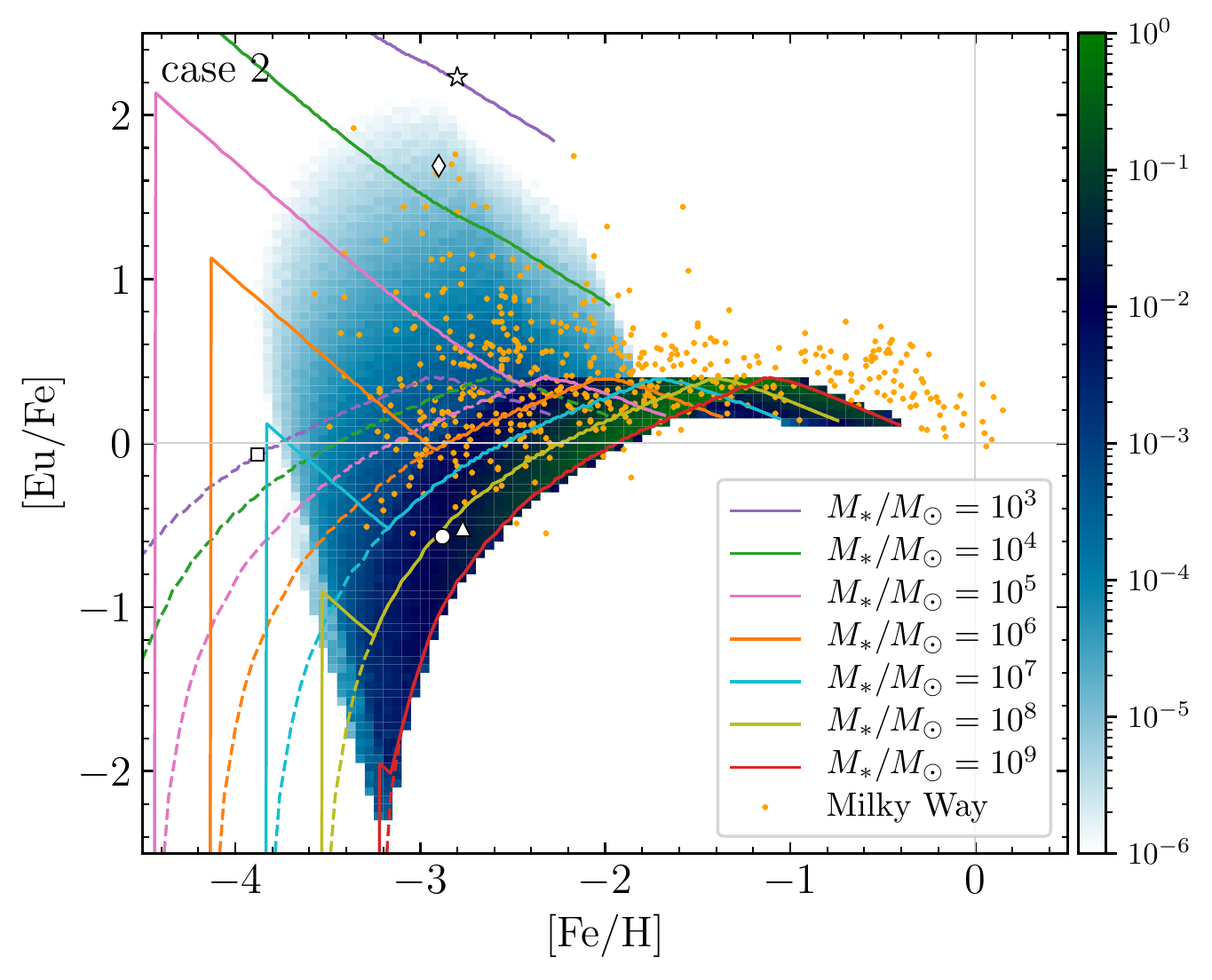}
	\includegraphics[width=0.86\columnwidth]{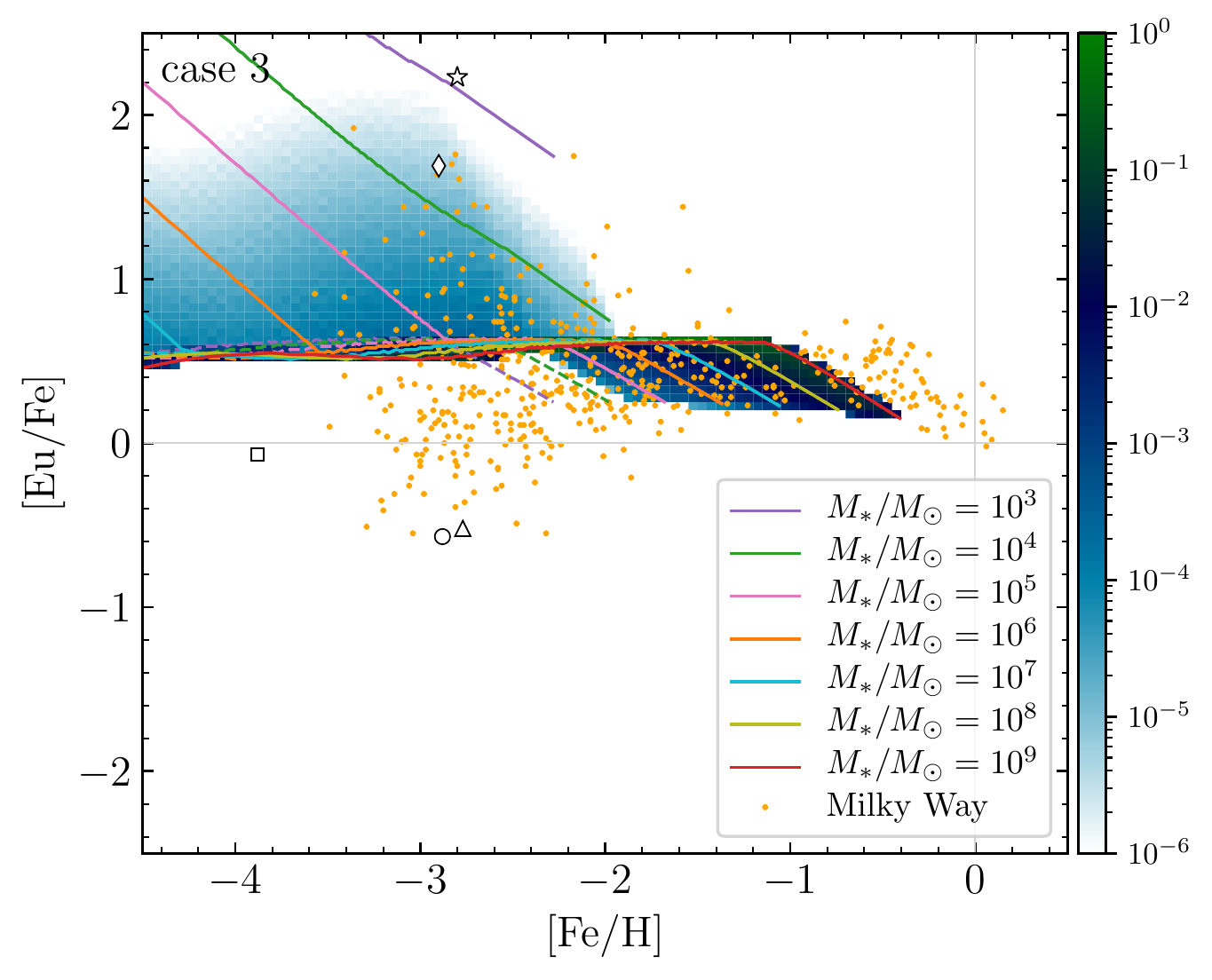}
	\includegraphics[width=0.86\columnwidth]{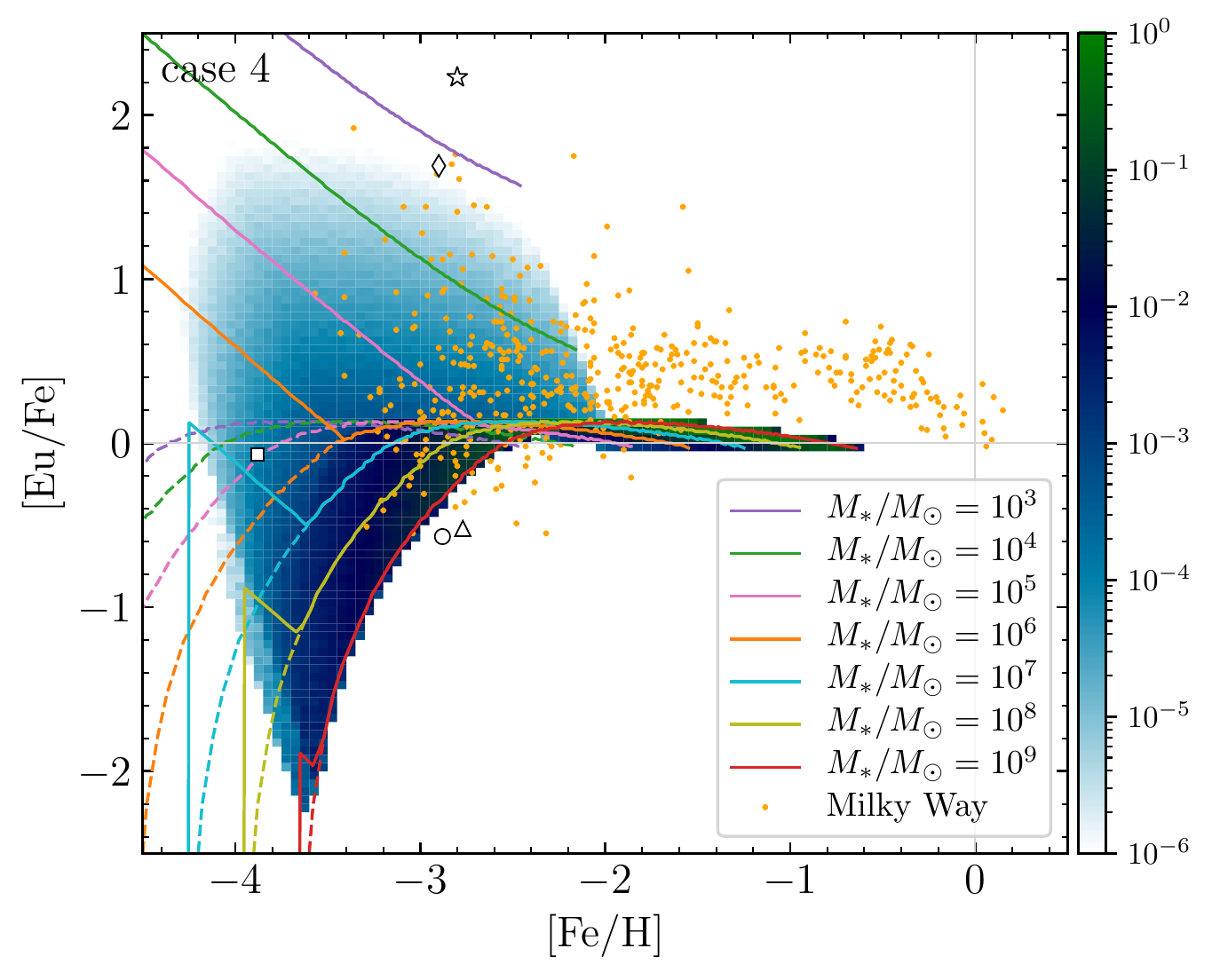}
    \caption{Same as Fig.~\ref{fig:mgfe_halo}, but for [Eu/Fe]. For each building-block galaxy with $M_*$ specified in the legend, the dotted curve indicates the evolutionary track for $\langle N_\mathrm{NSM} \rangle < 1$, while the solid curve shows the track when Eu/Fe is divided by $\langle N_\mathrm{NSM} \rangle$ for $\langle N_\mathrm{NSM} \rangle < 1$ (see the text). For the MW stars \citep[orange dots,][]{Suda2008,Suda2011}, carbon stars with [C/Fe] $> 1$ are excluded, which may be contaminated by the binary mass transfer from former asymptotic-giant-branch stars. Symbols indicate some of r-process-enhanced (diamond, CS 31082-001, \citealt{Hill2002};  star, J1521-3538, \citealt{Cain2020}) and r-process-deficient (triangle, HD 122563, \citealt{Honda2004}; circle, HD 4306, \citealt{Ishimaru2004}) stars. Note that J1521-3538 and HD 4306 are the stars with the highest and lowest [Eu/Fe] among available data, respectively. The square indicates the lowest-metallicity star with measured Eu \citep[CS 22891-200,][]{Roederer2014}.}
    \label{fig:eufe_halo}
\end{figure*}

We first focus on the result for (our fiducial) case 1, which is displayed in the top-left panel of Fig.~\ref{fig:eufe_halo}, and then compare with other cases. For each building-block galaxy with $M_*$ specified in the legend, the dotted line is the evolutionary track when the computed (average) number of NSMs, $\langle N_\mathrm{NSM} \rangle$, is less than 1. For instance, the building block of $M_* = 10^4$ reaches $\langle N_\mathrm{NSM} \rangle \approx 0.1$ at [Fe/H] $\approx -3$. This indicates that about one out of ten building blocks experiences a NSM. If a NSM occurs at this metallicity, the ratio Eu/Fe will become $1/\langle N_\mathrm{NSM} \rangle = 10$ times higher than the computed (averaged) value, while in the other nine building blocks no Eu will be produced. Here, we neglect the case of two or more mergers for $\langle N_\mathrm{NSM} \rangle < 1$ because of its small probability. In this way, the expected value of [Eu/Fe] when a single NSM occurs (for $\langle N_\mathrm{NSM} \rangle < 1$) is drawn by the solid line for each building block. For $\langle N_\mathrm{NSM} \rangle \ge 1$, the computed values of [Eu/Fe] are taken and thus the solid and dotted curves overlap.

For case 1, we find that most of the stellar values of the MW halo ([Fe/H] $< -1$) reside between the evolutionary tracks explored here and also in the coloured region. Note that the carbon stars are excluded in Fig.~\ref{fig:eufe_halo}, since they might have been contaminated by the binary mass transfer from former asymptotic-giant-branch stars. The smaller SFE for a less-massive galaxy leads to an increase of [Eu/Fe] at a lower [Fe/H]. It is notable that our model predicts that almost all highly r-process-enhanced stars ([Eu/Fe] $> 1$ at [Fe/H] $\sim -3$) originate from UFD-sized ($M_* < 10^5$) building blocks (but see a caution in \S~\ref{sec:discussion} for the applicability of the mass-metallicity relation of \citet{Kirby2013} to UFD-sized systems), as also suggested in \citet{Ishimaru2015,Ojima2018}. For instance, our model suggests that J1521-3538 \citep[star, the highest measured Eu/Fe,][]{Cain2020} and CS 31082-001 \citep[diamond,][]{Hill2002} were born in  building-block galaxies of $M_* \sim 10^3$ and $5\times 10^3$ (in $M_\odot$), respectively. This is a consequence of the fact that a less-massive galaxy contains a smaller amount of gas to be mixed with Eu from a NSM. In fact, the highly eccentric orbit of the former indicates that J1521-3538 originates from a building-block galaxy subsequently accreted by the MW \citep{Cain2020}. In contrast, the r-process-deficient stars may have been born in the most-massive building-block galaxies of $M_* \sim 10^8$  $M_\odot$, like HD 4306 \citep[circle, the lowest measured Eu/Fe,][]{Ishimaru2004} and HD 122563 \citep[triangle,][]{Honda2006}. Note that our model predicts the presence of even further r-process-deficient stars with [Eu/Fe] $< -1$ at [Fe/H] $\lesssim -3$. The absence of such stars with measured Eu is likely due to the current detection limit for Eu \citep{Ishimaru2004}. Overall, the result for case 1 (a mean delay of 0.1 Gyr in the log-normal term for NSMs) appears in qualitative agreement with those in \citet{Ishimaru2015,Ojima2018} with a fixed delay of 0.1 Gyr (see APPENDIX \ref{sec:appendex} for additional tests). 

The top-right panel of Fig.~\ref{fig:eufe_halo} shows the result for case 2, in which the delay-time distribution of NSMs is $\propto t^{-1}$. We find a slower increase of [Eu/Fe] with [Fe/H] and a smaller [Eu/Fe] at high [Fe/H] than those for case 1. This is due to an absence of the early (log-normal) component in the delay-time distribution. As a whole, however, the evolution of [Eu/Fe] including its scatter is similar between cases 1 and 2. 

The bottom-left panel of Fig.~\ref{fig:eufe_halo} shows the result for case 3, in which a CCSN-like, little delay time for NSMs is assumed (Fig.~\ref{fig:dtd}). We find that the evolution of [Eu/Fe] with [Fe/H] appears incompatible with the measured stellar abundance. The model predicts stars with [Fe/H] $< -3.5$, at which few stars with measured Eu have been found. Note this cannot be attributed to the detection limit for Eu, because the predicted stars have high [Eu/Fe] ($> 0.5$). It should also be noted that the reason cannot be  the rarity of stars with [Fe/H] $< -3.5$. The building blocks more massive than $M_* = 10^6$ (orange curve in the bottom-left panel of Fig.~\ref{fig:eufe_halo}), which generate the bulk of stars in the halo (the right panel of Fig.~\ref{fig:dmd_halo}), have already experienced NSM events ($\langle N_\mathrm{NSM} \rangle > 1$) at  [Fe/H] $< -3.5$. Moreover, the model does not predict stars with [Eu/Fe] $< 0$ at [Fe/H] $\sim -3$, at which a large number of stars with measured Eu exist (see also \S~\ref{subsec:massive} and APPENDIX \ref{sec:appendex}).

The result for case 4 is shown in the bottom-right panel of Fig.~\ref{fig:eufe_halo}. The smaller SFE ($K$ in Table~\ref{tab:dtd}) results in the appearance of Eu at too low metallicity ([Fe/H] $\sim -4$) to be compatible with the observation. However, the lowest metallicity at which Eu starts increasing depends on $t_\mathrm{min}$ in the delay-time distribution of NSMs (equation~(\ref{eq:dtdnsm}); see APPENDIX \ref{sec:appendex}). Another problem is that the contribution of SNe Ia starting at low metallicity inhibits the increase of Eu above [Eu/Fe] $\sim 0$. It should be noted, however, that the yield of Eu from NSMs and their event rate, the product of which determines the average [Eu/Fe], are somewhat uncertain. 

Overall, therefore, our model is (at least marginally) compatible with the observational trend of [Eu/Fe] in the MW halo, regardless of the explored variations in the delay-time distributions (except for case 3). Note that a reduction of the NSM (or rare CCSN-like) event rate (or $B$ in \S~\ref{subsec:yield}), which also reduces $\langle N_\mathrm{NSM} \rangle$, will enhance [Eu/Fe] at low metallicity in the low-mass building blocks with $\langle N_\mathrm{NSM} \rangle < 1$. However, this will not substantially change the probabilistic distribution of stars on the [Eu/Fe]--[Fe/H] plane (colour scale in Fig.~\ref{fig:eufe_halo}) as presented in APPENDIX~\ref{sec:appendex} (Fig.~\ref{fig:appendix_freq}) because of the resulting smaller number of such building blocks experiencing a NSM (or a rare CCSN-like event).


\subsection{Satellite dwarf galaxies}
\label{subsec:dwarf}

\begin{table*}
	\centering
	\caption{Reference dwarf galaxies (first column). The present-day stellar masses $M_*$ (second column) are adopted from \citep{McConnachie2012}, except for Reticulum II. For Reticulum II, $M_*$ is taken from \citet{Bechtol2015}. The SFE ($k_\mathrm{SF}$) of each galaxy (third and fourth columns for case 1 and case 4, respectively) is obtained from equation~(\ref{eq:sfof}) with $M_*$ and $K$ in Table~\ref{tab:dtd}. The initial gas mass of each galaxy, $M_0 = M_\mathrm{gas}(0)$ (fifth column), is obtained from equation~(\ref{eq:fgas}). The resultant metallicity at which the metallicity distribution peaks, [Fe/H]$_\mathrm{peak}$, for each reference galaxy is presented in the last column. Note that the values for cases 2 and 3 are the same as those for case 1.}
	\label{tab:dwarf}
	\begin{tabular}{lcccccc} 
		\hline
		Reference galaxy & $M_*$ & $k_\mathrm{SF}$ (case 1) & $k_\mathrm{SF}$ (case 4) & $M_0$ (case 1) & $M_0$ (case 4) & [Fe/H]$_\mathrm{peak}$ \\
		\hline
		Reticulum II  & $2.6\times 10^3$    & 0.0075 &  0.0029 & $1.0\times 10^6$ &  $1.4\times 10^6$ & $-2.9$ \\
		Ursa Minor    & $2.9\times 10^5$    & 0.031  &  0.012  & $2.8\times 10^7$ &  $3.7\times 10^7$ & $-2.3$ \\
		Sculptor      & $2.3\times 10^6$    & 0.058  &  0.022  & $1.2\times 10^8$ &  $1.6\times 10^8$ & $-2.0$ \\
		Fornax        & $2.0\times 10^7$    & 0.11   &  0.042  & $5.8\times 10^8$ &  $7.5\times 10^8$ & $-1.8$ \\
		\hline
	\end{tabular}
\end{table*}

In \S~\ref{subsec:halo}, we have modeled the MW halo as an ensemble of building-block galaxies that follow the stellar mass-metallicity relation in the Local Group \citep{Kirby2013}. It is a reasonable assumption, therefore, that the same relation in equation~(\ref{eq:sfof}) may be applied for the satellite dwarf galaxies (if not all), given those being surviving building blocks \citep[see][for such attempts]{Prantzos2008,Kirby2013}. For this reason, we use equation~(\ref{eq:sfof}) with the value of $K$ in Table~\ref{tab:dtd} to determine the SFE for each dwarf galaxy. The evolution of a given galaxy is computed as a single well-mixed system with gas outflow until $f_\mathrm{gas}(t) < 0.003$ is reached. As in \S~\ref{subsec:halo}, the OFE is fixed to be $k_\mathrm{OF} = 1.0$ (Gyr$^{-1}$).

In this study, we select a UFD Reticulum II \citep[$2.6\times 10^3\, M_\odot$,][]{Bechtol2015} and three dwarf spheroidals Ursa Minor, Sculptor and Fornax \citep[$2.9\times 10^5\, M_\odot$, $2.3\times 10^6\, M_\odot$ and $2.0\times 10^7\, M_\odot$, respectively,][]{McConnachie2012} as reference galaxies (Table~\ref{tab:dwarf}). What we refer for a given galaxy is only the present-day stellar mass $M_*$, which specifies the SFE ($k_\mathrm{SF}$) for each case according to equation~(\ref{eq:sfof}) as presented in Table~\ref{tab:dwarf}. In this sense, our model for the satellite dwarfs is parameter free. Nevertheless, the resulting metallicities at the peak of metallicity distribution, [Fe/H]$_\mathrm{peak}$ (last column in Table~\ref{tab:dwarf}), are in reasonable agreement with the mean metallicities estimated for these galaxies (except for Fornax): $\sim -2.6$ for Reticulum II \citep{Koposov2015,Simon2015,Walker2015}, $\sim -2.1$ for Ursa Minor, $\sim -1.7$ for Sculptor and $\sim -1.0$ for Fornax \citep{McConnachie2012}. For Sculptor, a photometric analysis indicates [Fe/H]$_\mathrm{peak} \sim -2.0$ \citep{deBoer2012}, being in accordance with our result. Note that the values of [Fe/H]$_\mathrm{peak}$ for cases 1 and 4 are approximately the same, because the effective SFE ($K$) has been adjusted to obtain the peak of metallicity distribution for the halo at the same metallicity. The initial gas mass, $M_0 = M_\mathrm{gas}(0)$, for a given galaxy estimated from equation~(\ref{eq:fgas}) is also presented in Table~\ref{tab:dwarf} for case 1 (fifth column) and case 4 (sixth column). Case 4 requires a larger amount of $M_0$ because of its smaller $k_\mathrm{SF}$. The outcomes for cases 2 and 3 are the same as those for case 1 in Table~\ref{tab:dwarf}.

It is emphasized that our purpose in this subsection is to test how our simple model, which is consistent with that of the MW halo, can (or cannot) represent the GCE of satellite dwarf galaxies. Our model may be too simplistic to describe the GCE of galaxies that have experienced episodic star formation and mass exchanges among those over a longer period of time. Moreover, the formation  and evolution of UFDs are poorly understood. This is still useful, however, for deeper understanding of the results in \S~\ref{subsec:halo} by analysing the GCE of single building-block-analogous galaxies. We refer the reader to the elaborate GCE studies of the r-process elements in classical dwarf spheroidals \citep{Hirai2015,Hirai2017,Hirai2017b} and UFDs \citep{Safarzadeh2017,Tarumi2020}. 



\subsubsection{Contributions of CCSNe, SNe Ia and NSMs}
\label{subsubsec:mht}

\begin{figure}
	\includegraphics[width=0.86\columnwidth]{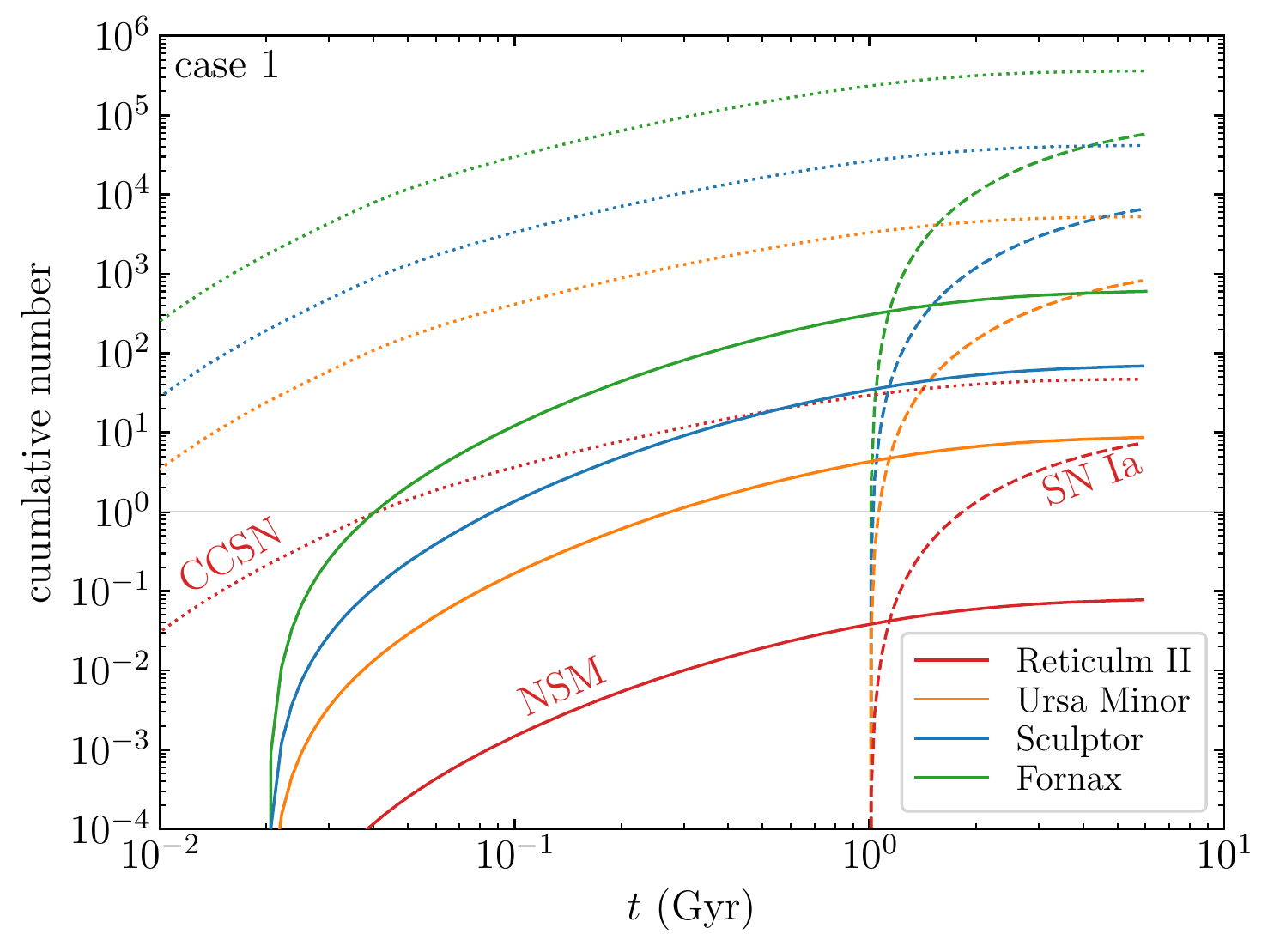}
    \caption{Cumulative numbers of NSMs (solid curves), SNe Ia (dashed curves) and CCSNe (dotted curves) for case 1. The colours indicate different reference galaxies specified in the legend. The gray horizontal line marks the cumulative number of unity.}
    \label{fig:num}
\end{figure}

We first inspect the contributions of CCSNe, SNe Ia and NSMs in GCE, which helps understand the subsequent results. Fig.~\ref{fig:num} shows the cumulative numbers of CCSNe (dotted curves), SNe Ia (dashed curves) and NSMs (solid curves) for case 1. 
Among the reference galaxies, the (average) cumulative number of NSMs, $\langle N_\mathrm{NSM} \rangle$, in Reticulum II (red curve) is below 1 all the way, reaching $\langle N_\mathrm{NSM} \rangle \sim 0.1$ at the end of evolution. This indicates that about 10 percent of similar-mass galaxies exhibit enhancement of Eu, being in reasonable agreement with the discovery of three r-process-enriched galaxies \citep{Ji2016,Roederer2016,Hansen2017,Hansen2020} out of 15 UFDs with detailed abundance measurements \citep[about 20 percent,][]{Simon2019}. All our modelled galaxies evolve up to about 6 Gyr, at which $f_\mathrm{gas}(t) < 0.003$ is reached. For a majority of UFDs, their star formation might have been terminated by reionization and thus within 1 Gyr \citep{Brown2014}. It is suggested, however, that some of UFDs resume star formation \citep{Weisz2014,Applebaum2021,Miyoshi2020}.

\begin{figure}
	\includegraphics[width=0.86\columnwidth]{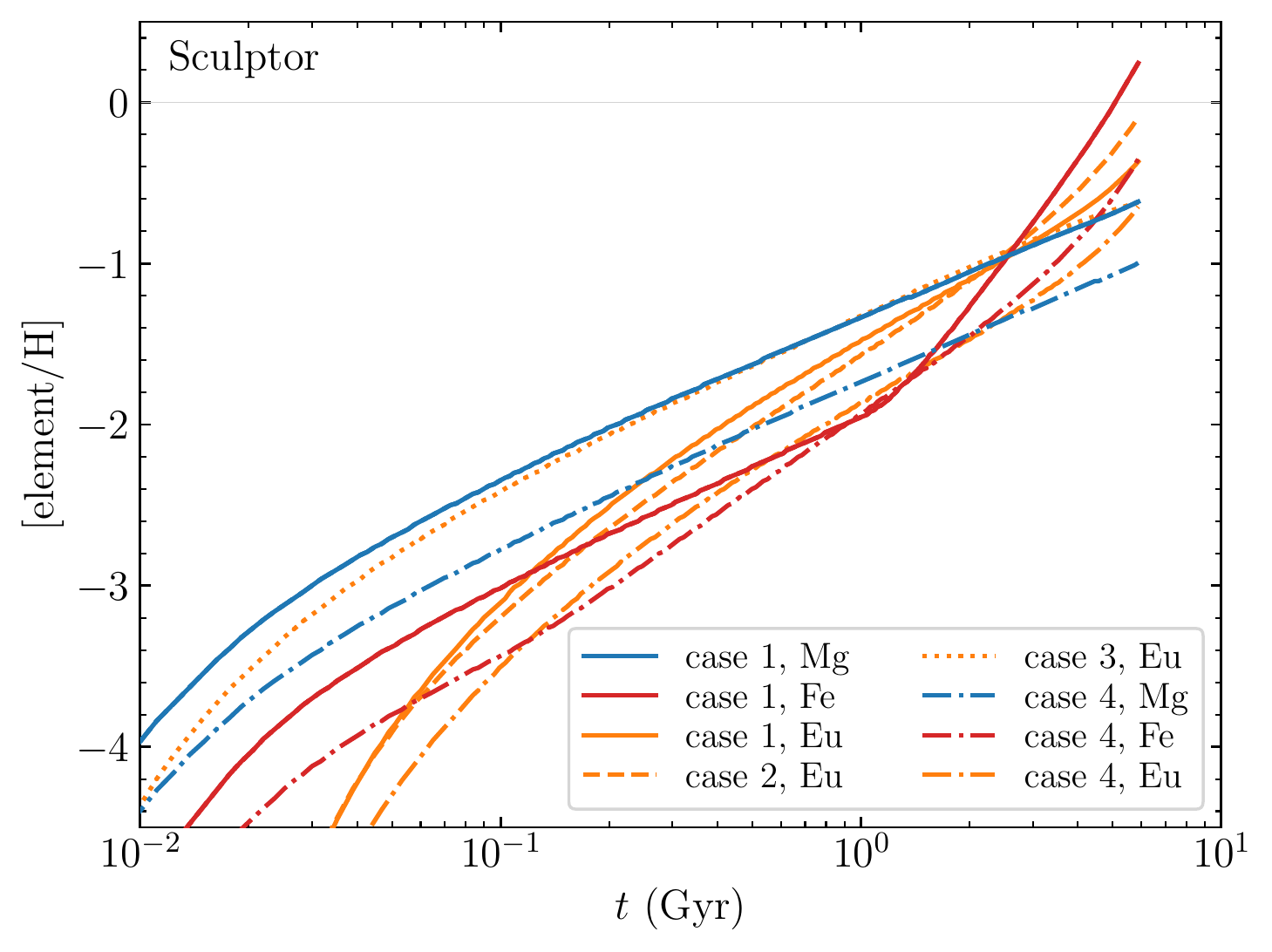}
    \caption{Temporal evolution of [Mg/H] (blue), [Fe/H] (red) and [Eu/H] (orange) for case 1 (solid), case 2 (dashed), case 3 (dotted) and case 4 (dash-dotted). Here, Sculptor is taken to be representative of reference dwarf galaxies. Note that the curves of Mg and Fe for cases 2 and 3 are the same as those for case 1.}
    \label{fig:mht}
\end{figure}

As can be seen in Fig.~\ref{fig:num}, CCSNe, SNe Ia and NSMs appear in the order of delay times from star formation. This is reflected in the temporal evolution of [Mg/H] (blue), [Fe/H] (red) and [Eu/H] (orange) as shown in Fig.~\ref{fig:mht}. Here, Sculptor is taken to be representative of reference galaxies. We find that the different forms of the delay-time distributions (cases 1--4 with different line styles) lead to a variation of the slopes as a function of time for these elements.

\subsubsection{Evolution of Mg}
\label{subsubsec:mg_dwarf}

\begin{figure*}
	\includegraphics[width=0.86\columnwidth]{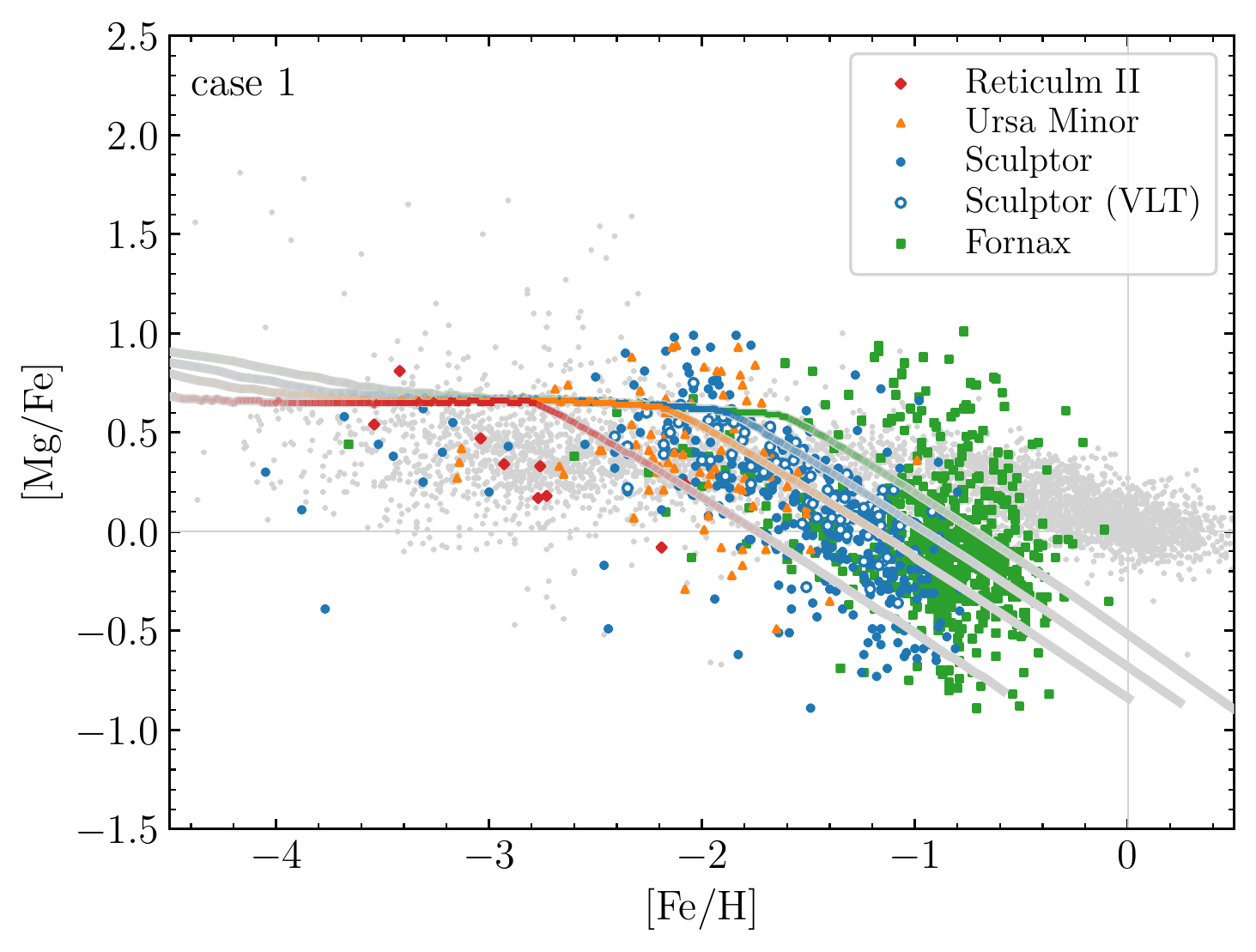}
	\includegraphics[width=0.86\columnwidth]{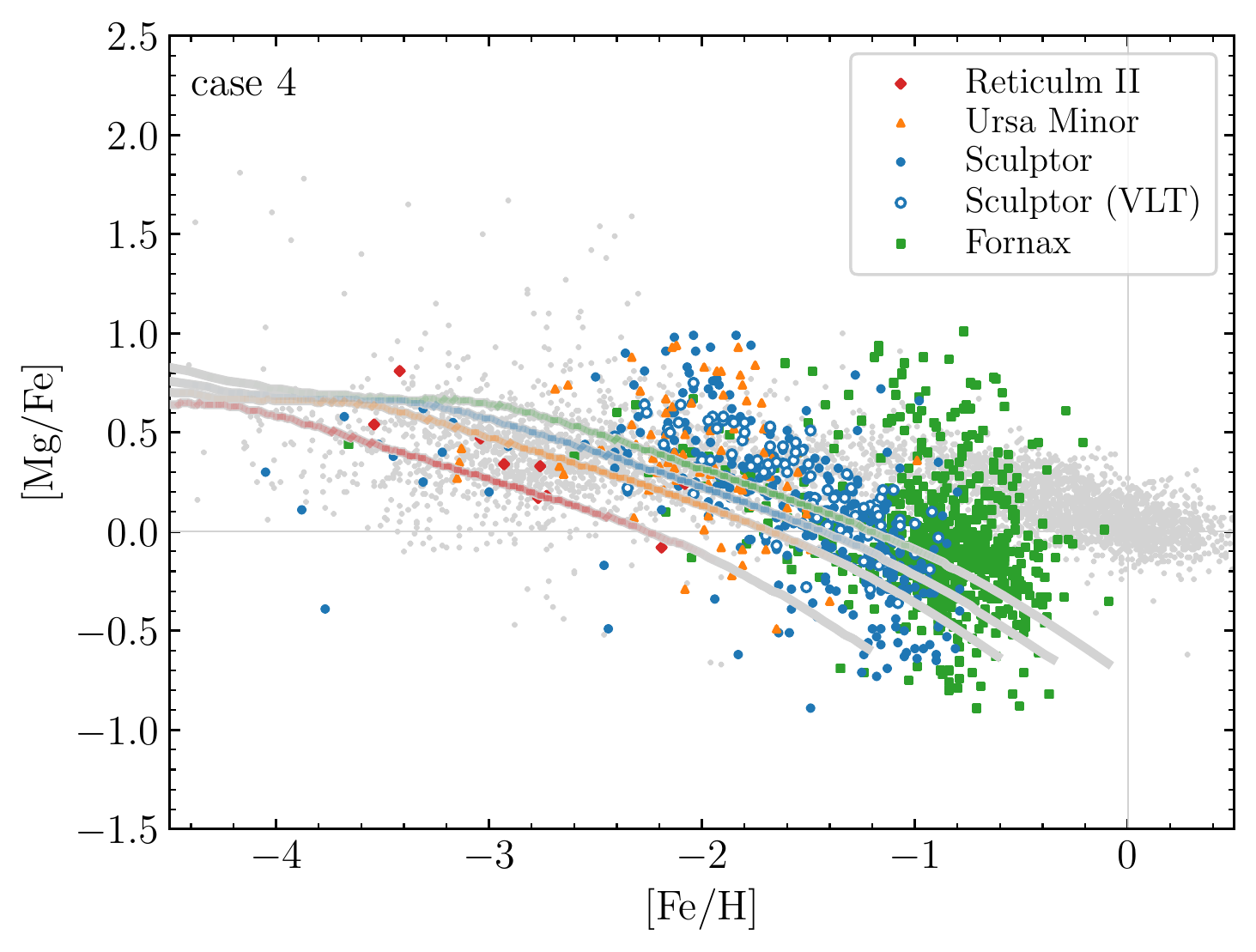}
    \caption{[Mg/Fe] as a function of [Fe/H] in satellite dwarf galaxies for case 1 (left) and case 4 (right). The results for cases 2 and 3 are the same as that for case 1. The evolutionary track of each galaxy is drawn by a gray curve with the (transparent) colour of the reference specified in the legend. The thickness of colour is proportional to the number of stars formed in each bin, i.e., $df_\mathrm{star}/d$[Fe/H]. The coloured symbols show the measured values for the galaxies specified in the legend, which are taken from the SAGA data base \citep{Suda2008,Suda2011}. For Sculptor, the values measured by VLT \citep{Skuladottir2019} are overplotted by blue open circles. The gray dots are the measured values in the MW \citep{Suda2008,Suda2011}.}
    \label{fig:mgfe}
\end{figure*}

Fig.~\ref{fig:mgfe} displays the evolution of [Mg/Fe] as a function of [Fe/H] in dwarf galaxies for case 1 (left) and case 4 (right). Note that the results for cases 2 and 3 are the same as that for case 1. The evolutionary track of each galaxy is drawn by a gray curve with the (transparent) colour of the reference in the legend. The measured stellar values are plotted by coloured symbols specified in the legend. For Sculptor, the values measured by VLT (\citealt{Skuladottir2019}; see also \citealt{Hill2019}) are overplotted by blue open circles.

It has been known that the knee position of [$\alpha$/Fe] varies depending on the luminosity (and thus stellar mass) of dwarf galaxies \citep{Tolstoy2009,Reichert2020}. In fact, our model predicts the knee position of [Mg/Fe] at lower metallicity for a less-massive galaxy, in particular for case 1, as also found in the reference galaxies. This is due to the smaller $k_\mathrm{SF}$ for a less-massive galaxy, which is determined by equation~(\ref{eq:sfof}). Note that we assume a constant OFE, $k_\mathrm{OF} = 1.0$ (Gyr$^{-1}$), in equation~(\ref{eq:sfof}). We also can obtain a similar metallicity distribution of the MW halo when $k_\mathrm{SF}$ is fixed to a constant value. However, a constant $k_\mathrm{SF}$ leads to the same knee position for the galaxies with different stellar masses \citep[][see also APPENDIX~\ref{sec:appendex}]{Ishimaru2015}. We consider, therefore, a constant $k_\mathrm{OF}$ in equation~(\ref{eq:sfof}) is a reasonable assumption to appreciably differentiate the values of $k_\mathrm{SF}$.

It is, however, non-trivial to specify the exact position of the knee from the measured abundances in dwarf galaxies. For instance, Fornax exhibits large star-to-star scatter in [Mg/Fe], which could be due to difficulties in measurements for a distant galaxy. This galaxy also is suggested to have experienced multiple (or prolonged) star formation episodes \citep{deBoer2012b,Hendricks2014}. If this is the case, our model may be too simple to describe the evolution of Fornax. Conversely, Sculptor is suggested to have had a simple star formation history \citep{deBoer2012,Hill2019}, which may justify a use of our simple model to examine its GCE. In particular, this galaxy exhibits small scatter in [Mg/Fe] among the values measured by VLT \citep{Hill2019,Skuladottir2019} with the knee position of [Fe/H] $\sim -2$. If we restrict ourselves to a comparison of our model with the VLT data for Sculptor, the result for case 1 with a long delay ($t_\mathrm{min} = 1.0$ Gyr) of SNe Ia is in good agreement with the stellar values. This is in line with the conclusion obtained from the analysis of star formation history in Sculptor \citep[$2\pm 1$ Gyr,][]{deBoer2012}. On the contrary, the knee position with $t_\mathrm{min} \ll 1$ Gyr (at [Fe/H] $\sim -3$ for case 4, $t_\mathrm{min} = 0.1$ Gyr) appears incompatible with that for Sculptor.

\subsubsection{Evolution of Eu}
\label{subsubsec:eu_dwarf}

\begin{figure*}
	\includegraphics[width=0.86\columnwidth]{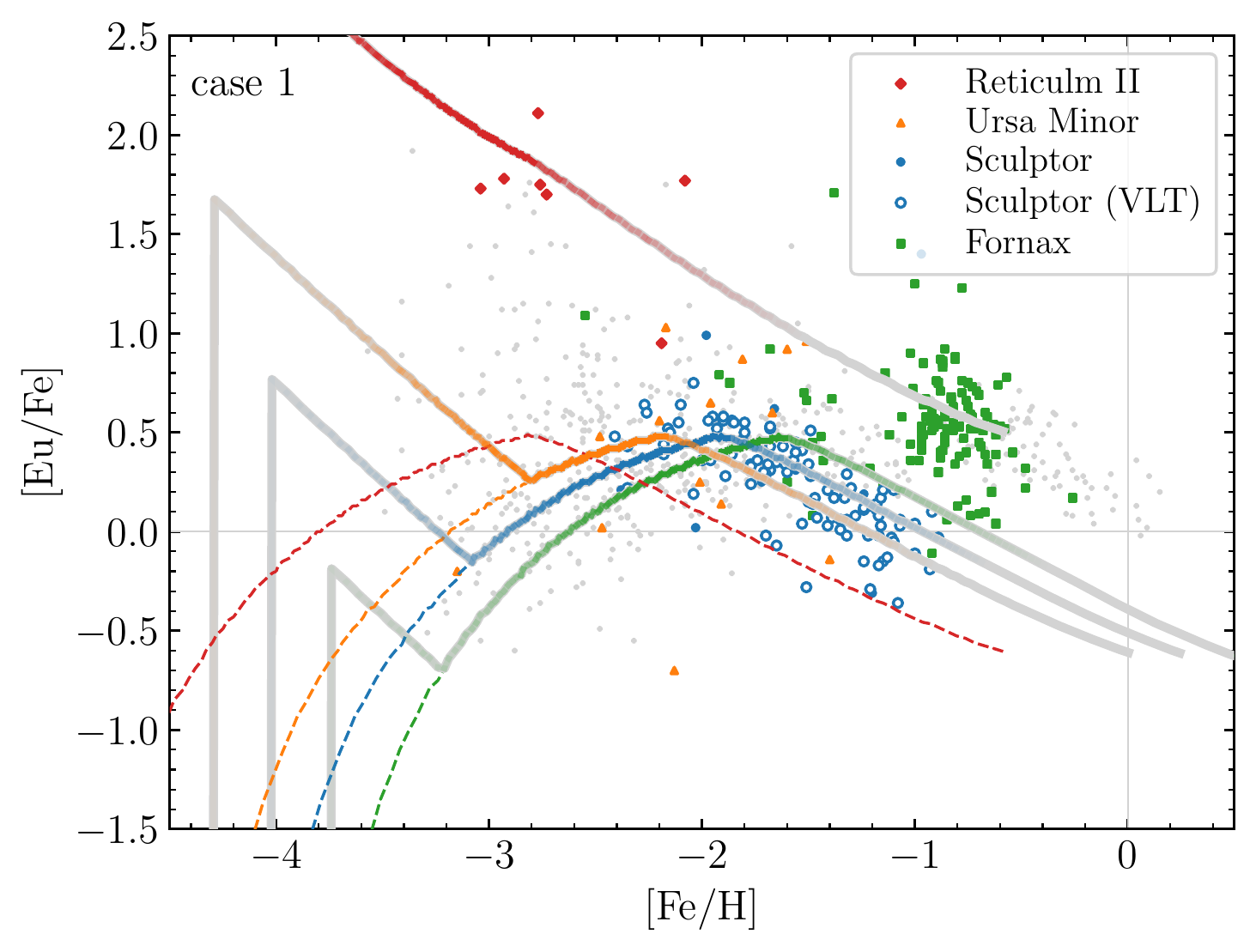}
	\includegraphics[width=0.86\columnwidth]{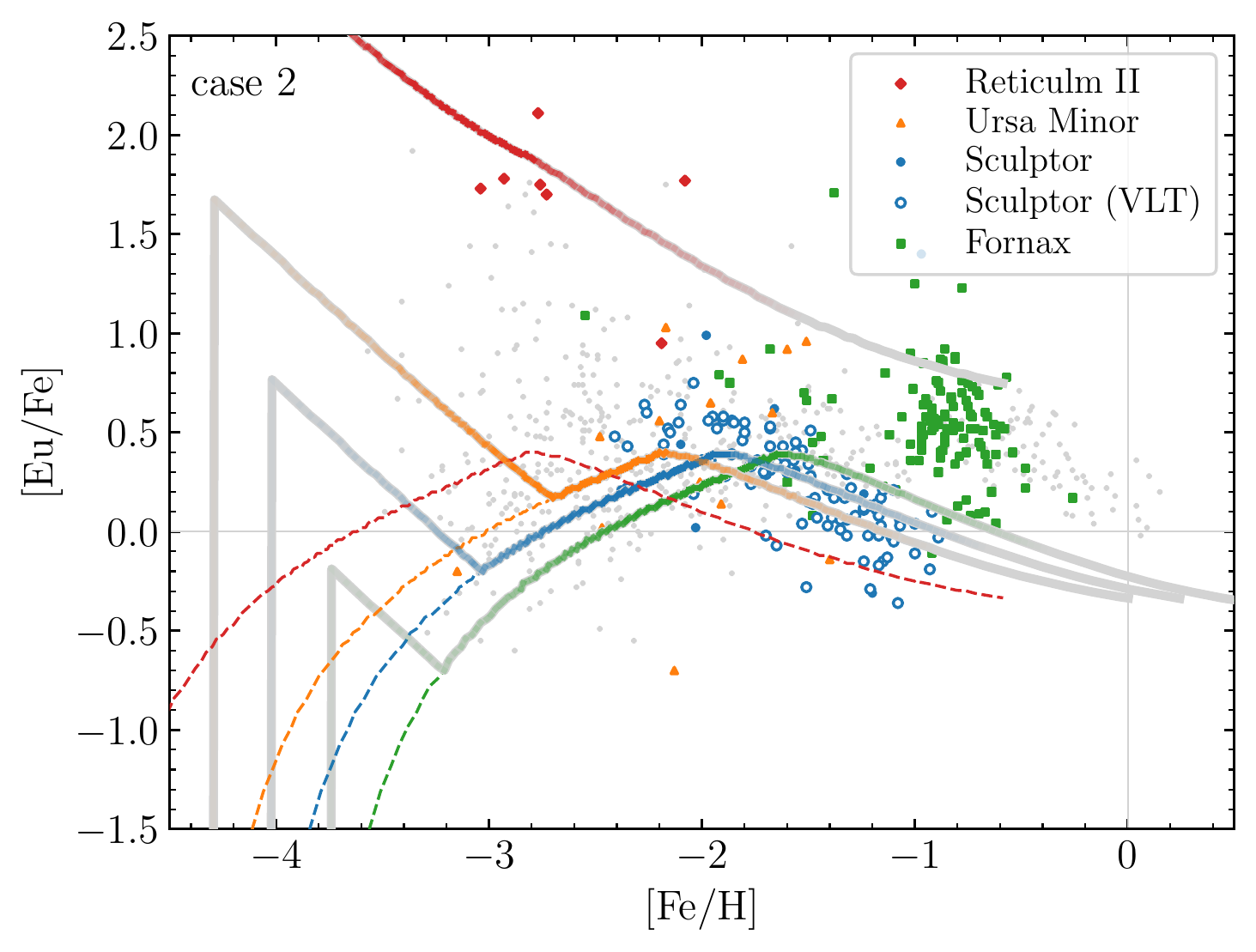}
	\includegraphics[width=0.86\columnwidth]{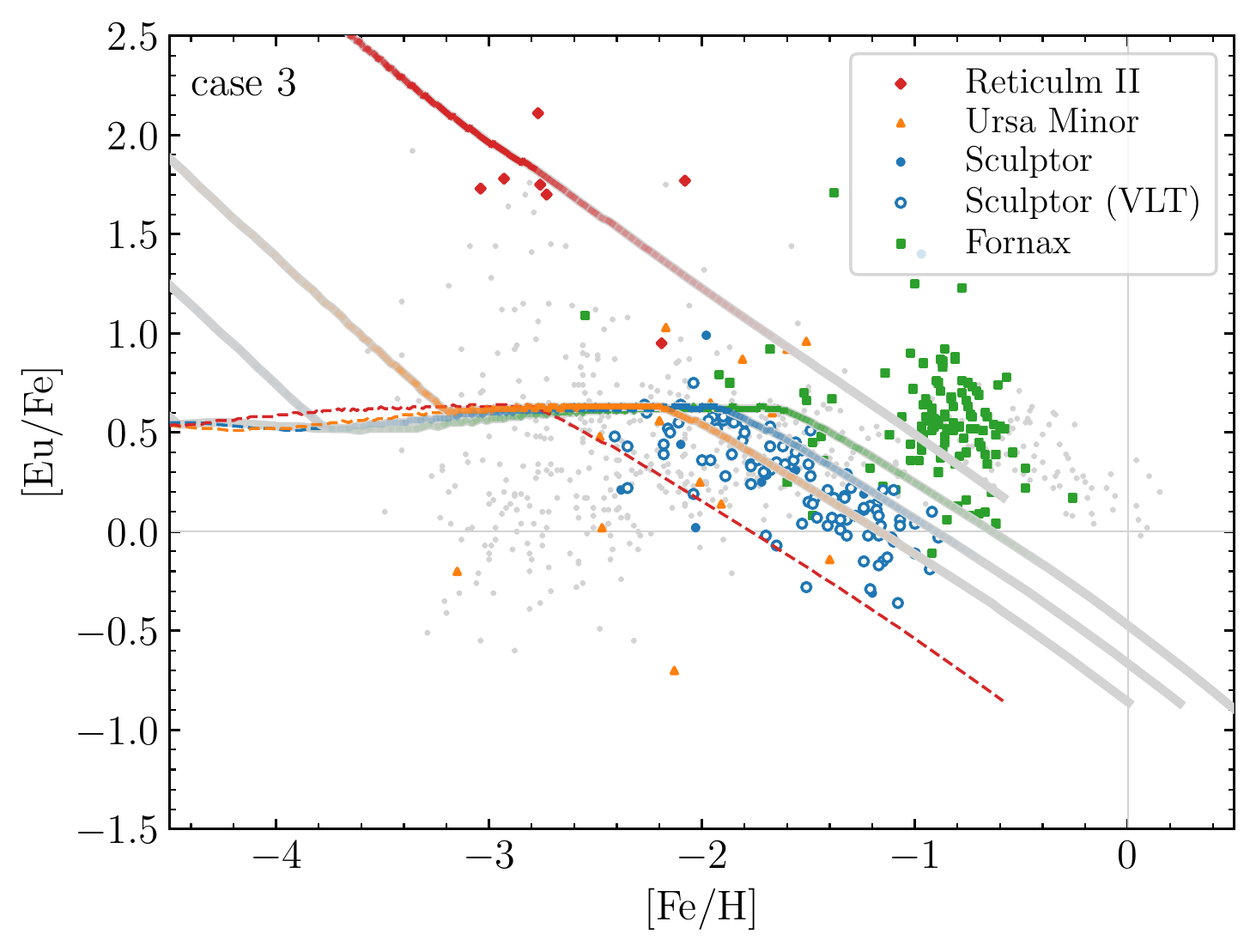}
	\includegraphics[width=0.86\columnwidth]{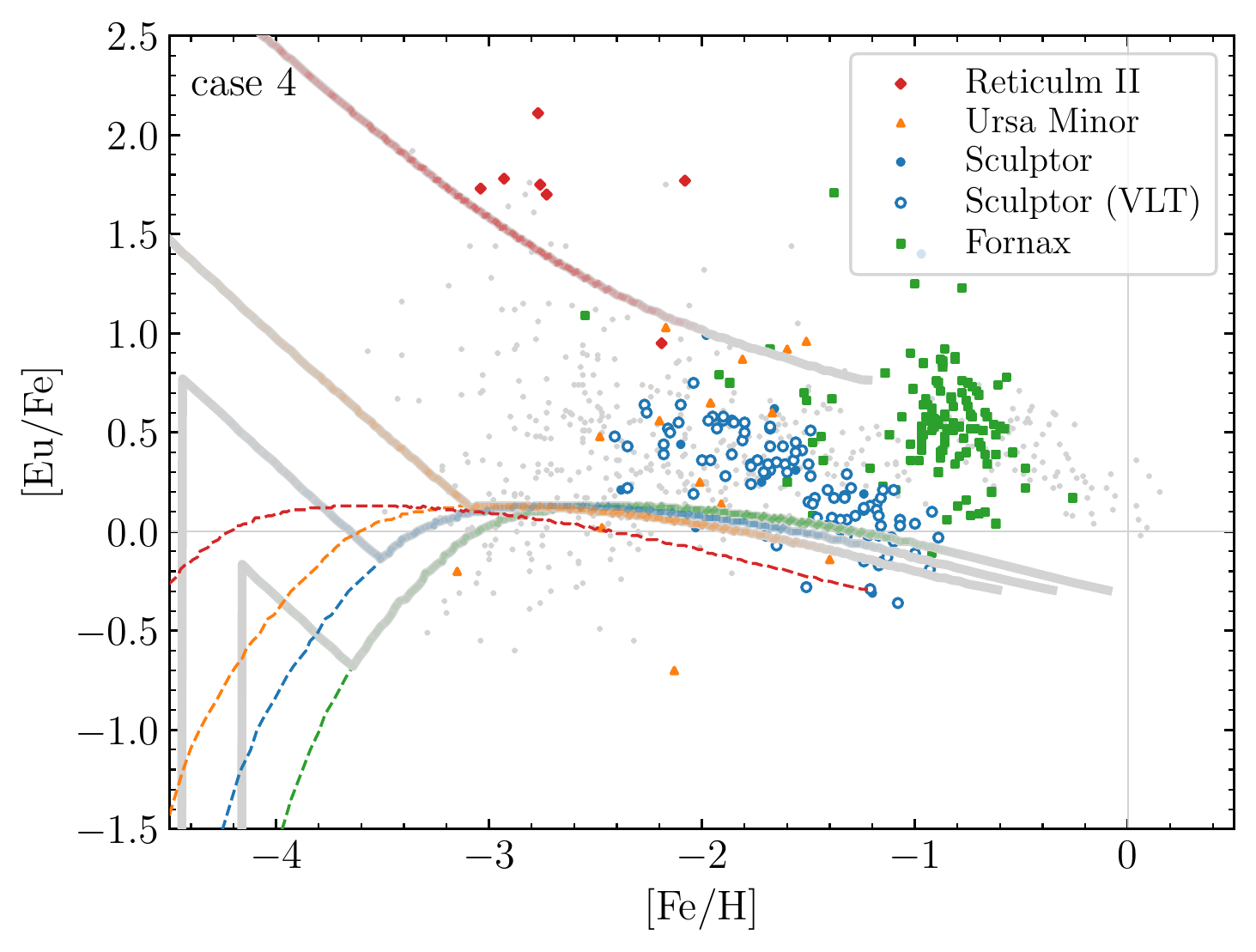}
    \caption{Same as Fig.~\ref{fig:mgfe}, but for [Eu/Fe]. For each reference galaxy, the dashed curve with the colour specified in the legend indicates the evolutionary track  for $\langle N_\mathrm{NSM} \rangle < 1$, while the gray curve with the (transparent) colour shows the track when Eu/Fe is divided by $\langle N_\mathrm{NSM} \rangle$ for $\langle N_\mathrm{NSM} \rangle < 1$ (see the text). For the MW stars \citep[gray dots,][]{Suda2008,Suda2011}, carbon stars with [C/Fe] $> 1$ are excluded, which may be contaminated by the binary mass transfer from former asymptotic-giant-branch stars.}
    \label{fig:eufe}
\end{figure*}

\begin{figure*}
	\includegraphics[width=0.86\columnwidth]{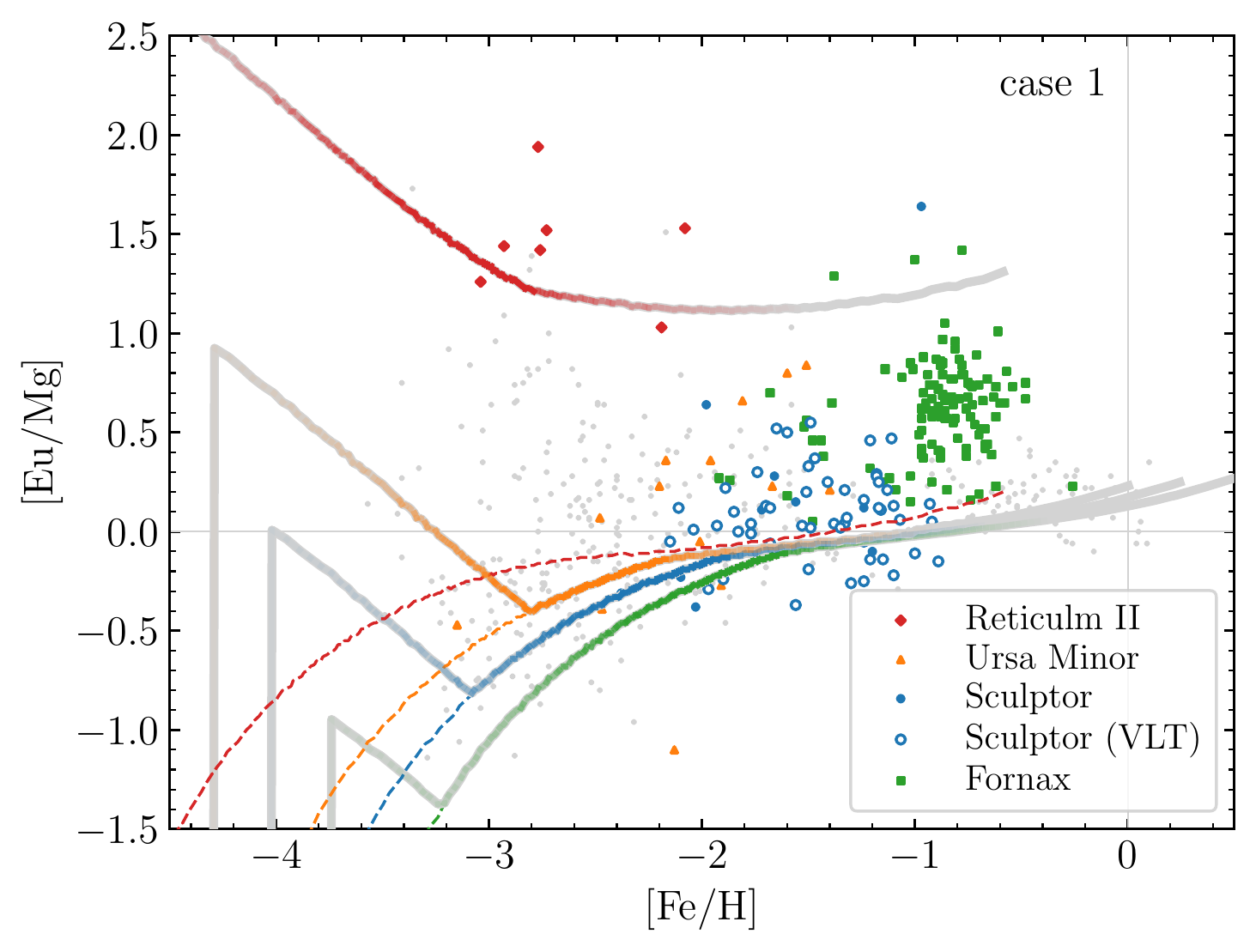}
	\includegraphics[width=0.86\columnwidth]{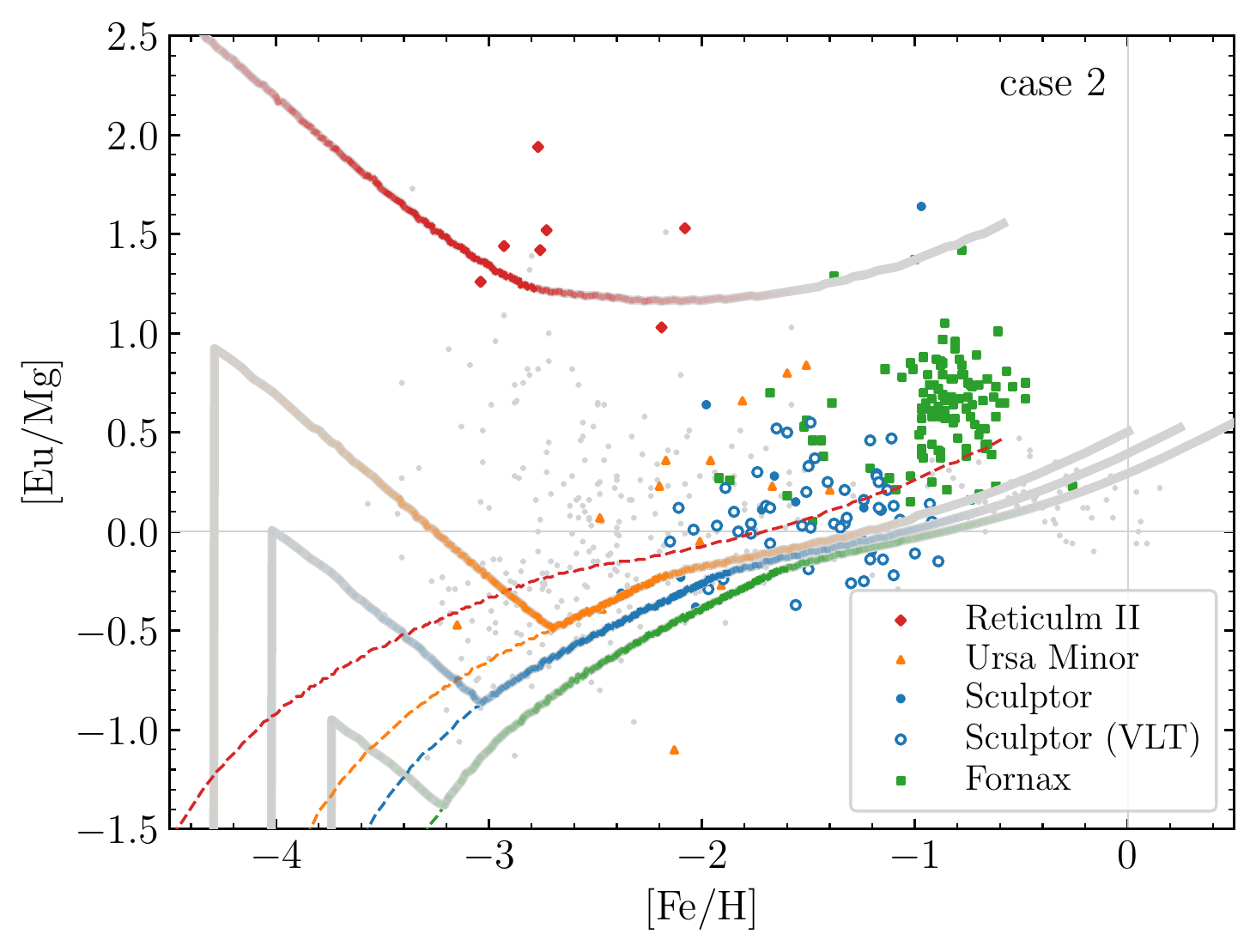}
	\includegraphics[width=0.86\columnwidth]{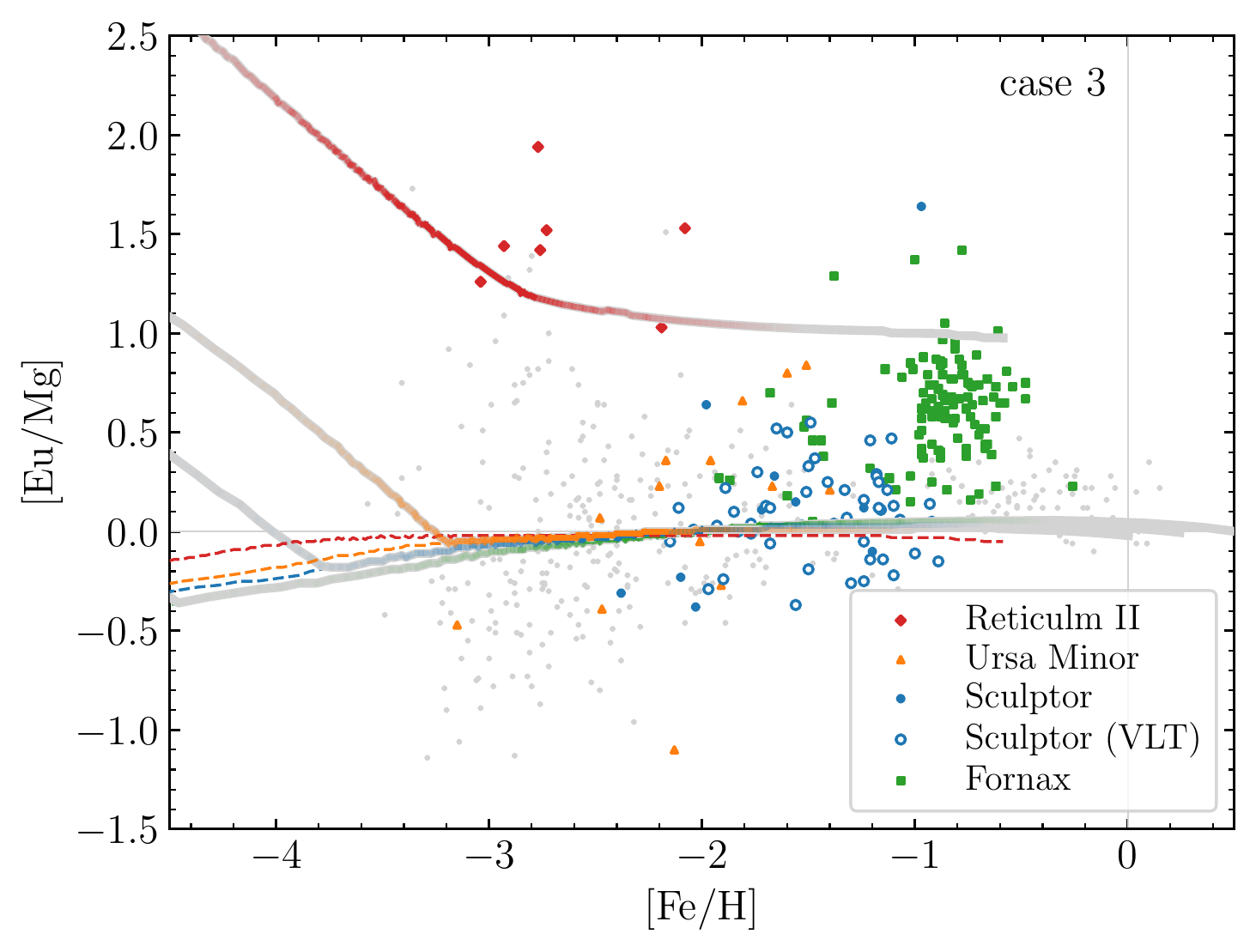}
	\includegraphics[width=0.86\columnwidth]{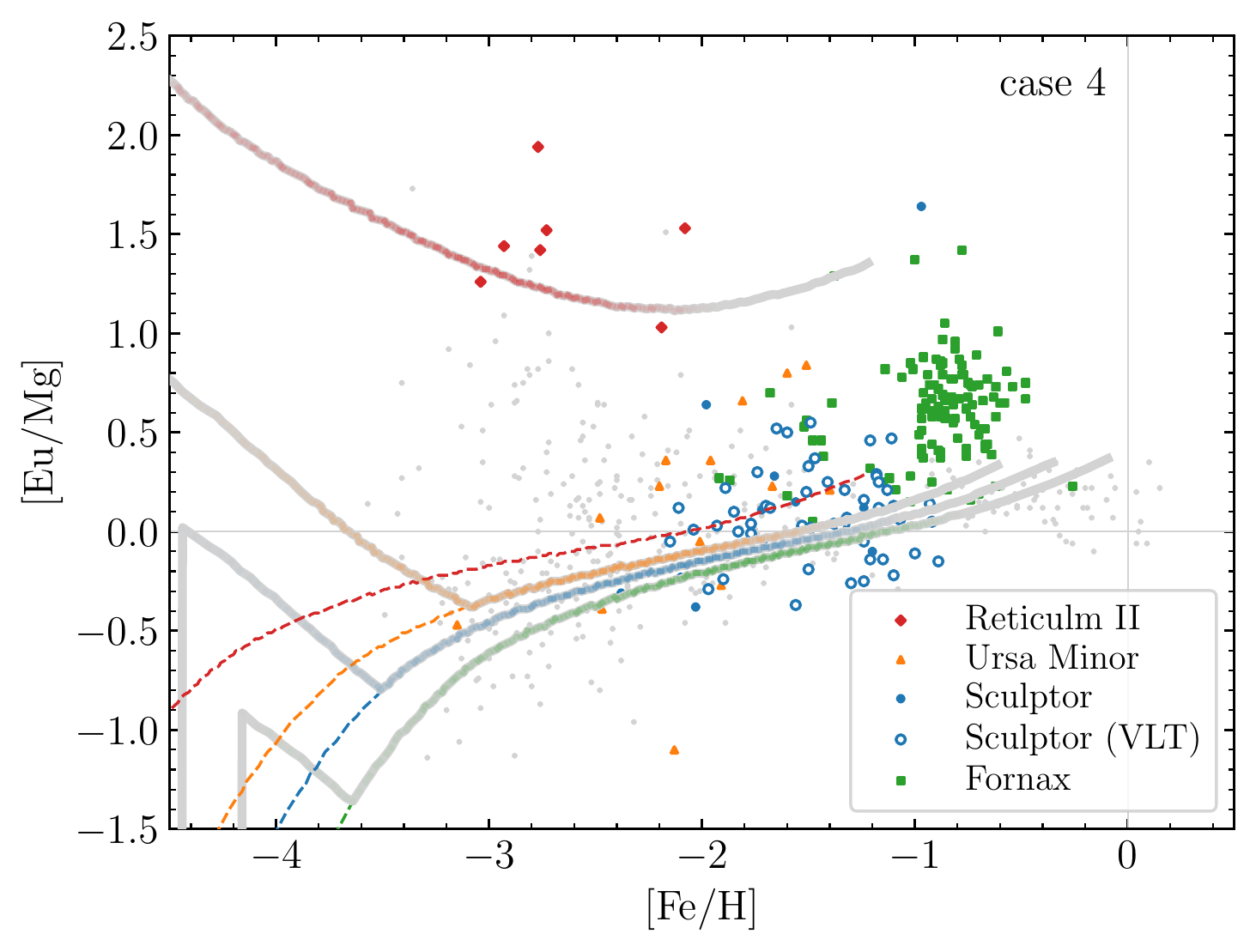}
    \caption{Same as Figs.~\ref{fig:mgfe} and \ref{fig:eufe}, but for [Eu/Mg].}
    \label{fig:eumg}
\end{figure*}

The evolution of [Eu/Fe] as a function of [Fe/H] for the reference dwarf galaxies is shown in Fig.~\ref{fig:eufe} for cases 1--4. The computed average Eu/Fe (indicated by the dashed curve) for each galaxy is divided by $\langle N_\mathrm{NSM} \rangle$ for $\langle N_\mathrm{NSM} \rangle < 1$, which is shown by the gray curve with the (transparent) colour of the reference galaxy in the legend. The smaller number of stars with Eu (than of those with Mg) makes it difficult to judge if our model accounts for the observational trend of the Eu evolution. Therefore, we focus on Reticulum II and Sculptor as representative of UFDs and classical dwarf spheroidals, respectively, for comparison.

The evolution of our Reticulum II-like galaxy is similar among all explored cases 1--4 (reddish-gray curves in Fig.~\ref{fig:eufe}). As shown in Fig.~\ref{fig:num}, $\langle N_\mathrm{NSM} \rangle$ never exceeds unity throughout its evolution. As a result, the value of [Eu/Fe] calculated as Eu/Fe divided by $\langle N_\mathrm{NSM} \rangle$ becomes $\sim 2$. This is in qualitative agreement to the rarity of UFDs exhibiting enhancement of the r-process elements as noted in \S~\ref{subsubsec:mht}. The fact that Eu has been detected only in the stars of [Fe/H] $\ge -3.0$ \citep{Ji2016,Roederer2016} suggests that our Reticulum II-like galaxy experiences the first NSM at $t \approx 0.7$ Gyr (for cases 1--3), making a jump of the [Eu/Fe] value from [Eu/Fe] $= -\infty$ \citep[or a small pre-enriched value,][]{Tsujimoto2015,Ojima2018}. In addition, the presence of Eu-enhanced stars up to [Fe/H] $= -2.1$ implies that the galaxy evolves up to $t \approx 2$ Gyr (for cases 1--3). This indicates that Reticulum II has experienced prolonged star formation beyond reionization \citep[e.g.,][]{Weisz2014,Applebaum2021,Miyoshi2020}. It should be noted, however, that another choice of OFE, e.g., $k_\mathrm{OF} = 2.0$ (Gyr$^{-1}$), and thus a factor of two greater $k_\mathrm{SF}$ for a fixed value of $K$ in equation~(\ref{eq:sfof}), leads to the same metallicities above at $t \approx 0.35$ Gyr and $t \approx 1.4$ Gyr. Alternatively, its star formation has been terminated by reionization and the increase of [Fe/H] is due to chemical inhomogeneity \citep{Safarzadeh2017,Tarumi2020} rather than the GCE. If this is the case, a reasonable accordance of our results with the measured abundance trend is merely a coincidence.

For our Sculptor-like galaxy, $\langle N_\mathrm{NSM} \rangle$ exceeds unity at the metallicity below [Fe/H] $\sim -3$. The evolution of [Eu/Fe] above this metallicity for each case can be thus interpreted by referring Fig.~\ref{fig:mht}. At the metallicity below [Fe/H] $\approx -2$ ($t \approx 1$ Gyr), [Eu/Fe] for case 1 increases with [Fe/H] because of the steeper curve of Eu than that of Fe in Fig.~\ref{fig:mht} (solid curves). At $t = 1$ Gyr ([Fe/H] $\approx -2$), the slope of Fe becomes steeper than that of Eu and thus [Eu/Fe] starts decreasing, being in accordance with the observational trend. The same holds true for case 2 (dashed curves). For case 1, however, the presence of the early (log-normal) component in the delay-time distribution of NSMs leads to a shallower (i.e., more Mg-like) curve in Fig.~\ref{fig:mht}. This leads to a steeper decrease of [Eu/Fe] for case 1 as can be seen in Fig.~\ref{fig:eufe} (top panels). For case 3 (that mimics little delay for collapsars or MRSNe), the slope of Eu is similar to that of Mg all the way (dotted curves in Fig.~\ref{fig:mht}). As a result, the evolution of [Eu/Fe] is similar to that of [Mg/Fe] as well as that for the other galaxies (except for Reticulum II). For case 4, the short delay of SNe Ia results in the similar slope of Eu to that of Fe as can be seen in Fig.~\ref{fig:mht} (dash-dotted curves). This leads to a flat trend of [Eu/Fe] in our Sculptor-like galaxy (the bottom-right panel in Fig.~\ref{fig:eufe}), being incompatible with the measurements.

To minimize the complexity arising from various combinations of the delay-time distributions, we also present the plots of [Eu/Mg] for all cases in Fig.~\ref{fig:eumg}. This removes the contribution of SNe Ia from the vertical axis in the plots. For Sculptor, the measured values show a flat trend of [Eu/Mg], being in agreement with the result for case 3 (bottom left). However, our Sculptor-like galaxy for case 1 also exhibits a relatively flat trend between the measured range of [Fe/H] $\sim -2.4$--$-1$. Moreover, we may not be able to exclude the slowly increasing trend of [Eu/Mg] in cases 2 and 4 because of some scatter in the measured values.
More observational data of Eu at lower metallicity will be needed to obtain meaningful constraints.

\subsection{MW disc}
\label{subsec:disc}

In this subsection, we present the results of GCE in the MW disc by using standard (one-) and two-infall models described in \S~\ref{subsec:infall}. A possible role of the natal kicks of binary neutron stars is also discussed in \S~\ref{subsubsec:kick}.

\subsubsection{Evolution of Mg and Eu}
\label{subsubsec:mgeu_disc}

\begin{figure*}
	\includegraphics[width=0.67\columnwidth]{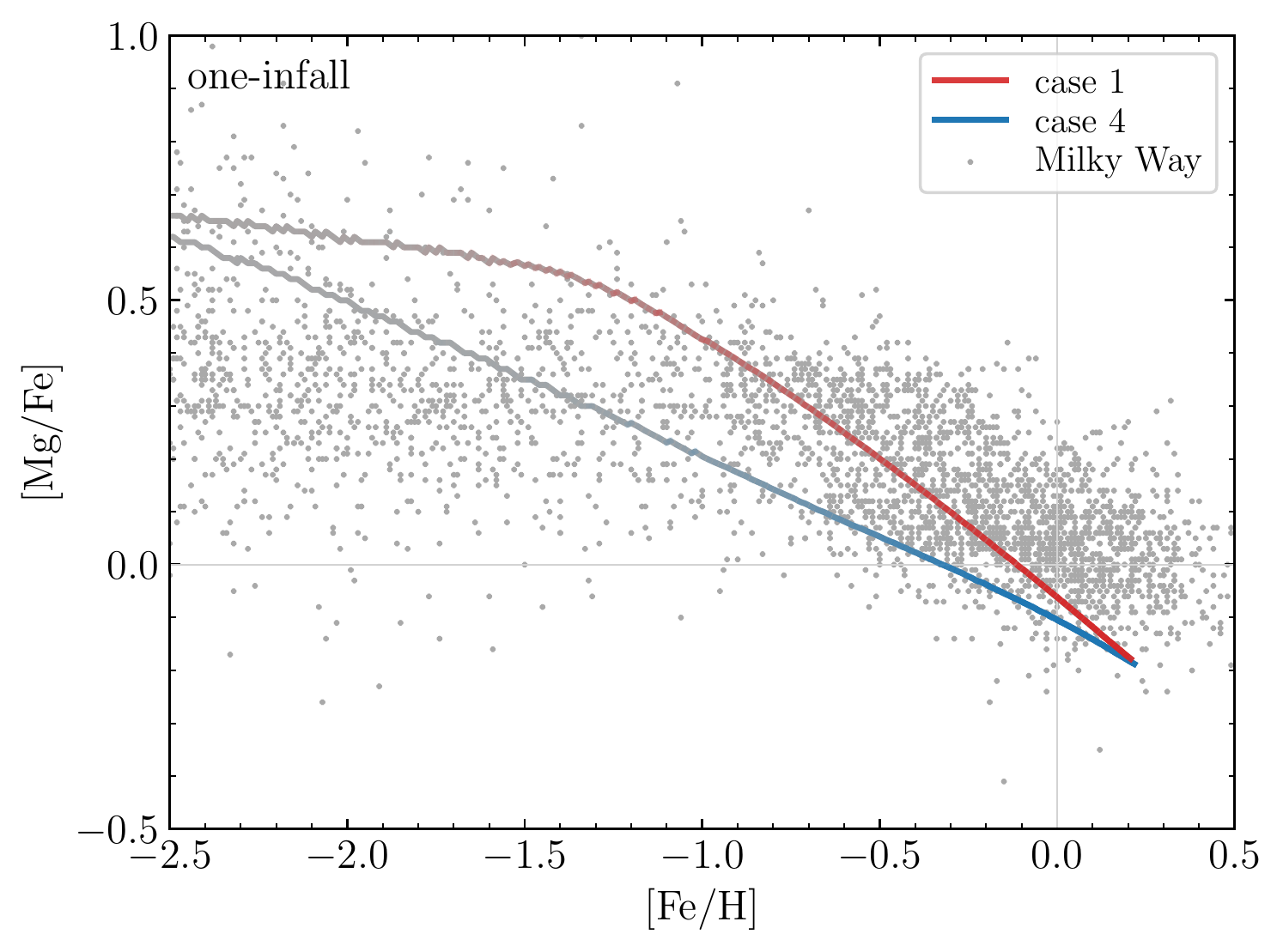}
	\includegraphics[width=0.67\columnwidth]{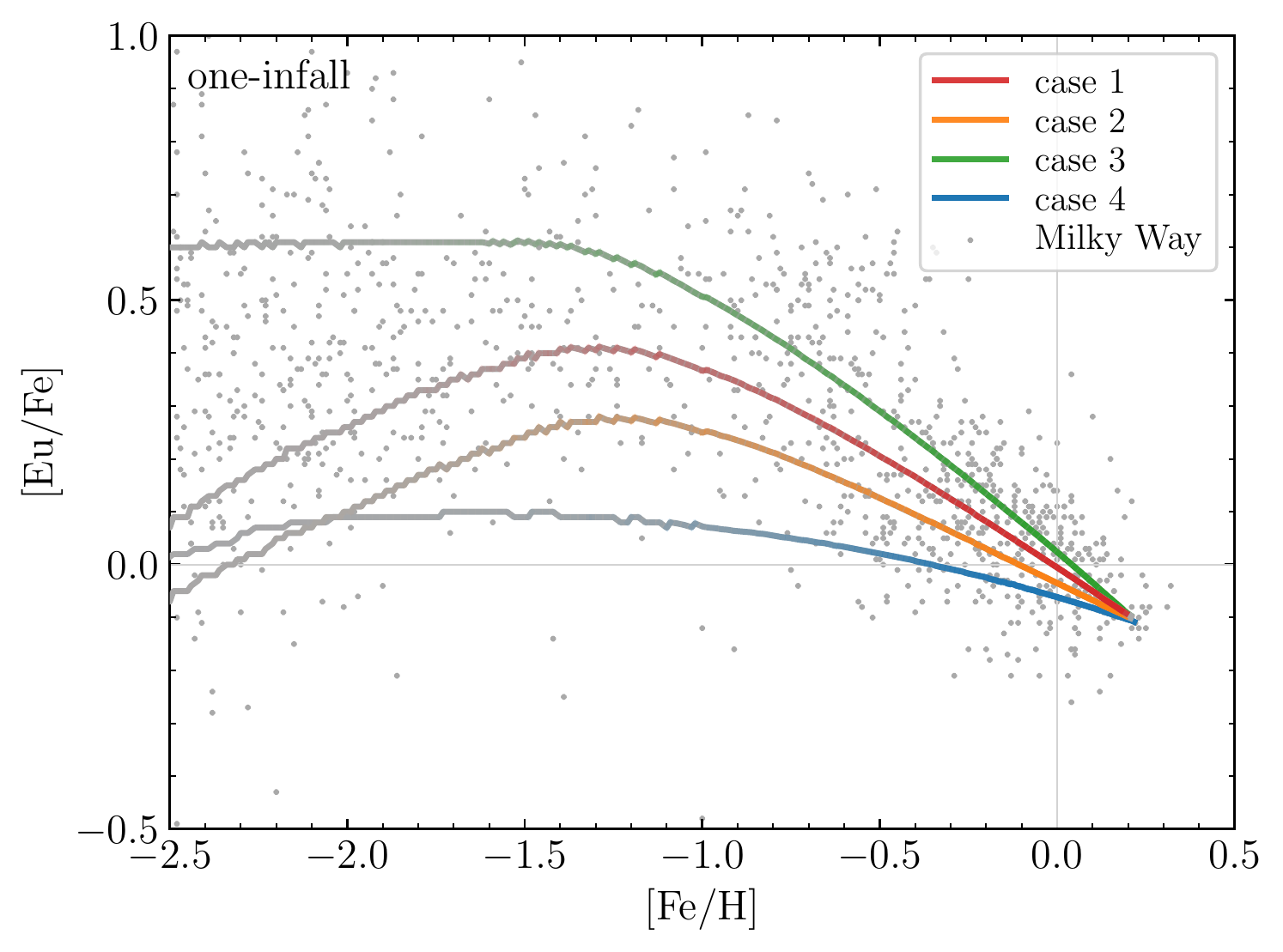}
	\includegraphics[width=0.67\columnwidth]{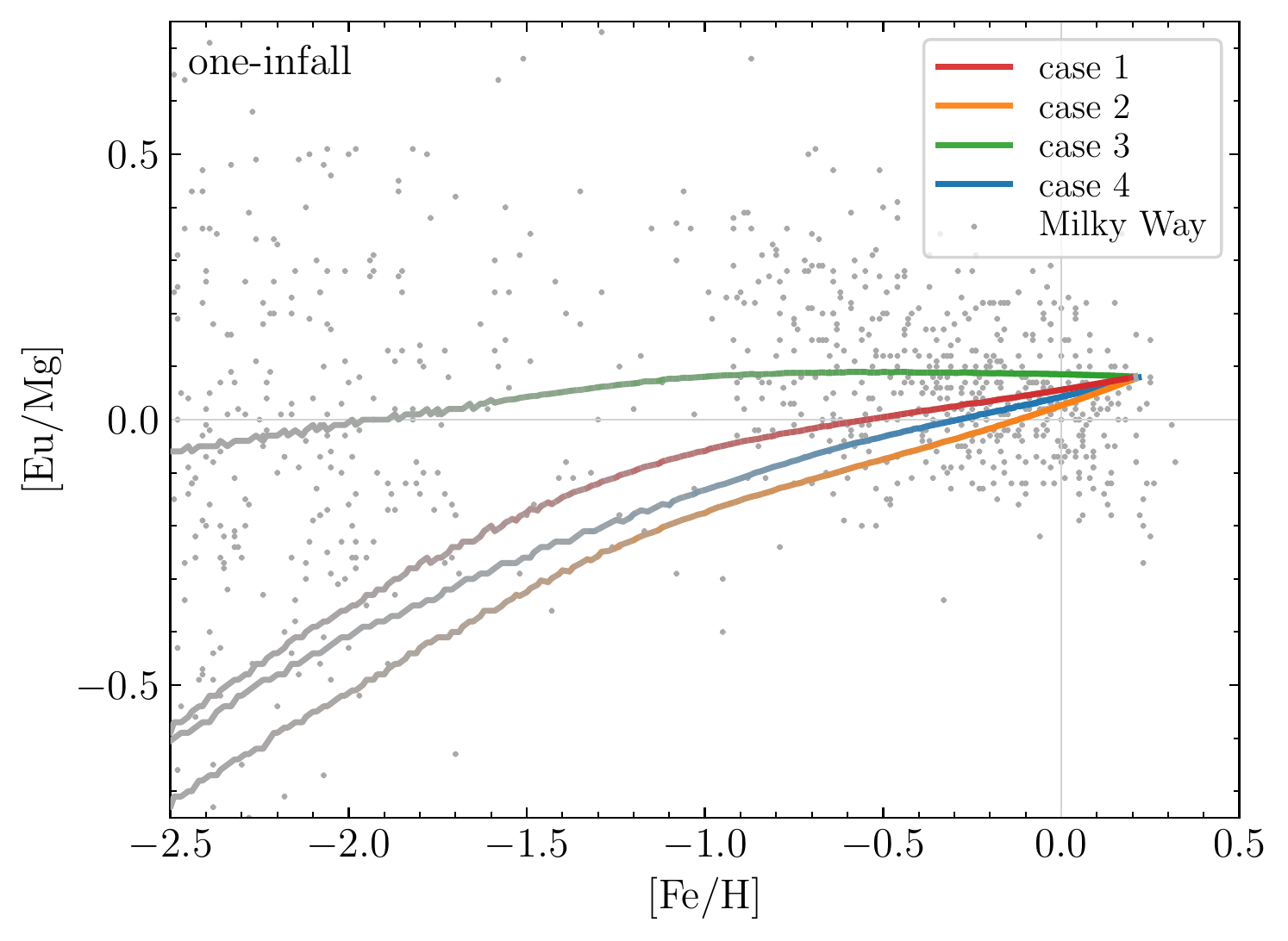}
	\includegraphics[width=0.67\columnwidth]{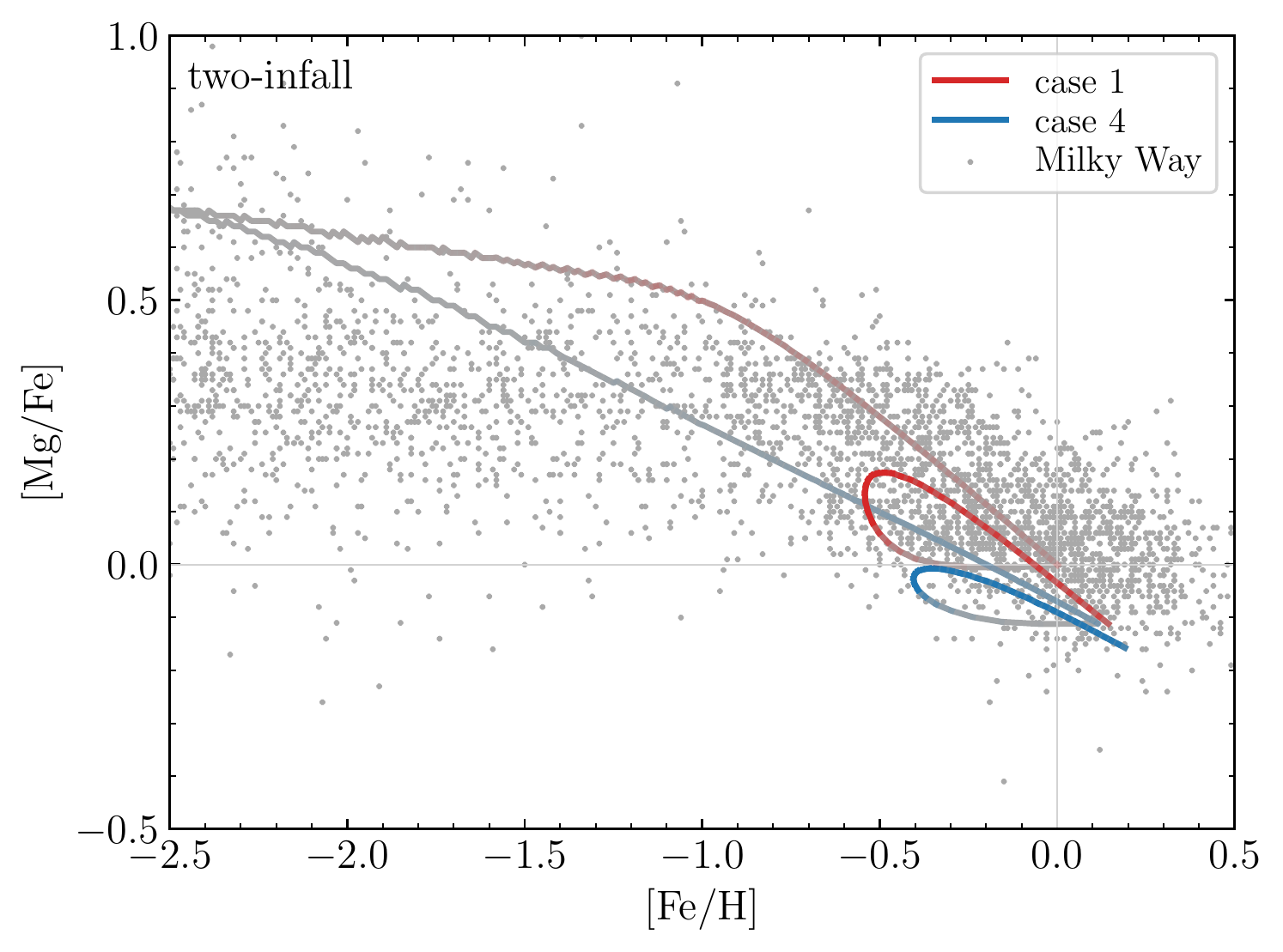}
	\includegraphics[width=0.67\columnwidth]{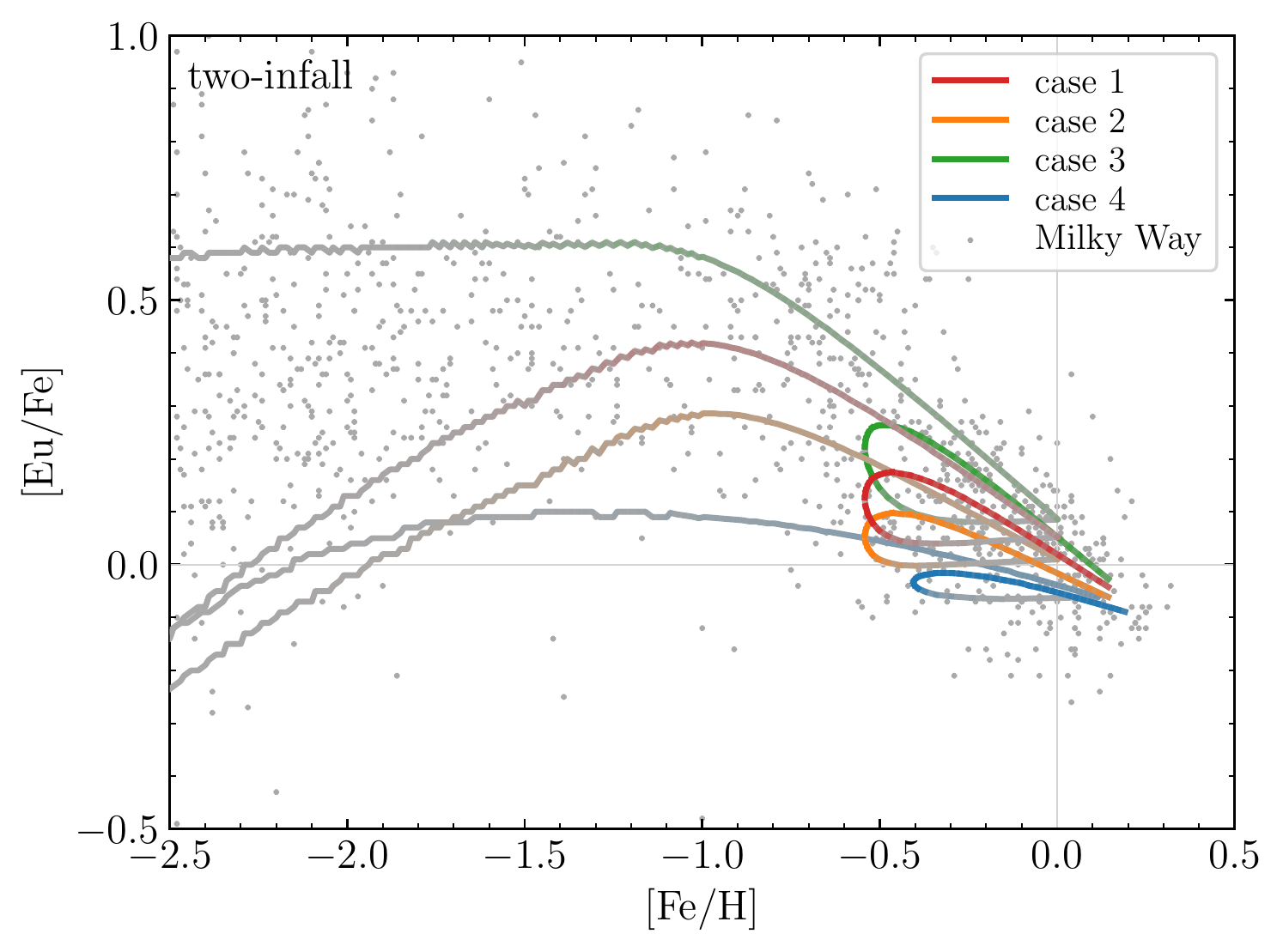}
	\includegraphics[width=0.67\columnwidth]{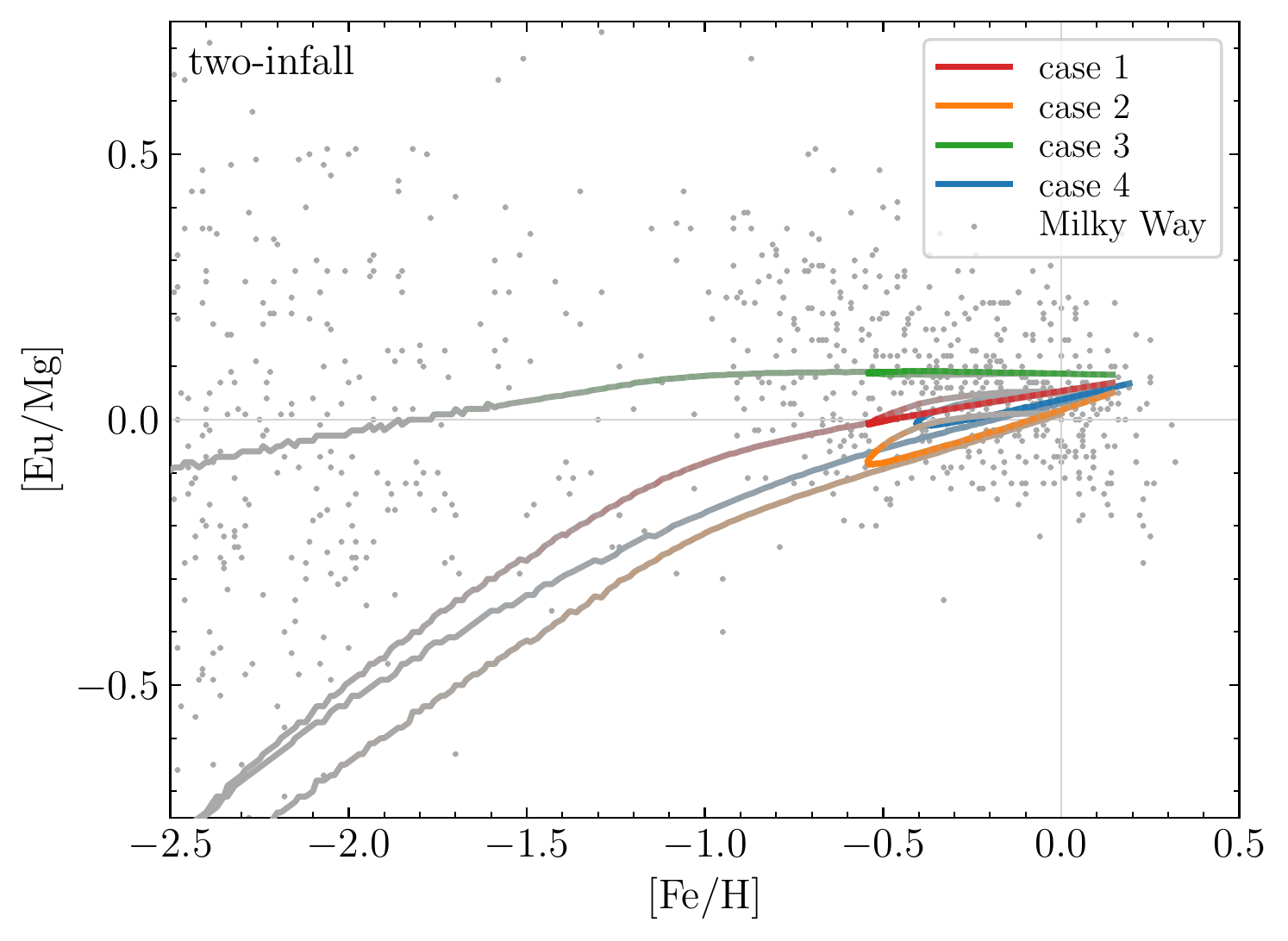}
	    \caption{Evolution of [Mg/Fe] (left), [Eu/Fe] (middle) and [Eu/Mg] (right) as a function of [Fe/H] in the MW disc. The results for (standard) one- and two-infall models are shown in the top and bottom panels, respectively. The evolutionary track for each case is drawn by a gray curve with the (transparent) colour specified in the legend. The thickness of the colour is proportional to the number of stars formed in each bin, i.e., $df_\mathrm{star}/d$[Fe/H]. The gray dots are the measured values in the MW taken from the SAGA database \citep{Suda2008,Suda2011}.}
    \label{fig:mgeu_disc}
\end{figure*}

Fig.~\ref{fig:mgeu_disc} shows the evolution of [Mg/Fe] (left), [Eu/Fe] (middle) and [Eu/Mg] (right). As the same as in \S~\ref{subsec:dwarf}, the evolutionary track of each case is drawn by a gray curve with the (transparent) colour specified in the legend. The thickness of colour is proportional to the number of stars formed within each bin, i.e., $df_\mathrm{star}/d$[Fe/H]. Note that the results of [Mg/Fe] for cases 2 and 3 are the same as that for case 1. The measured stellar abundances are taken from the SAGA database \citep{Suda2008,Suda2011}. Here, we focus only the disc stars of [Fe/H] $> -1$.

The results with a standard infall model are displayed in the top panels of Fig.~\ref{fig:mgeu_disc}. The evolution of [Mg/Fe] is compatible with the observed knee position of [Fe/H] $\sim -1$ for case 1 (red) but not for case 4 (blue). This suggests that the delay of SNe Ia should be sufficiently long ($t_\mathrm{min} \sim 1$ Gyr), the same conclusion obtained for the halo (\S~\ref{subsec:halo}) and the satellite dwarfs (\S~\ref{subsec:dwarf}). As the observational trend of [Eu/Fe] in the disc is similar to that of [Mg/Fe], the result for case 3 (greenish curve in the top-middle panel of Fig.~\ref{fig:mgeu_disc}) seems to be the best, which mimics little delay for subsets of CCSNe. However, the result for (our fiducial) case 1 appears also to be marginally consistent with the observational trend of [Eu/Fe] for [Fe/H] $> -1$. This is due to the sufficiently long delay of SNe Ia \citep[$t_\mathrm{min} = 1$ Gyr, see also][]{Hotokezaka2018} as well as the early (log-normal) merging component of binary neutron stars, as also found in the GCE of dwarf satellites (\S~\ref{subsec:dwarf}). The only slightly increasing trend of [Eu/Mg] for case 1 is also marginally consistent with the flat trend of measured values. In fact, the result for case 1 approaches that of case 3 by increasing $t_\mathrm{min}$ ($> 1$ Gyr) for SNe Ia or the early merging component ($A < 0.5$ in equation~(\ref{eq:dtdnsm})). 

The bottom panels of Fig.~\ref{fig:mgeu_disc} present the results for the two-infall model. We find the dichotomy of [Mg/Fe] as the thick (upper thin red) and thin (lower thick red) disc components (for case 1), being in reasonable agreement with recent spectroscopic analyses \citep[][although the dichotomy is not evident from the SAGA data in Fig. \ref{fig:mgeu_disc}]{Adibekyan2013,Haywood2013,Bensby2014,Hayden2015,Queiroz2020}. For case 1, the evolutionary track of the early thick-disc phase reaches [Fe/H] $\approx 0$ at $t = 5$ Gyr (Table~\ref{tab:infall}) and then switches to that of the thin-disc phase by drawing a loop. The evolutionary track goes back to [Fe/H] $\approx -0.55$ ($t \approx 6.5$ Gyr) and returns toward high metallicity up to [Fe/H] $\approx 0.15$ ($t \approx 12$ Gyr). In reality, such a loop will not be observed because of a short returning period ($\approx 1.5$ Gyr) and thus a small number of stars (illustrated by its fading colour on the gray curve). Moreover, the evolution of thick and thin discs may not be sequential but independent and coeval as suggested in some cosmological simulations \citep[][]{Agertz2021,Renaud2021,Renaud2021b}. If this is the case, the loop of the evolutionary track is merely an artefact. In any case, a dichotomy of [Eu/Fe] is also predicted when considering both thick and thin disc components \citep{Griffith2019}. 

However, the evolutionary tracks of thick and thin discs merge into a single curve in the plot of [Eu/Mg] (the bottom-right panel of Fig.~\ref{fig:mgeu_disc}), because the effects of SNe Ia are cancelled and do not appear in the vertical axis. In this regard, the ratio [Eu/Mg] is a better diagnostic of the models, being nearly independent of the complex formation histories of thick and thin discs. In fact, the result for the two-infall model is almost the same as that for a standard infall model (the top-right panel) for each case (but see \S~\ref{subsubsec:kick}). 

\subsubsection{A possible effect of the natal kicks of binary neutron stars}
\label{subsubsec:kick}

\begin{figure*}
	\includegraphics[width=0.67\columnwidth]{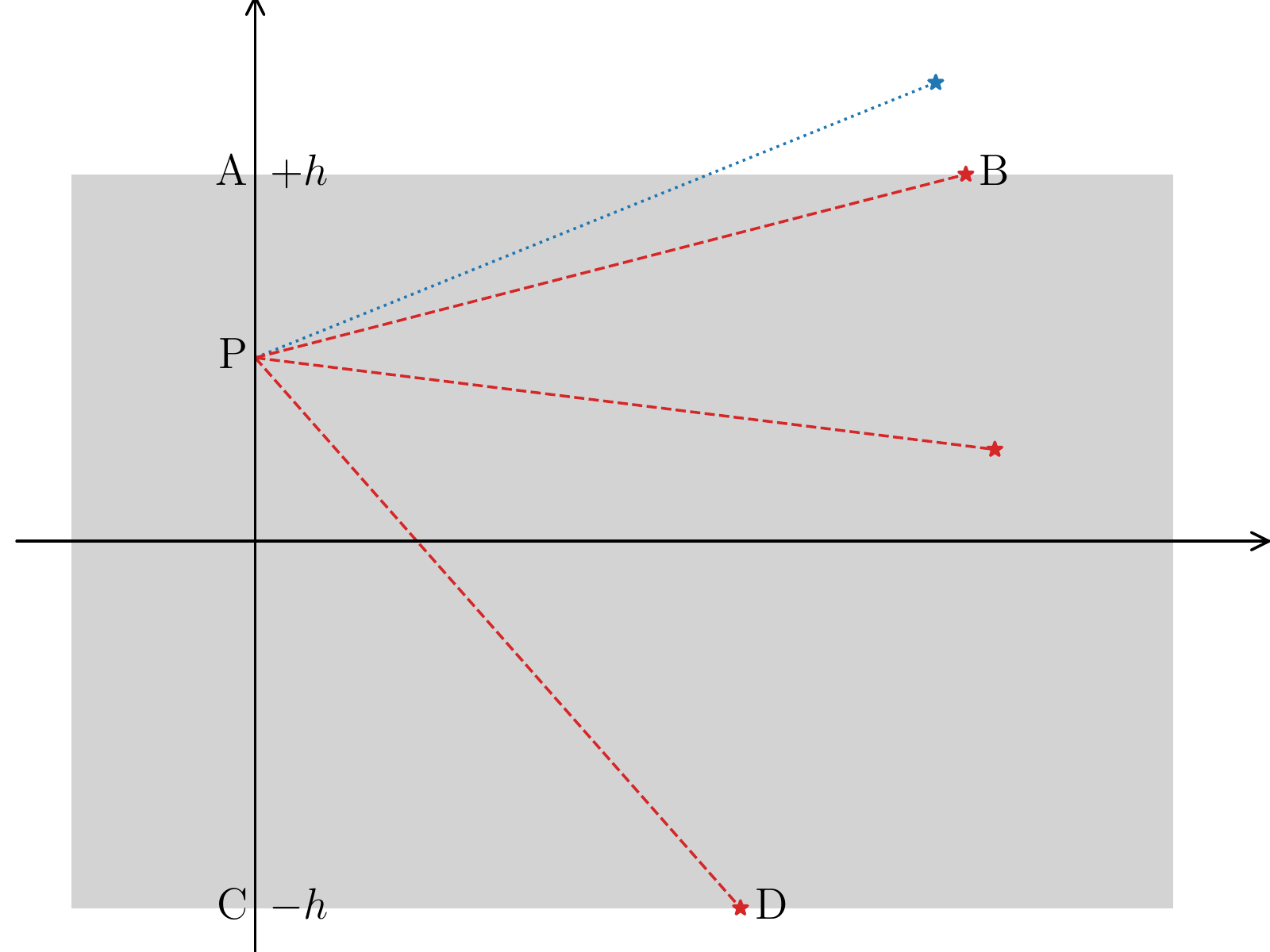}
	\includegraphics[width=0.67\columnwidth]{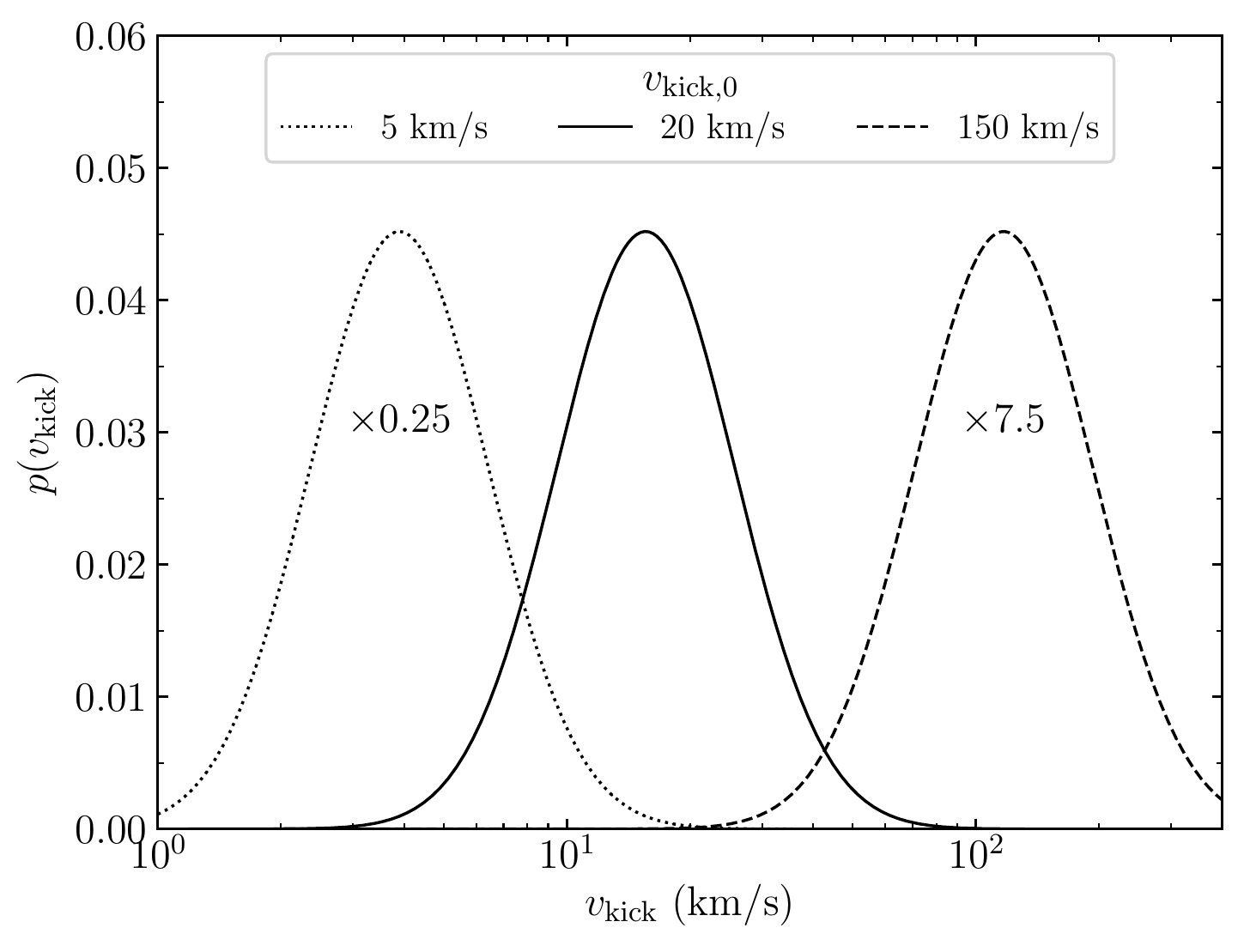}
	\includegraphics[width=0.67\columnwidth]{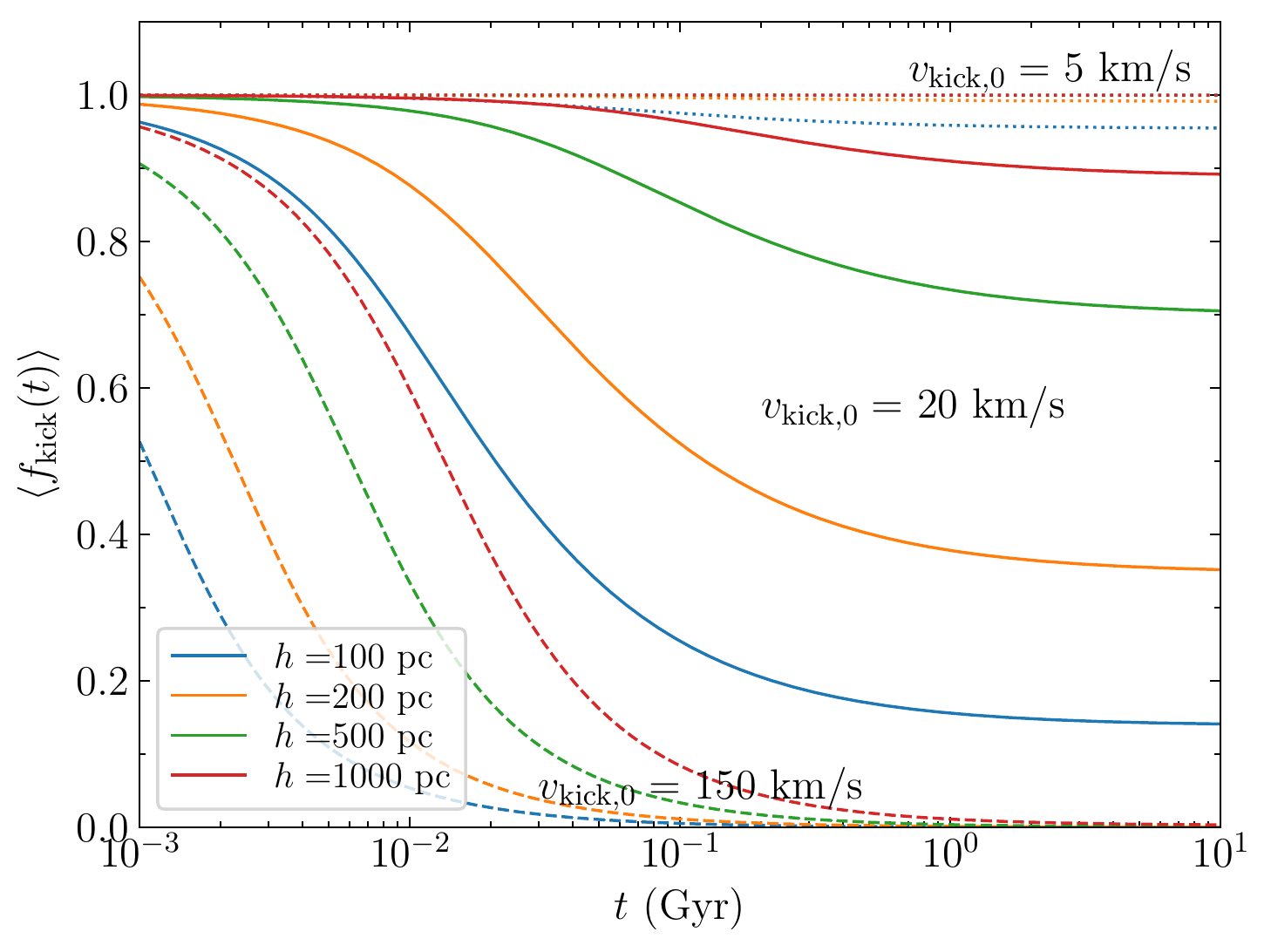}
    \caption{\textit{Left}: Schematic view for modeling the natal kicks of binary neutron stars. The gray region is the MW disc with  half-thickness $h$ (distance from the midplane). A neutron star binary born at P$(0, z)$ travels with a constant velocity and merges at a star symbol. 
    The NSMs that reside in the disc (red) contribute to the GCE, while others do not (blue). \textit{Middle}: Assumed probability distribution of kick velocities $v_\mathrm{kick}$ defined by equation~(\ref{eq:vdist}) for $v_\mathrm{kick,0} = 5$ km/s (dotted curve), 20 km/s (solid curve) and 150 km/s (dashed curve). \textit{Right}: 
    Resulting average fraction of binary neutron stars that remain in the disc at the merger as a function of time. The colours indicate the results for different values of $h$ specified in the legend.
    }
    \label{fig:kick0}
\end{figure*}

\begin{figure*}
	\includegraphics[width=0.67\columnwidth]{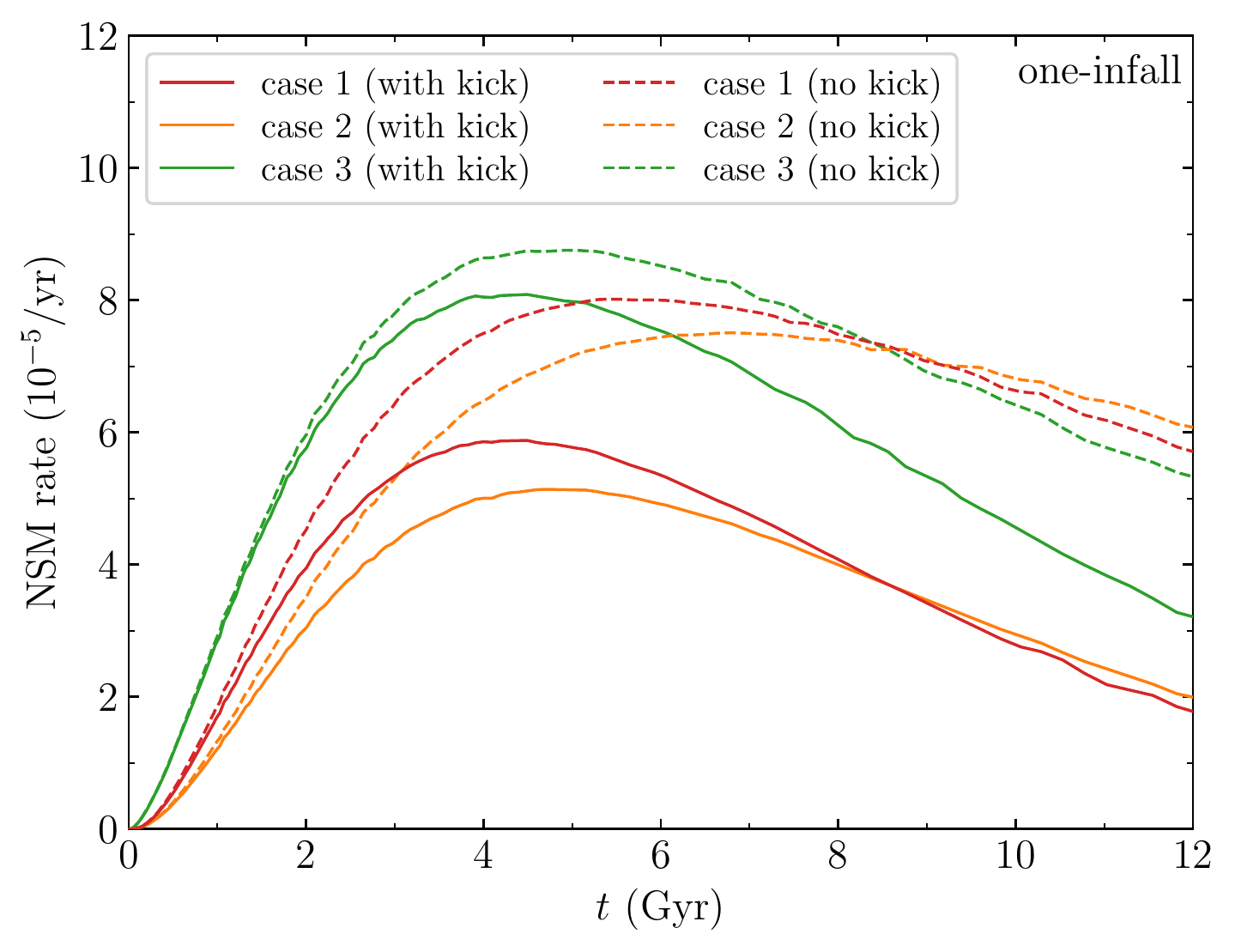}
	\includegraphics[width=0.67\columnwidth]{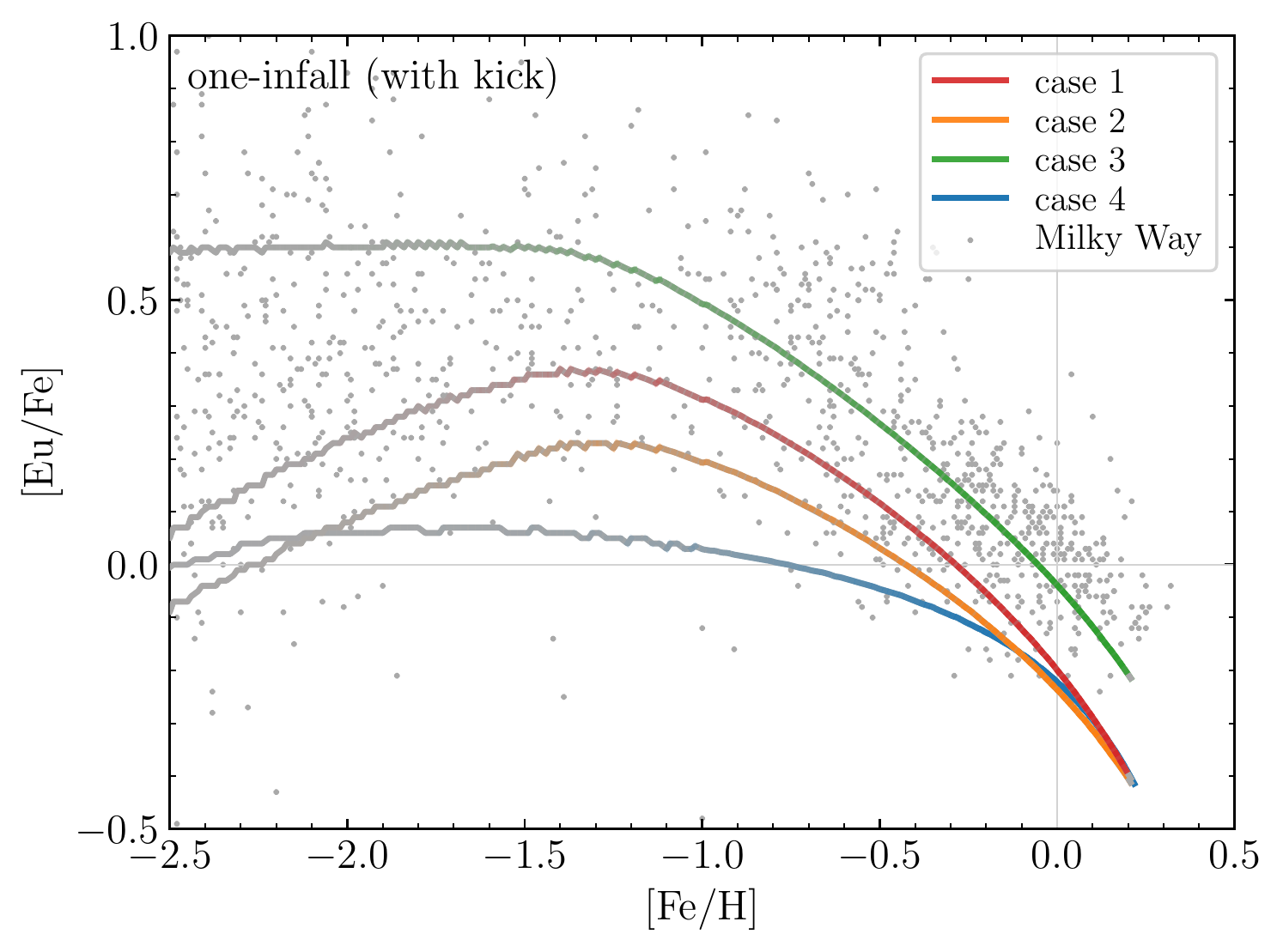}
	\includegraphics[width=0.67\columnwidth]{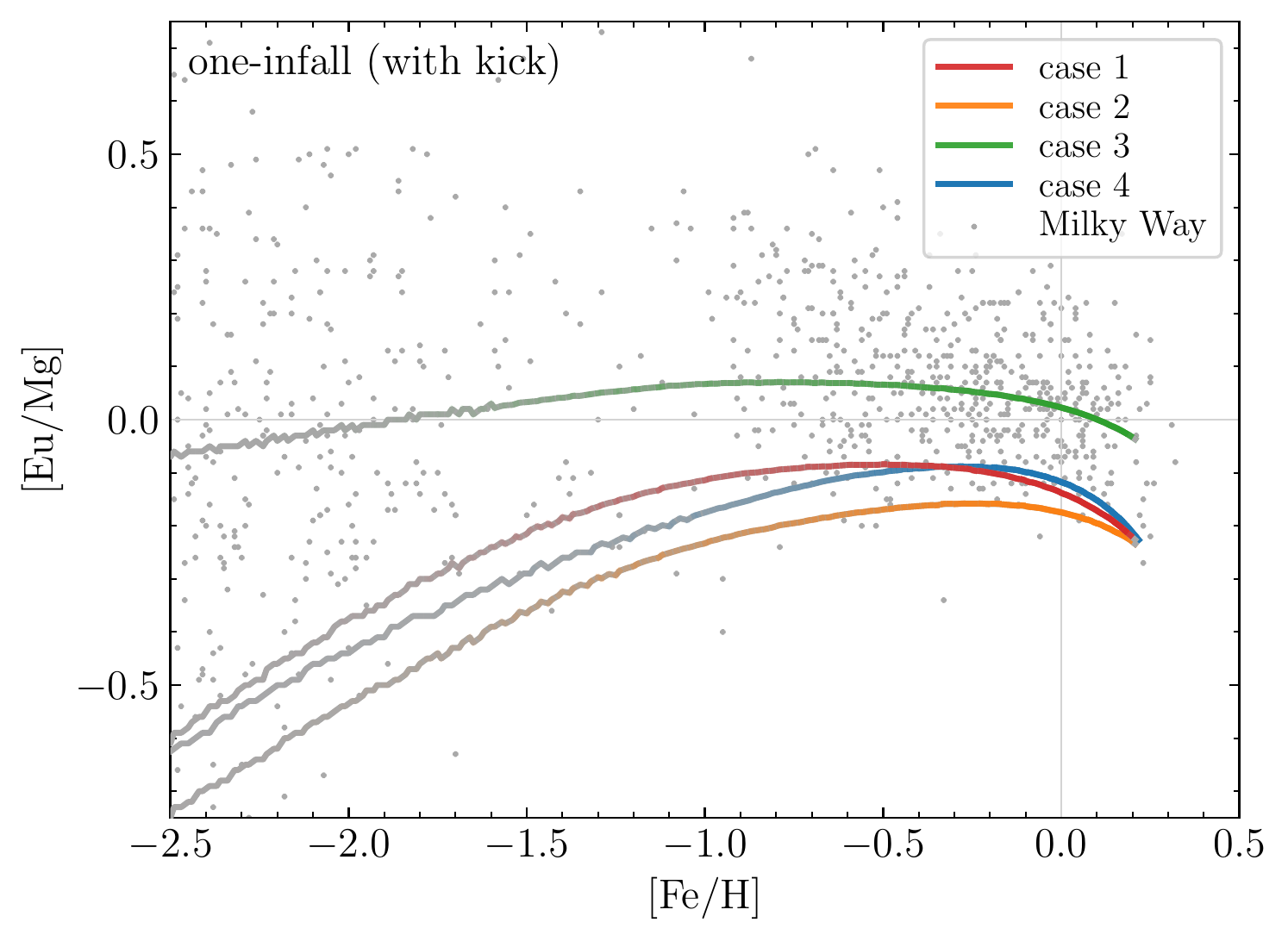}
	\includegraphics[width=0.67\columnwidth]{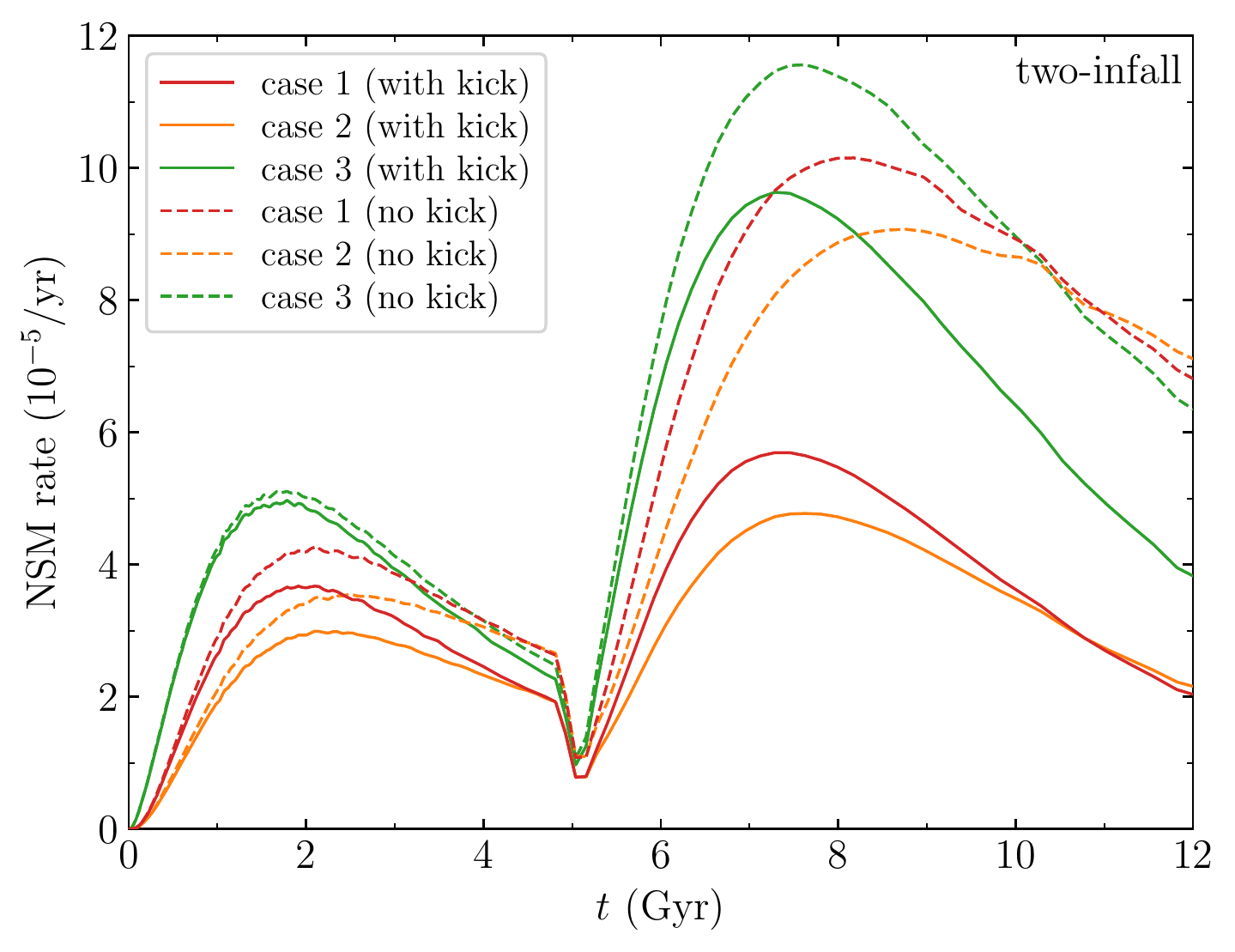}
	\includegraphics[width=0.67\columnwidth]{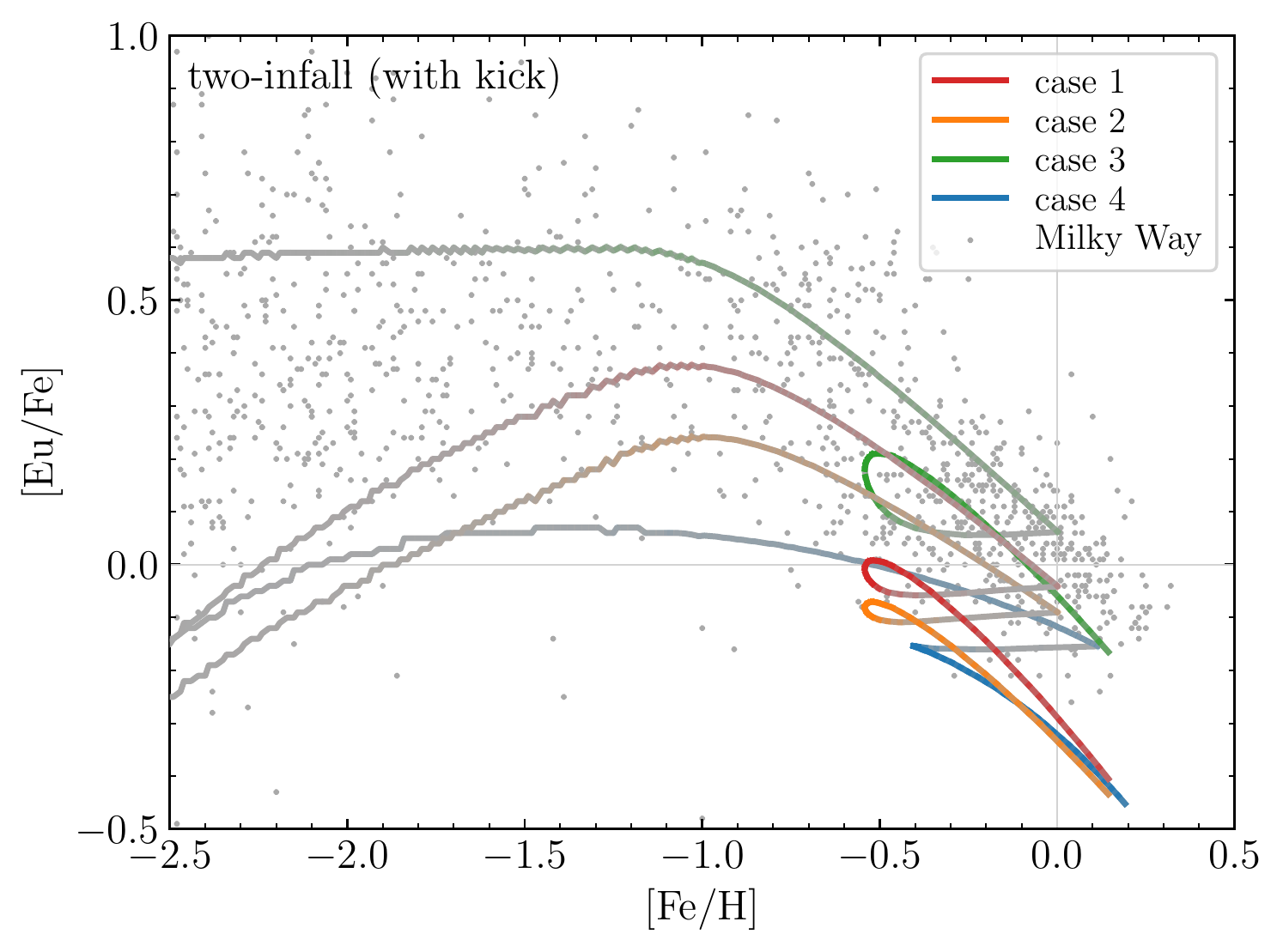}
	\includegraphics[width=0.67\columnwidth]{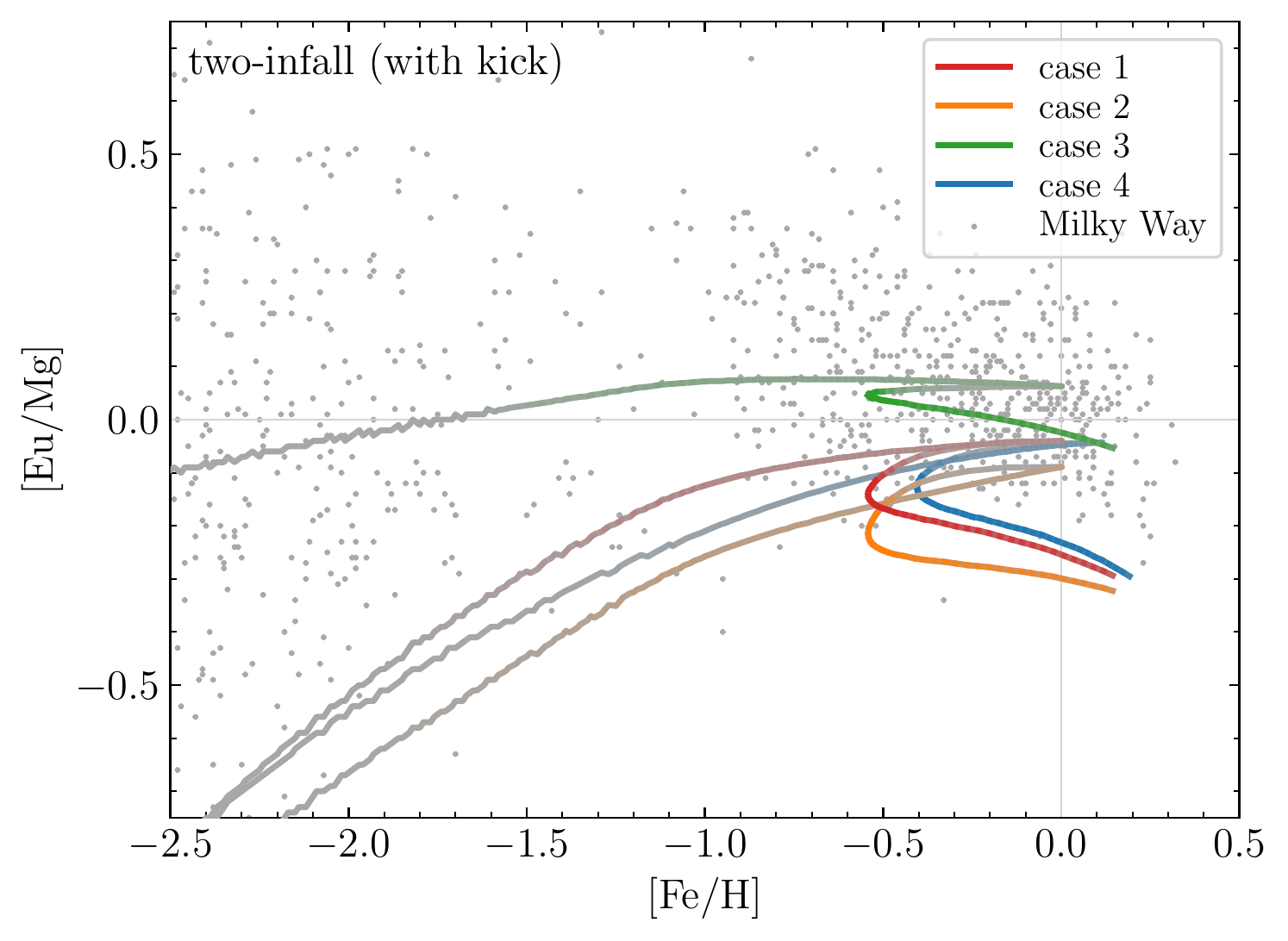}
    \caption{\textit{Left}: Galactic NSM rate as a function of time for standard (top) and two-infall (bottom) models with $M_0 = 6\times 10^{10}\, M_\odot$ \citep[the stellar mass of the MW today,][]{McMillan2011}. 
    \textit{Middle and right}: same as the middle and right panels in Fig.~\ref{fig:mgeu_disc}, but with the natal kicks of binary neutron stars for $v_\mathrm{kick,0} = 20$ km/s.}
    \label{fig:kick}
\end{figure*}


The presence of r-process-enriched UFDs (with shallow gravitational potential) indicates small natal kick velocities $v_\mathrm{kick}$ for binary neutron stars at birth \citep[$< 15$ km/s,][]{Beniamini2016b}. In fact, the analyses of observed binary neutron stars in the MW suggest that the majority of the systems receive small kicks of $v_\mathrm{kick} < 30$ km/s \citep{Beniamini2016}, although a high velocity population of $v_\mathrm{kick} > 150$ km/s also exists \citep[$\sim 20\%$,][]{Behroozi2014}. 

In the MW disc, the velocity needed to reach the height $h$ from the midplane, $v_\mathrm{out}$ ($\sim$ velocity dispersion of the disc stars), is about
\begin{equation}
    v_\mathrm{out} = \sqrt{2\pi G \Sigma_\mathrm{disc} h} \approx 12 \left(\frac{\Sigma_\mathrm{disc}}{50\, M_\odot\, \mathrm{pc}^{-2}}\right)^\frac{1}{2} \left(\frac{h}{100\, \mathrm{pc}}\right)^\frac{1}{2}\, \mathrm{km/s},
    \label{eq:vout}
\end{equation}
where the present-day column density and the scale height (for gaseous layer) of the thin disc \citep{Prantzos2008b} are applied to the denominators. Thus, the binary neutron stars with $v_\mathrm{kick} \gg v_\mathrm{out}$ may not contribute to the GCE. In reality, such binaries may oscillate about the midplane and some of them reside in the disc at the merger. These NSMs are, however, not necessarily in a star forming region because of its clumpiness in the disc. As a limiting case, we ignore the contribution of the binary neutron stars  once they leave the disc.

We consider a homogeneous and infinite disc plane, in which a  neutron star binary forms at P$(0, z)$ with its natal kick velocity $v_\mathrm{kick}$ as illustrated in Fig.~\ref{fig:kick0} (left). For simplicity, we ignore any influences from stars, gas and dark matter such that a binary neutron star travels with the constant velocity, $v_\mathrm{eff} \equiv v_\mathrm{kick} - v_\mathrm{out}$ (for $v_\mathrm{kick} > v_\mathrm{out}$, which is needed for a binary to escape from the disc; $v_\mathrm{eff} = 0$ otherwise), and keeps the original direction all the way. Thus, the distance that the binary travels until the merger with a delay time $t$ is $v_\mathrm{eff}t$. Given that the directions of kicks are isotropic, the probability that a NSM occurs in the disc within $|z| < h$ is given by
\begin{eqnarray}
    g_\mathrm{kick}(z, t) & = & \frac{\pi - \angle \mathrm{APB} - \angle \mathrm{CPD}}{\pi} \nonumber \\
                      & = & 1 - \frac{1}{\pi} \arccos \frac{h - z}{v_\mathrm{eff}t} - \frac{1}{\pi} \arccos \frac{h + z}{v_\mathrm{eff}t},
                      \label{eq:fnsm}
\end{eqnarray}
where A, B, C and D are the positions shown in Fig. \ref{fig:kick0}. Assuming that the number density of newly forming binaries is independent of $z$, we obtain the average fraction of NSMs in the disc for $t > 2h/v_\mathrm{eff}$ as
\begin{eqnarray}
    f_\mathrm{kick}(t) & = & \frac{1}{2h} \int_{-h}^{h} g_\mathrm{kick}(z, t)\, dz \nonumber \\
                                 & = & 1 - \frac{v_\mathrm{eff}t}{\pi h} \left[1 - \sqrt{1 - \left(\frac{2h}{v_\mathrm{eff}t}\right)^2} \right] - \frac{2}{\pi} \arccos \frac{2h}{v_\mathrm{eff}t},
                   \label{eq:fnsm_dgth}
\end{eqnarray}
where the formula
    $\int \arccos x\, dx = x \arccos x - \sqrt{1 - x^2} + \mathrm{constant}$
is utilized. For $t \gg 2h/v_\mathrm{eff}$, equation~(\ref{eq:fnsm_dgth}) is reduced to
$f_\mathrm{kick}(t) \sim (2h/\pi v_\mathrm{eff})\, t^{-1}$,
which can be interpreted as the fraction of the arc of half the circle with a radius $v_\mathrm{eff}\, t$ across the disc.
For $t \leq 2h/v_\mathrm{eff}$, a similar calculation gives
\begin{equation}
    f_\mathrm{kick}(t) = 1 - \frac{v_\mathrm{eff}}{\pi h}\, t
\end{equation}
(notice that the ranges of integration become $h - v_\mathrm{eff}t < z < h$ and $-h < z < v_\mathrm{eff}t - h$ for the second and third terms in the right-hand side of equation.~(\ref{eq:fnsm}), respectively). 

We adopt a log-normal probability function of binary neutron stars \citep[similar to that in][]{Beniamini2019},
\begin{equation}
    p(v_\mathrm{kick}) = \frac{1}{\sqrt{2\pi {\sigma_\mathrm{kick}}^2}\, v_\mathrm{kick}} \exp \left[-\frac{(\ln v_\mathrm{kick} - \mu_\mathrm{kick})^2}{2{\sigma_\mathrm{kick}}^2} \right],
	\label{eq:vdist}
\end{equation}
where $\exp\, (\mu_\mathrm{kick}) \equiv v_\mathrm{kick, 0}$ is the mean velocity and $\sigma_\mathrm{kick} = 0.5$ (Fig.~\ref{fig:kick0}; middle). In the right panel of Fig.~\ref{fig:kick0}, the average of $f_\mathrm{kick}(t)$, $\langle f_\mathrm{kick}(t) \rangle = \int_0^\infty f_\mathrm{kick}(t)\, p(v_\mathrm{kick})\, dv_\mathrm{kick}$, is displayed for different values of $h$ and $v_\mathrm{kick,0}$ specified in the legend. We find that the number of NSMs in the disc progressively decreases with time, in particular for a smaller $h$. To mimic the evolution of the thick and thin discs at early and late times, respectively, we assume that the disc was thicker in the past such that $h(t) = h_0\, \exp \left(-t/\tau_\mathrm{col}\right)$ with a collapsing timescale of $\tau_\mathrm{col} = 6$ Gyr and the initial height of $h_0 = 750$ pc. This gives $h(12\, \mathrm{Gyr}) \approx 100$ pc. These values are taken such that the resulting slope of [Eu/Fe] at high metallicity for case 1 (with $v_\mathrm{kick,0} = 20$ km/s) becomes similar to that for case 3 in the standard infall model (Fig. \ref{fig:kick}; top middle).

\citet{Beniamini2016} have postulated the single and bimodal $v_\mathrm{kick}$ distributions that result in $v_\mathrm{kick,0} = 20$ km/s and (5, 150) km/s, respectively. According to their implication, we plot $\langle f_\mathrm{kick}(t) \rangle$ for these $v_\mathrm{kick,0}$ in the right panel of Fig.~\ref{fig:kick0}. It is anticipated that, for $v_\mathrm{kick,0} = 5$ km/s (dotted curves) and $v_\mathrm{kick,0} = 150$ km/s (dashed curves), most of the binaries remain in the disc and escape from the disc, respectively. We consider, therefore, only the case for $v_\mathrm{kick,0} = 20$ km/s (solid curves), in which one may expect a reasonable role of natal kicks.

The GCE of the disc is re-computed with the above prescription of the natal kicks with $D(t)$ replaced by the effective delay-time distribution of NSMs, $D(t) \langle f_\mathrm{kick}(t) \rangle$. The resulting Galactic NSM rates (left; assuming $M_0 = 6\times 10^{10}\, M_\odot$) as well as the evolutionary tracks of [Eu/Fe] (middle) and [Eu/Mg] (right) are shown in Fig.~\ref{fig:kick} for the standard (top) and two-infall (bottom) models. For the standard infall model, we find that the slope of [Eu/Fe] becomes steeper than that without natal kicks (the top-middle panel of Fig.~\ref{fig:mgeu_disc}), being in agreement with the observational trend (except for case 4). The evolution of [Eu/Mg] also becomes flat as can be seen in the measured stellar abundances. This is due to the progressively decreasing NSM rate compared to that without kicks (Fig.~\ref{fig:kick}; top left). In fact, the effective delay-time distribution becomes $D(t) \langle f_\mathrm{kick}(t) \rangle \propto t^{-2}$ after a sufficiently long time (except for case 3). This leads to a similar evolution of Eu to that of Mg. 

Provided that the deficiencies of Eu (by about a factor of two except for case 3) can be attributed to the uncertainties in the Eu yield or the NSM rate, the natal kicks appear to reconcile the modelled [Eu/Fe] (or [Eu/Mg]) trend with that of measurements at high metallicity. It should be noted, however, this mechanism works only when the MW disc is treated as a single system. As found in the bottom-middle panel of Fig.~\ref{fig:kick}, the two-infall model results in a clear separation of the evolutionary curves of [Eu/Fe] between the thick and thin disc components, which cannot be seen in the measured stellar abundances. The curve of [Eu/Mg] also leads to a bifurcation between the thick and thin disc components (the bottom-right panel), which disagrees with its observational flat trend. This is a consequence of the fact that the steepening of [Eu/Fe] (or the flattening of [Eu/Mg]) owes the small $h$ of the thin disc and thus the resulting smaller NSM rate (the bottom-left panel of Fig.~\ref{fig:kick} for $t > 5$ Gyr), which is not the case for the thick disc ($t < 5$ Gyr). This implies that the effect of the natal kicks on the GCE is subdominant because of, e.g., overall small kick velocities ($\ll 20$ km s$^{-1}$) in reality.

\section{Discussion}
\label{sec:discussion}

Our GCE model for the MW halo is constructed on the basis of the observationally well-established mass-metallicity relation of galaxies in the Local Group \citep[][see \S~\ref{subsec:sfof}]{Kirby2013}. Recent semi-analytical models also indicate that the same relation holds for $M_*/M_\odot = 10^3$--$10^8$ \citep[][]{Xia2019a,Xia2019b}. However, \citet{Simon2019} has shown that additional data for UFDs exhibit some scatter in metallicity for a given $M_*$ with a tendency of higher $\langle \mathrm{[Fe/H]} \rangle$ than that from the relation of \citet{Kirby2013}. This may indicate that the mass-metallicity relation of \citet{Kirby2013} is inapplicable to UFD-sized galaxies. Alternatively, this may be due to the fact that such small systems are prone to be affected by tidal stripping (that reduces $M_*$ without changing $\langle \mathrm{[Fe/H]} \rangle$). Because of such uncertainties, our results, in which UFD-sized building blocks play important roles (\S~\ref{subsec:halo}), should be regarded as to be qualitative.

Another limitation of our models is that each single (building block, dwarf or disc) component is treated as a one-zone homogeneous gaseous system with outflow or inflow. This inhibits us to quantify any intrinsic chemical inhomogeneity in single systems \citep[such as that pointed out by][]{Leaman2012}.

Keeping in mind these major limitations of our models, we comprehensively discuss the following issues across the different components of the MW.

\subsection{What makes NSMs to occur at low metallicity?}
\label{subsec:metallicity}

In this study, the smaller SFE for a building-block galaxy with smaller $M_*$ has resulted in the appearance of Eu at low metallicity ([Fe/H] $\lesssim -3$) in the MW halo, as suggested in \cite{Prantzos2006} and showed in previous studies with a similar approach \citep{Ishimaru2015,Komiya2016,Ojima2018}. This is due to the slower increase of [Fe/H] in a less-massive building block, leading to an appearance of NSMs at lower metallicity. Given that the satellite dwarfs share similar evolutionary histories to those of building blocks, this $M_*$-dependent SFE prescription reconciles our model with the $M_*$-dependent knee position of [$\alpha$/Fe] observed in dwarf spheroidals \citep{Tolstoy2009,Reichert2020}.

Currently, the latest cosmological zoom-in simulations of MW-analogous galaxies barely resolve the UFD-sized structures \citep[][]{Applebaum2021} but not the GCE in their sub-scales. It is not clear, therefore, that the same interpretation holds in the previous hydrodynamical simulations of MW-analogous galaxies \citep{Shen2015,Voort2015,Naiman2018,Haynes2019,Voort2020}. Some of these studies rather attribute the reason to the gas inhomogeneity owing to large-scale (but incomplete) metal mixing beyond the scale of building blocks owing to, e.g., galactic wind or accretion flow \citep{Shen2015,Voort2015}, an effect not considered in this study. For instance, mixing between a nearly primordial gas (i.e., with little Fe and Eu) and that already enriched in Fe and Eu may lead to an appreciable reduction of [Fe/H] but not of [Eu/Fe]. Therefore, two different effects, $M_*$-dependent SFE and large-scale metal mixing, may be invoked to justify the presence of NSMs at low metallicity. Cosmological simulations with higher resolution will be needed to elucidate which of these effects (or both) plays a more important role.

\subsection{What gives rise to star-to-star scatter in Eu?}
\label{subsec:scatter}

There are two reasons that give rise to scatter in Eu as also found in previous similar studies \citep{Ishimaru2015,Komiya2016,Ojima2018}. In our halo model, almost all r-process-enhanced stars ([Eu/Fe] $> 1$) originate from UFD-sized building blocks ($M_* < 10^5$). This is in agreement with the implication from the semi-analytic analysis of dark-matter cosmological simulations \citep[][]{Brauer2019}. For r-process-deficient stars ([Eu/Fe] $< 0.5$), the $M_*$-dependent SFE for massive building blocks (probably including the in situ stellar halo) leads to different values of [Eu/Fe] at a given metallicity. According to these two reasons (here we refer these to $M_*$-dependent [Eu/Fe]), our model predicts a scatter of [Eu/Fe] over  four orders of magnitude, including currently undetected stars with [Eu/Fe] $< -0.6$. 

As previous cosmological simulations \citep{Shen2015,Voort2015,Naiman2018,Haynes2019,Voort2020} do not resolve the GCE of UFD-sized building blocks, the source of Eu-scatter in these studies is probably different from ours. One of possible sources is limited metal mixing, which is induced by supernovae (or NSMs) as well as turbulence in reality. Here, we refer to it as small-scale metal mixing (e.g., within each building block), an effect not considered in our model. In fact, some early work has suggested that the small-scale metal mixing induced by CCSNe (i.e., the average of the nucleosynthesis product and the swept-up inter-stellar medium) leads to large scatter in [Eu/Fe] \citep{Ishimaru1999,Tsujimoto1999}. \citet{Argast2004} showed, however, that the small-scale metal mixing induced by NSMs (as well as that by CCSNe) led to too large scatter in [Eu/Fe] to be compatible with the measurements \citep[see also][]{Qian2000}. Therefore, even more efficient mixing (but within a building block) owing to, e.g., turbulence should exist, given that NSMs are the dominant sources of the r-process elements \citep{Beniamini2020,Dvorkin2020}. Both mechanisms, $M_*$-dependent [Eu/Fe] and small-scale metal mixing, are conceivable for the cause of scatter in Eu. We consider, however, that the former plays a more important role, because small-scale metal mixing should also work in the disc and dwarf galaxies as well. Currently, no large scatter in [Eu/Fe] has been found among measured stars in the disc and dwarf galaxies \citep[even though small-scale metal mixing still exists,][]{Leaman2012}. 

\subsection{Can NSMs be the unique site of the r-process?}
\label{subsec:merger}

It appears that NSMs can be the predominant sources of the r-process elements when we adopt our fiducial combination of the delay-time distributions for SNe Ia and NSMs (case 1). Our model requires  a fairly limited, but not necessarily null, contribution of SNIa before 1 Gyr as shown in APPENDIX \ref{sec:appendex} (cases 4A--4C). 

For the MW halo, $M_*$-dependent SFR (\S~\ref{subsec:metallicity}) and $M_*$-dependent [Eu/Fe] (\S~\ref{subsec:scatter}) can explain the appearance of Eu at low metallicity and its star-to-star scatter, respectively. We emphasize that such effects cannot be included when the halo is treated as a single system \citep[e.g.,][]{Argast2004}. These effects also reconcile our model of satellite dwarfs with the presence of r-process-enriched UFDs. For the classical dwarf spheroidals and the MW disc, our models (case 1) are (marginally) consistent to the observational trends of [Eu/Fe] and [Eu/Mg]. The accordance will be even better if a greater fraction of the early component, i.e., a smaller $A$ in equation~(\ref{eq:dtdnsm}), is adopted. In fact, it is suggested that the fraction of the early component can be up to 60--80\% \citep{Beniamini2019,Galaudage2021}. 

\subsection{Can collapsars or MRSNe be the dominant sources of the r-process elements?}
\label{subsec:massive}

We have modeled the possible contribution of subsets of CCSNe (e.g., collapsars or MRSNe)  with little delay from star formation (case 3). For the MW halo, our model predicts no r-process-deficient stars ([Eu/Fe] $< 0$) at [Fe/H] $\sim -3$, being in disagreement with measurements. This is in line with few measured stars with [Mg/Fe] $< 0$ at low metallicity, when considering the origin of Mg being CCSNe. Large- or small-scale mixing, which is absent in this study, may yield some stars with low [Eu/Fe]. As pointed out by \citet[][see their fig.~1]{Argast2002}, such incomplete metal mixing also can lead to a large scatter of [Mg/Fe] at low metallicity, reflecting the adopted mass-dependent CCSN yields. This implies that metal mixing in the MW halo (or the building blocks) was fairly efficient. Note that, however, the intrinsic scatter of [Mg/Fe] owing to the mass-dependent yields is expected to be much smaller if the stars above, e.g., $25\, M_\odot$ collapse to black holes \citep[as adopted in][]{Prantzos2018}. 

It should be noted that we intrinsically assume a rarity of such events (2\% of all CCSNe; see also \citealt{Tsujimoto2015}). Thus, the above interpretation may be inadequate if the rarity is due to the limited mass range of progenitors (in addition to, e.g., rapid rotation or strong magnetic field), because the delay for case 3 ($t_\mathrm{min} = 0.005$ Gyr) corresponds to the lifetime of $\sim 40\, M_\odot$ stars. In APPENDIX \ref{sec:appendex}, we also test the same case 3 but with longer delay, $t_\mathrm{min} = 0.01$ Gyr (model 3A) and $t_\mathrm{min} = 0.03$ Gyr (model 3B), which correspond to the lifetimes of $\sim 20\, M_\odot$ and $\sim 9\, M_\odot$ stars, respectively (note that the result with $t_\mathrm{min} < 0.005$ Gyr, corresponding to the stellar life for $> 40\, M_\odot$, will be the same as that of case 3 for [Fe/H] $> -4$; see Fig.~\ref{fig:appendix_tmin} in APPENDIX \ref{sec:appendex}). We find that the result for the latter is in good agreement with the observational trend of [Eu/Fe] in the MW halo. Solely from a point of view of GCE, therefore, low-mass CCSNe can be the dominant sources of the r-process elements, a conclusion already reached in earlier work \citep[e.g.,][]{Ishimaru2004}. Note that the diffusion time of r-process elements in gas (not considered in this study) owing to the low frequency of events can also lead to a delay in r-process enrichment, although the delay from star formation ($t_\mathrm{min}$) appears more important \citep{Tarumi2021}. 

In this regard, collapsars (but not necessarily low-mass MRSNe), which are expected to result from rather massive stars, are probably excluded from the candidates for the major r-process site. Conversely, we find no reason to exclude a possibility that such subsets of CCSN play a substantial role for the enrichment of Eu in the dwarf satellites and the MW disc. If the dwarf satellites are the surviving building blocks of the MW halo, the above problem will also arise in these galaxies. However, a lack of stars with measured Eu at low metallicity in dwarfs hampers such a comparison.

\subsection{Do the natal kicks of binary neutron stars affect the Eu evolution at high metallicity?}
\label{subsec:kick}

We have shown that the decreasing number of binary neutron stars in the thin disc owing to the natal kicks may make the slopes of [Eu/Fe] and [Eu/Mg] steeper and flatter, respectively. A recent work by \citet{Banerjee2020} also suggests that the natal kicks combined with the so-called ``inside-out evolution" of the thin disc leads to a similar result. Such mechanisms may work, however, only in the thin disc but not in the thick disc. Moreover, the trend of decreasing [Eu/Fe] (and flat [Eu/Mg]) at high metallicity can be commonly seen in satellite dwarfs \citep[][see also Figs. \ref{fig:eufe} and \ref{fig:eumg}]{Reichert2020} and the MW bulge \citep{Johnson2012}. Therefore, an inherent mechanism of the thin disc cannot be the cause of settling the evolutionary trend of Eu. In addition, the bulk of neutron star binaries should have relatively small kick velocities ($< 15$ km/s) to account for the presence of r-process-enriched UFDs \citep{Beniamini2016b}. Note that other possibilities have been postulated, e.g., different conditions of star formation in the regions nearby supernovae and NSMs \citep{Schoenrich2019} or radial migration \citep{Tsujimoto2019}. Our model (for case 1) without natal kicks also gives a result marginally consistent with the observational trends of [Eu/Fe] and [Eu/Mg].

\section{Conclusions}
\label{sec:conclusion}

We have examined the enrichment histories of Eu as representative of the r-process elements in the halo, the disc and the satellite dwarf galaxies of the MW by assuming that NSMs are the unique r-process site. Particular attention was payed to the functional forms of delay-time distribution for both SNe Ia and NSMs by exploring  modifications to the commonly used time dependence of $\propto t^{-1}$. The Galactic halo was modeled as an ensemble of one-zone well-mixed building blocks. The satellite dwarf galaxies were assumed to be surviving building blocks. As for the disc, simple one- and two-infall models were adopted, the latter mimicking the evolution of thick and thin discs.


On the basis of our results, we argue that NSMs can be the predominant sources of the r-process elements throughout the enrichment history of the MW and its satellites.  The, still poorly known,  delay-time distributions of SNe Ia and NSMs play key roles in the evolution of Eu; we find that the required conditions are a subdominant contribution of the former until $\sim 1$ Gyr and an appreciable contribution (providing $\gtrsim 50\%$) of the latter at $\sim 0.1$ Gyr after star formation. 
In the MW halo, r-process-enhanced ([Eu/Fe] $> 1$) and r-process-deficient ([Eu/Fe] $< 0.5$) stars likely originate from UFD-sized ($M_* < 10^5\, M_\odot$) and massive ($M_* > 10^5\, M_\odot$) building-block galaxies, respectively; the latter includes the in situ halo. Subsets of CCSNe, e.g., collapsars or MRSNe are not favored as candidates for the major r-process site in the MW halo but cannot be excluded as those in the satellite dwarfs or the disc. Finally, the natal kicks of NSMs appear to play a subdominant role for steepening [Eu/Fe] (and flattering [Eu/Mg]) at high metallicity.

More progress is needed in both theoretical and observational studies before drawing a firm conclusion on the origin of the r-process elements.   In simplistic approaches as this one, a number of non-trivial assumptions are introduced, e.g., regarding the assembly history of building blocks, the evolution of UFD-like systems and metal mixing. Such simplistic approaches should be testified by the future cosmological zoom-in simulations that resolve the UFD-sized structures of MW-like galaxies. More measurements of stellar abundances in terms of both the number of stars and their accuracy will serve to refine the modelling of the enrichment history of the r-process elements in the MW (e.g., R-Process Alliance, \citealt{Hansen2018,Sakari2018,Holmbeck2020,Gudin2021}; GALAH, \citealt{Buder2021,Aguado2021,Matsuno2021}; LAMOST, \citealt{Chen2021}). In particular, more data are highly desired for UFDs and classical dwarf spheroidals as survivors from the early Galactic merging history.

\section*{Acknowledgements}

The authors acknowledge the contribution of the late Y. Ishimaru to this project: she initiated it, she developed the original version of the numerical code \texttt{iGCE} that was used throughout this study and she provided critical insight in various physical aspects of it. SW is supported by the RIKEN iTHEMS Project. YH is supported by the Special Postdoctoral Researchers (SPDR) program at RIKEN and JSPS KAKENHI Grant Numbers 20K14532, 19H01933, 21H04499, 21K03614 and 21J00153.

\section*{Data availability}

The data underlying this article will be shared on reasonable request to the corresponding author.





\bibliographystyle{mnras}
\bibliography{reference} 





\appendix

\section{Additional tests for the evolution in the MW halo}
\label{sec:appendex}

\begin{table}
	\centering
	\caption{Parameters in the additional tests to original cases (Table~\ref{tab:dtd}) with replacements specified in bold. For some cases, the values of $K$ in equation~(\ref{eq:sfof}) are also changed (specified in italic) to obtain [Fe/H]$_\mathrm{peak} \approx -1.55$. The units of time is in Gyr.}
	\label{tab:appendix}
	\begin{tabular}{lcccccc} 
		\hline
		         & $t_\mathrm{min}$ (SN Ia) & $p$          & $t_\mathrm{min}$ (NSM) & $K$            & $k_\mathrm{SF}$     & $k_\mathrm{OF}$     \\
		\hline
		case 4A  & \textbf{0.50}         & ---          & 0.020                     & \textit{0.059} & eq. (\ref{eq:sfof}) & 1.0                 \\
		case 4B  & 0.10                  & \textbf{0.0} & 0.020                     & \textit{0.043} & eq. (\ref{eq:sfof}) & 1.0          		 \\
		case 4C  & 0.10                  & \textbf{1.0} & 0.020                     & \textit{0.053} & eq. (\ref{eq:sfof}) & 1.0          	     \\
		case 1A  & 1.0                   & ---          & \textbf{0.050}            & 0.045          & eq. (\ref{eq:sfof}) & 1.0                 \\
		case 1B  & 1.0                   & ---          & \textbf{0.10}             & 0.045          & eq. (\ref{eq:sfof}) & 1.0                 \\
		case 1C  & 1.0                   & ---          & 0.020                     & \textit{0.039} & eq. (\ref{eq:sfof}) & \textbf{2.0}        \\
		case 1D  & 1.0                   & ---          & 0.020                     & 0.045          & \textbf{0.18}       & eq. (\ref{eq:sfof}) \\
		case 3A  & 1.0                   & ---          & \textbf{0.010}            & 0.045          & eq. (\ref{eq:sfof}) & 1.0                 \\
		case 3B  & 1.0                   & ---          & \textbf{0.030}            & 0.045          & eq. (\ref{eq:sfof}) & 1.0                 \\
		case 3C  & 1.0                   & ---          & \textbf{0.0030}           & 0.045          & eq. (\ref{eq:sfof}) & 1.0                 \\
		\hline
	\end{tabular}
\end{table}

\begin{figure}
	\includegraphics[width=0.86\columnwidth]{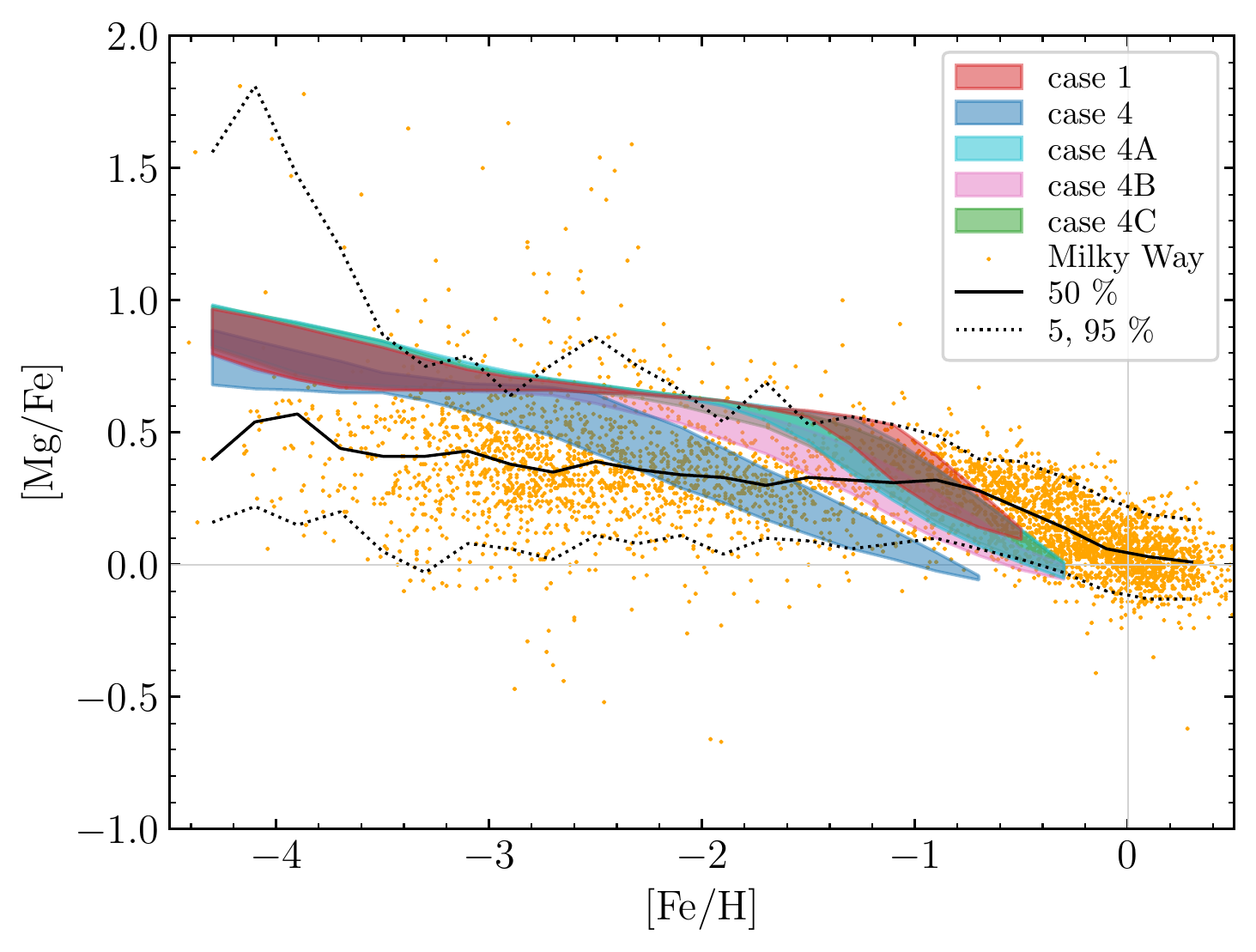}
    \caption{Evolution of [Mg/Fe] for cases 4A (cyan), 4B (magenta) and case 4C (green) in Table~\ref{tab:appendix} as well as those in \S~\ref{subsubsec:mg_halo} (cases 1 and 4; red and blue, see also Fig.~\ref{fig:mgfe_halo}). The range of colour indicates the 90\% coverage of stars at a given [Fe/H] for each model (note that the results for cases 4A and 4C are mostly overlapped). The same range for the MW stars (orange dots) is drawn by the dotted curves (between 5\% and 95\% with the 50\% count indicated by the solid curve).}
    \label{fig:appendix_mg}
\end{figure}

\begin{figure*}
	\includegraphics[width=0.67\columnwidth]{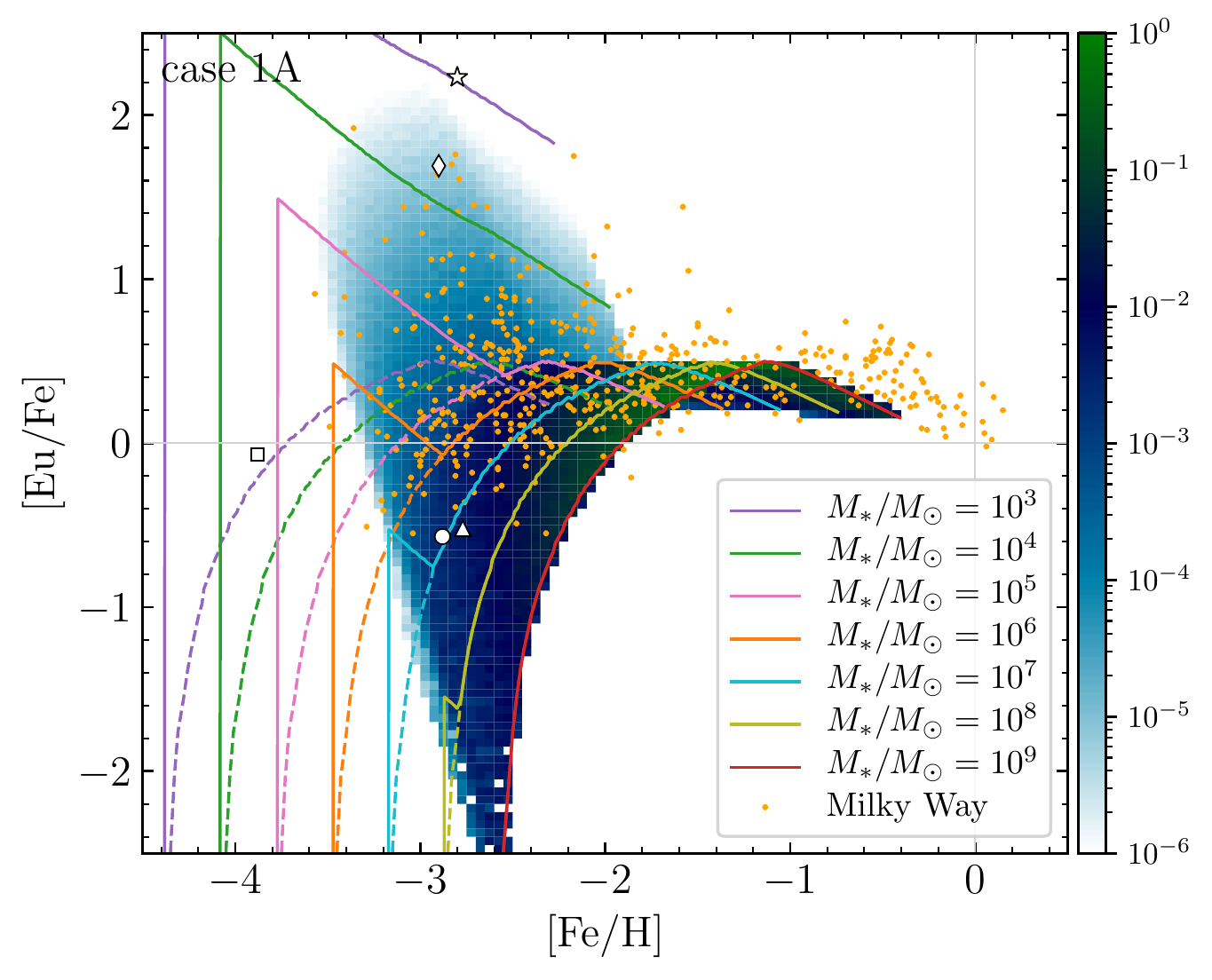}
	\includegraphics[width=0.67\columnwidth]{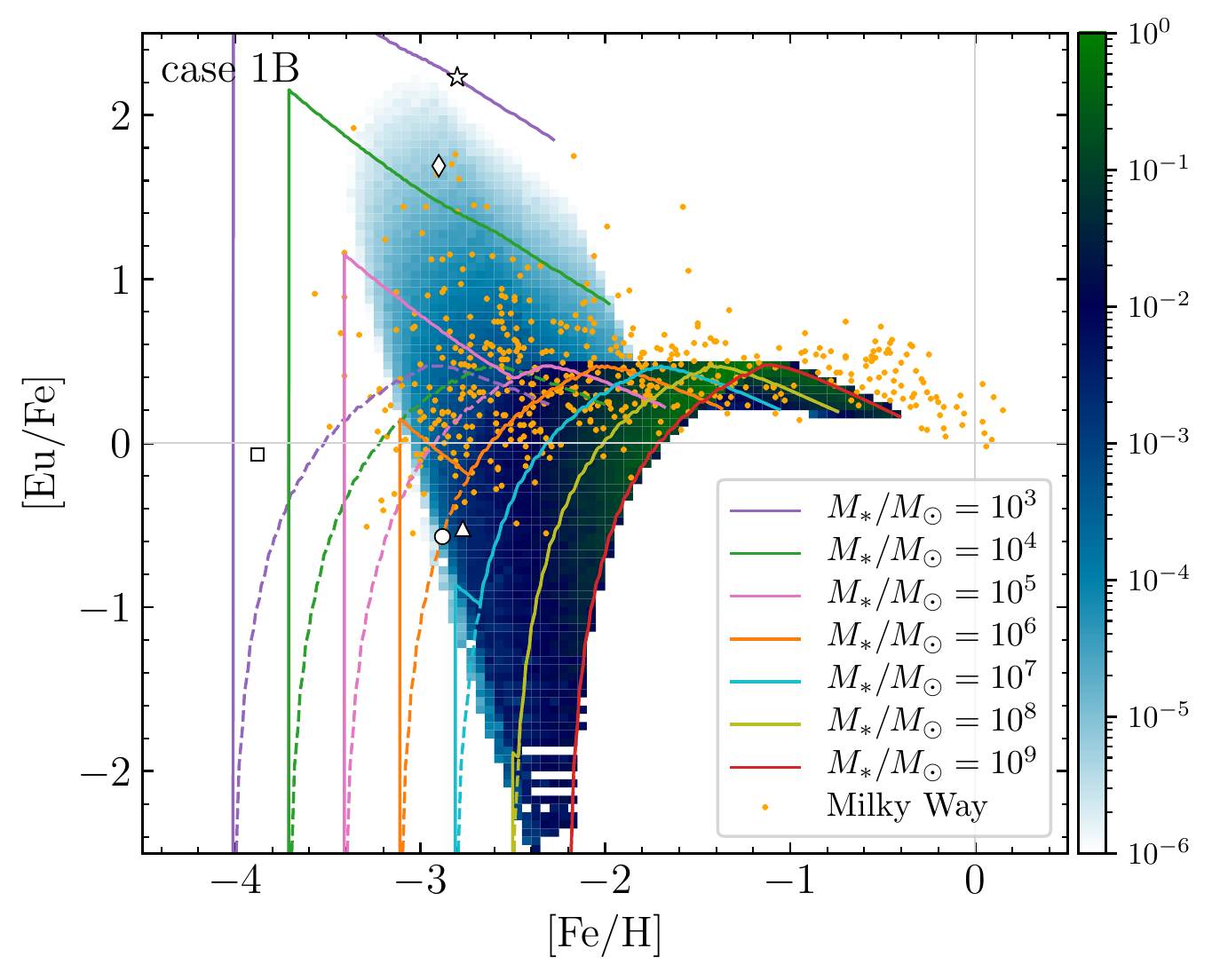}
	\includegraphics[width=0.67\columnwidth]{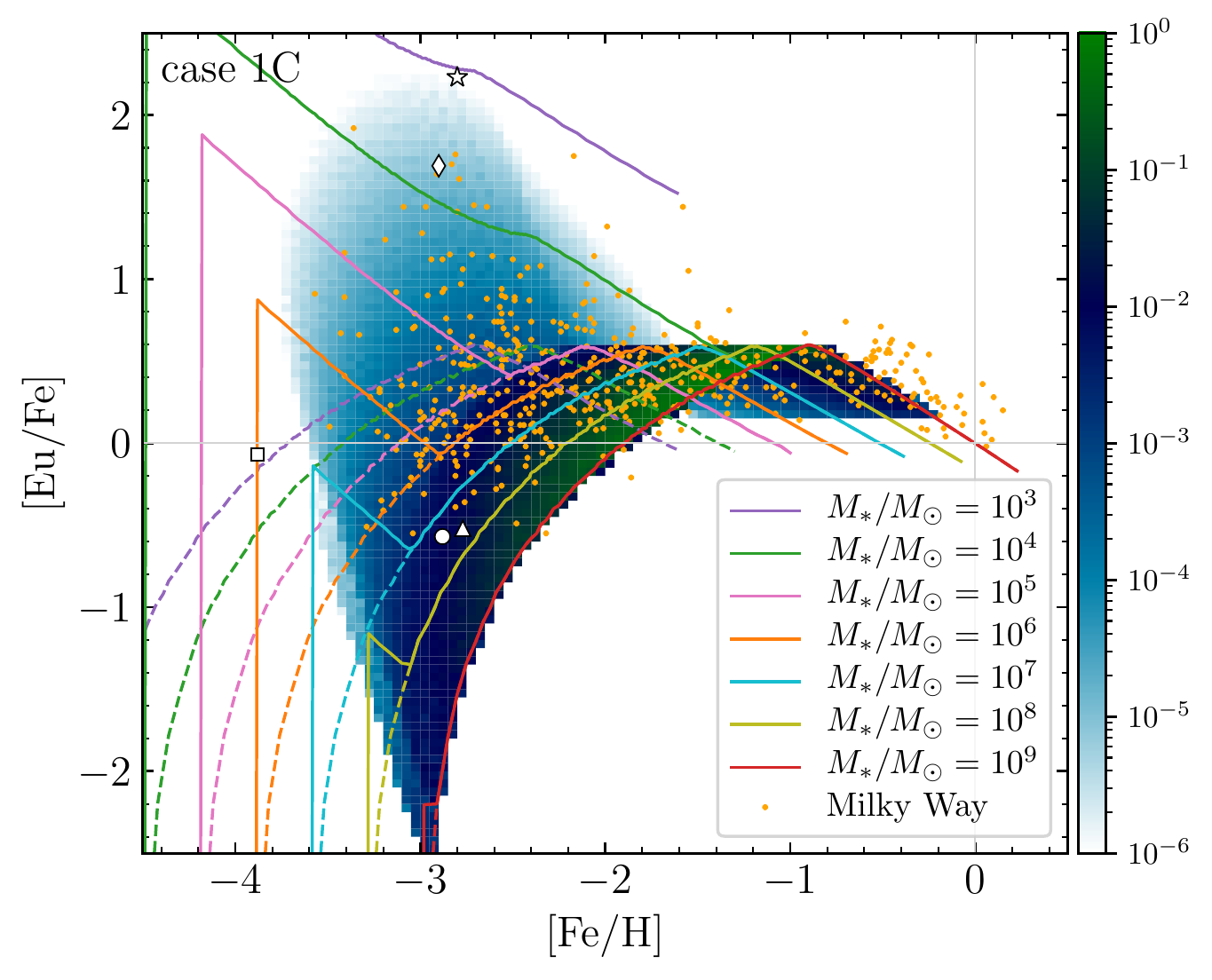}
	\includegraphics[width=0.67\columnwidth]{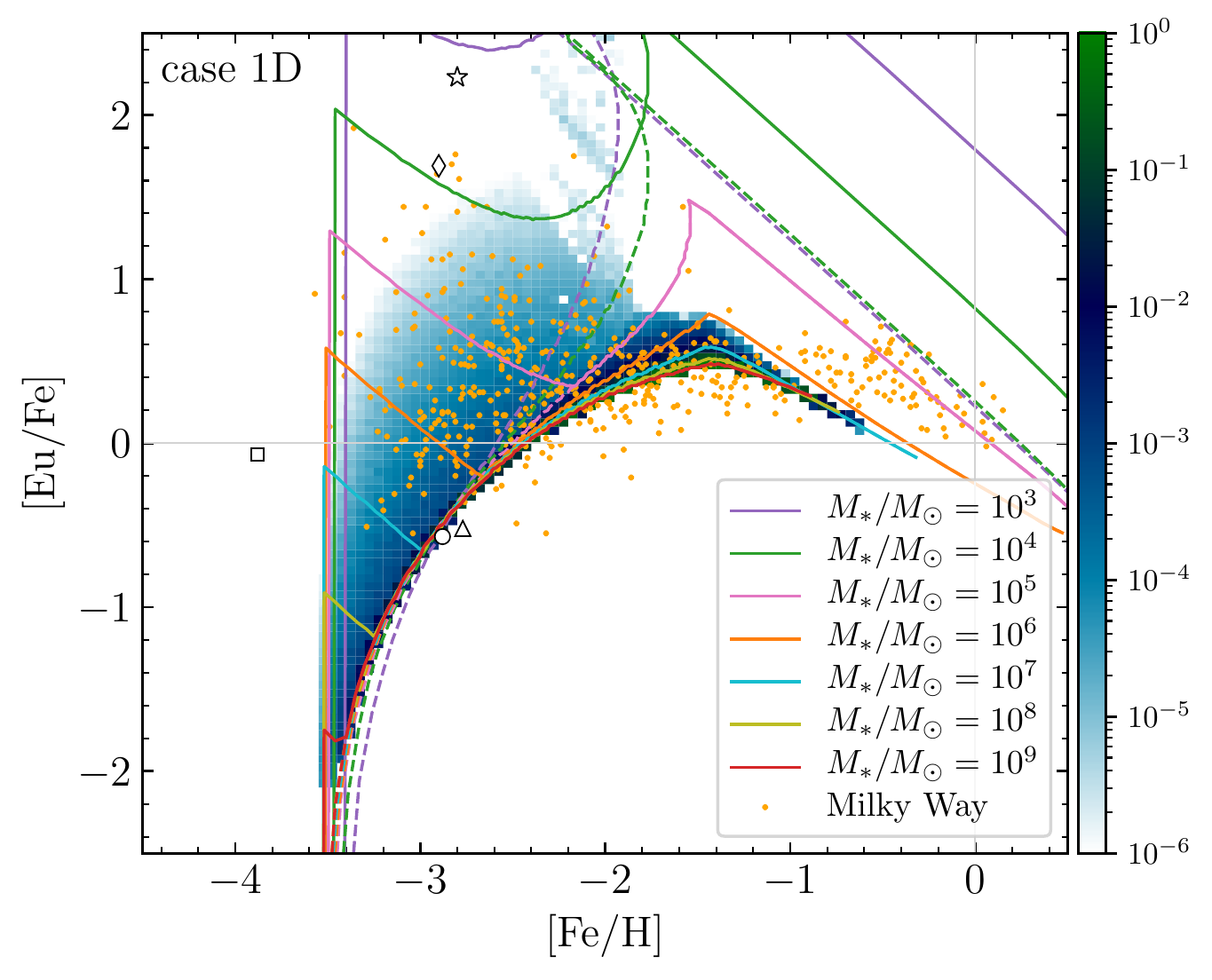}
	\includegraphics[width=0.67\columnwidth]{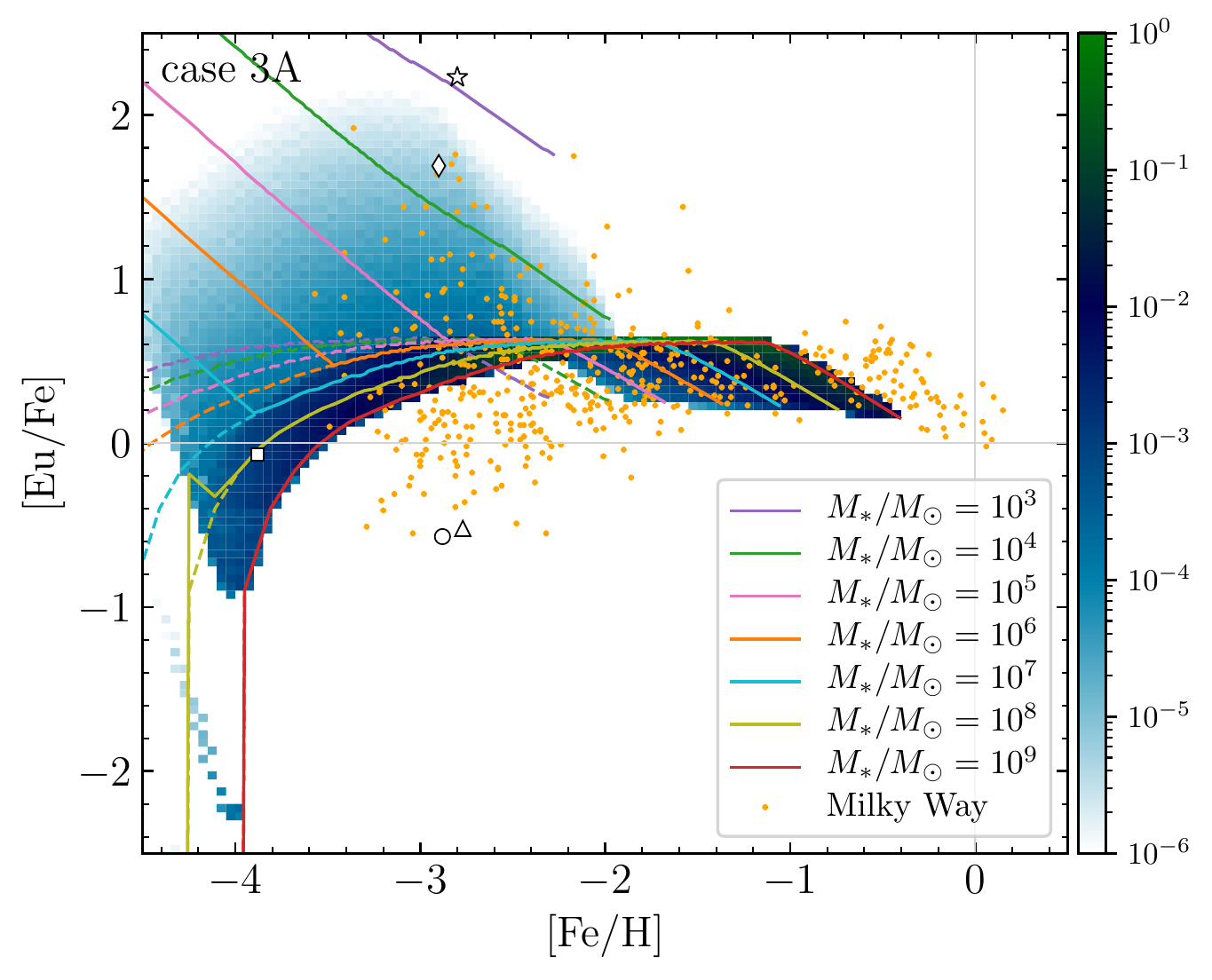}
	\includegraphics[width=0.67\columnwidth]{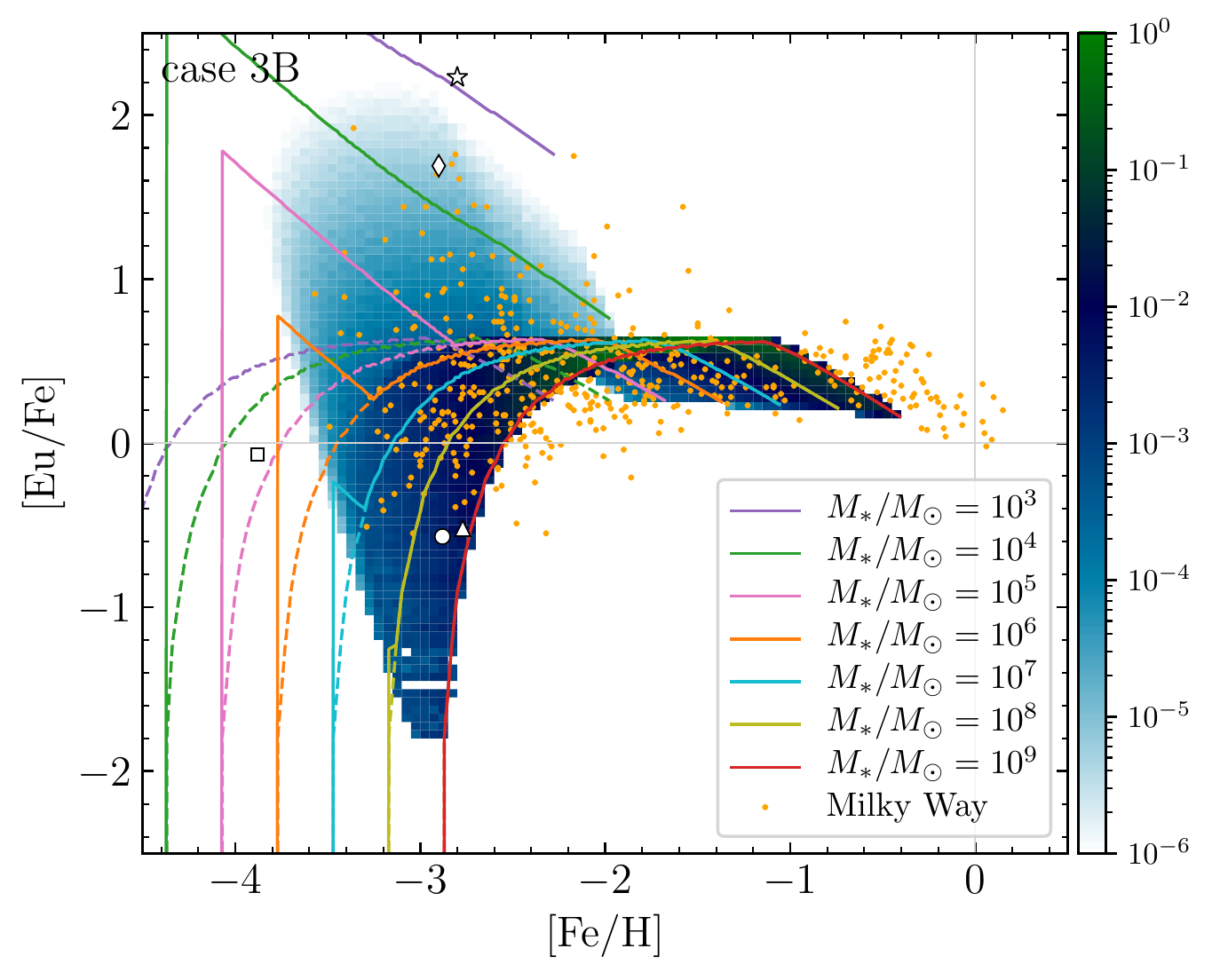}
    \caption{Same as Fig.~\ref{fig:eufe_halo}, but for cases 1A, 1B, 1C, 1D, 3A and 3B in Table~\ref{tab:appendix}.}
    \label{fig:appendix_eu}
\end{figure*}

We present some additional tests for the evolution in the MW halo, in which one parameter in our original case is changed as indicated by the bold face in Table~\ref{tab:appendix}.

Fig.~\ref{fig:appendix_mg} shows the evolution of [Mg/Fe] for each model, in which the 90\% coverage of stars at a given [Fe/H] is indicated by the colour specified in the legend. The same range for the MW stars (orange dots) is drawn by the dotted curves (between 5\% and 95\% with the 50\% count indicated by the solid curve). As described in \S~\ref{subsubsec:mg_halo}, the descending [Mg/Fe] at low metallicity for case 4 (blue) is inconsistent with the overall flat trend for the MW stars (up to [Fe/H] $\sim -1$). Case 4A (cyan) is the same as case 4 but with the minimum delay for SNe Ia replaced by $t_\mathrm{min} = 0.5$ Gyr. The [Mg/Fe]-knee emerges from only slightly lower metallicity ($\Delta$[Fe/H] $\sim -0.2$) than that in our fiducial case 1. Cases 4B (magenta) and 4C (green; mostly overlapped with cyan) test a different form of delay-time distribution such as $D(t) = A\, t^p$ ($A$ is the normalization constant and $p$ is the value in the third column of Table~\ref{tab:appendix}) for $0.1 < t < 1.0$ and $D(t) = A\, t^{-1}$ for $1.0 < t < 10$ (in Gyr). The [Mg/Fe]-knee emerges from a slightly low metallicity compared to that in case 1 ($\Delta$[Fe/H] $\sim -0.4$ and $-0.2$ for cases 4B and 4C, respectively) as well, although its transition becomes smoother for case 4B. Overall, the evolutionary trends of these additional cases (and case 1) appear to be consistent with that of MW stars. 

Cases 1A and 1B (the top left and middle panels of Fig.~\ref{fig:appendix_eu}) are the same as case 1 but with the minimum delay for NSMs replaced by $t_\mathrm{min} = 0.05$ Gyr \citep[the minimum value estimated from the observation of binary neutron stars,][]{Stovall2018} and 0.1 Gyr, respectively. These cases are still compatible to the highly r-process-enhanced stars with [Eu/Fe] $> 1$ at [Fe/H] $\sim -3$. However, more stars with [Eu/Fe] $\sim -0.5$--1 at [Fe/H] $\lesssim -3$ get out of the predicted bluish region as $t_\mathrm{min}$ increases. 

In this study, we have fixed the OFE to be $k_\mathrm{OF} = 1.0$ (Gyr$^{-1}$). Case 1C tests a higher value of $k_\mathrm{OF} = 2.0$ (Gyr$^{-1}$), which increases $k_\mathrm{SF}$ according to equation~(\ref{eq:sfof}). Fig.~\ref{fig:eufe_halo} (top left) and Fig.~\ref{fig:appendix_eu} (top right) show similar outcomes to each other, although the latter leads to a more outstanding effect of SNe Ia for [Fe/H] $> -2$. 

In case 1D, SFE, instead of OFE, is fixed to be $k_\mathrm{SF} = 0.18$ (Gyr$^{-1}$) as to give $k_\mathrm{OF} = 1.0$ (Gyr$^{-1}$) for $M_* = 10^8$ \citep[similar to Case 2 in][]{Ishimaru2015}. As displayed in the bottom-left panel of Fig.~\ref{fig:appendix_eu}, the building block galaxies of $M_* > 10^5$ lead to similar evolutionary tracks with the knees (owing to SNe Ia) at [Fe/H] $\sim -1.4$. This appears incompatible with the observation of satellite dwarfs that generally indicate the knee position at smaller metallicity for a less-massive galaxy (\S~\ref{subsec:dwarf}). In addition, the model predicts few stars with a large enhancement in [Eu/Fe] $> 1.5$ at [Fe/H] $\sim -3$ owing to a greater $k_\mathrm{OF}$ and thus an earlier gas consumption for a smaller-$M_*$ building block (equation~(\ref{eq:sfof})).

Cases 3A and 3B (the bottom middle and right panels of Fig.~\ref{fig:appendix_eu}) are the same as (CCSN-like) case 3 but with the minimum delay replaced by $t_\mathrm{min} = 0.01$ Gyr and 0.03 Gyr, respectively. The former and the latter approximately correspond to the lifetimes of $20\, M_\odot$ and $9\, M_\odot$ stars, respectively. The results indicate the latter being in good agreement with the observational trend of [Eu/Fe] in the MW halo.

\begin{figure*}
	\includegraphics[width=0.67\columnwidth]{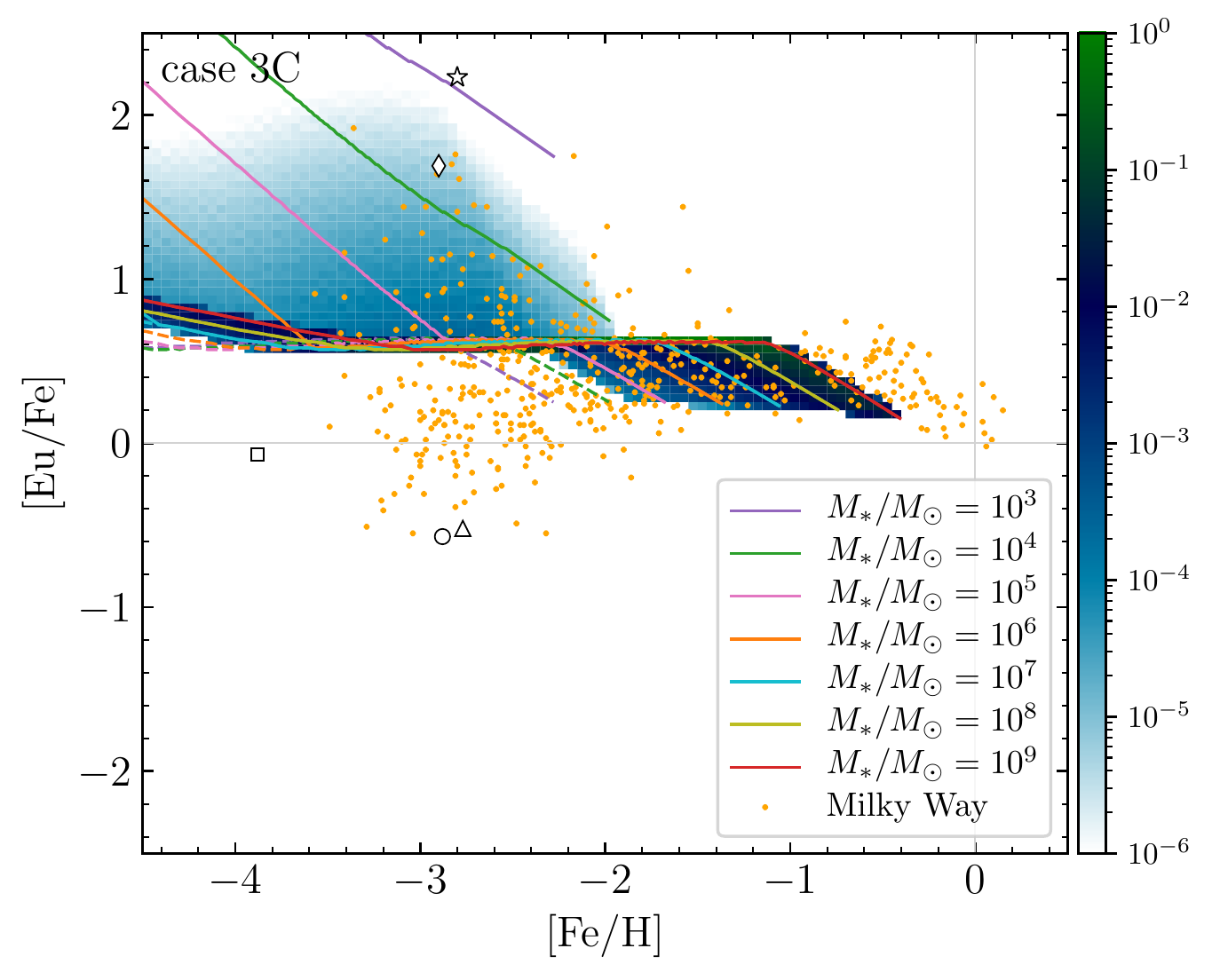}
	\includegraphics[width=0.67\columnwidth]{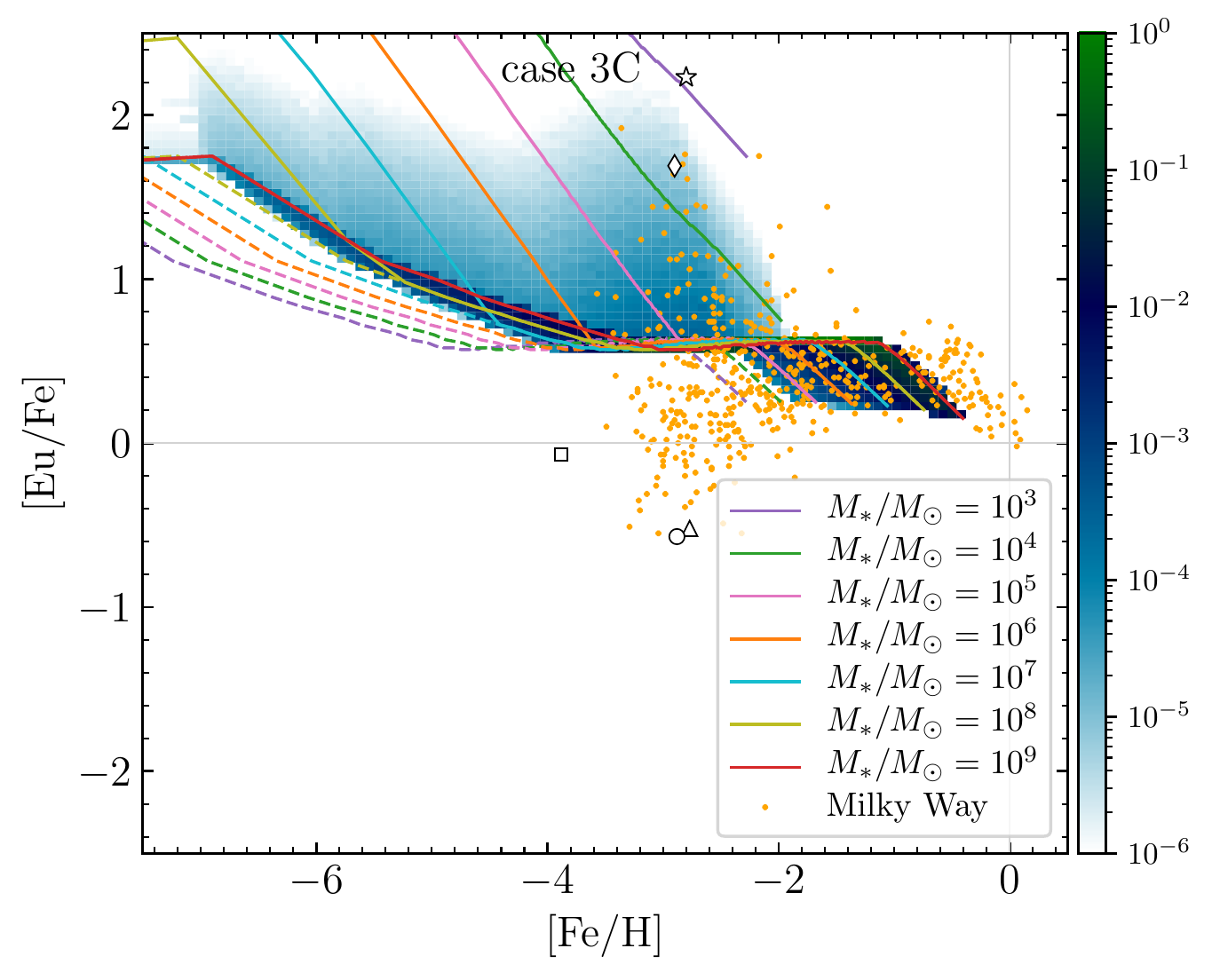}
	\includegraphics[width=0.67\columnwidth]{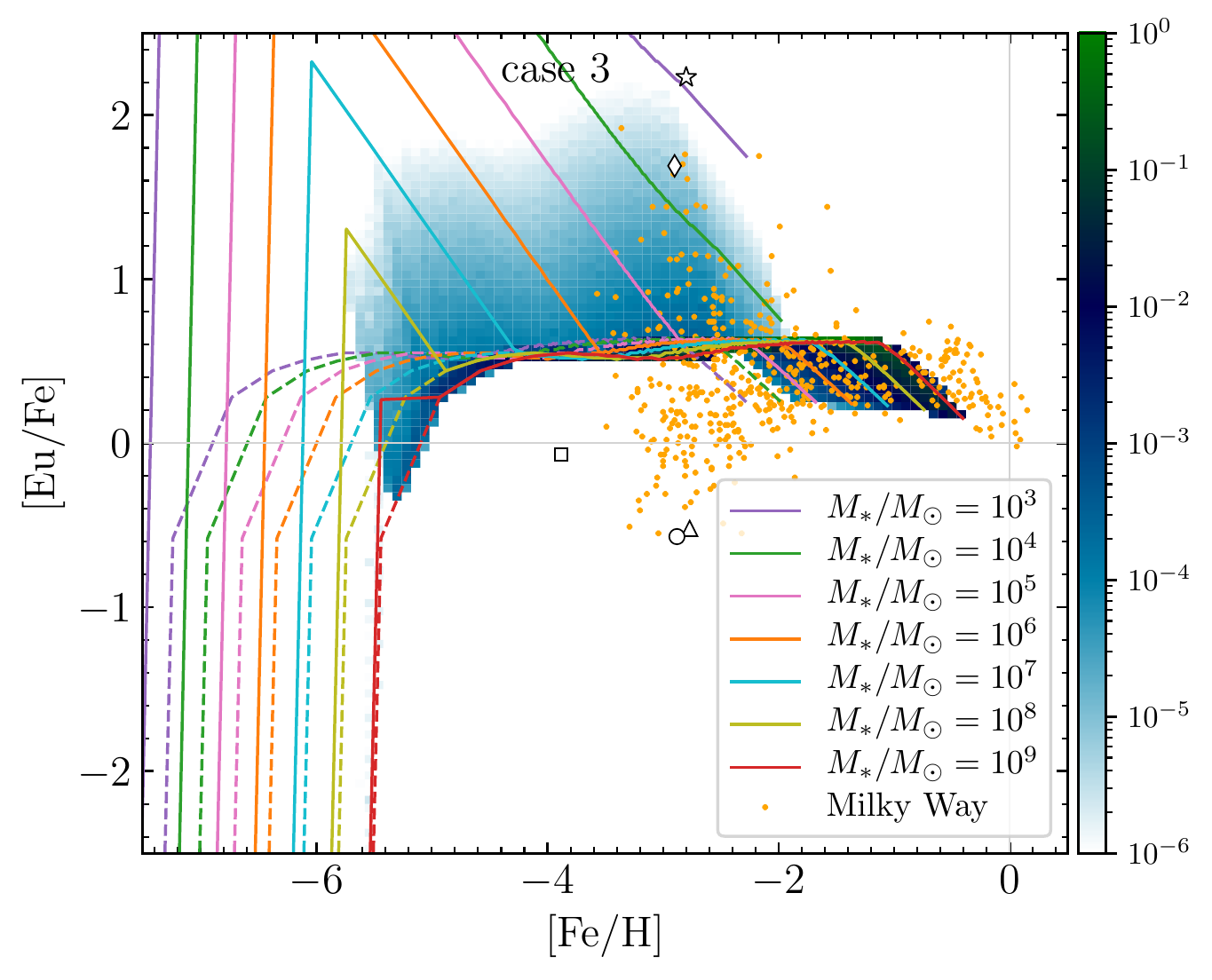}
    \caption{Same as Fig.~\ref{fig:eufe_halo}, but for case 3C (left panel). The middle and right panels are the same as those for cases 3C and 3, but plotting over a wider range towards low metallicity.}
    \label{fig:appendix_tmin}
\end{figure*}

Case 3C is the same as case 3 in Fig.~\ref{fig:eufe_halo} but with $t_\mathrm{min} = 0.003$ Gyr (corresponding to the stellar life of $\sim 100\, M_\odot$). As can be found in the left panel of Fig.~\ref{fig:appendix_tmin}, the result is similar to that of case 3 (Fig.~\ref{fig:eufe_halo}; bottom left). The middle and right panels in Fig.~\ref{fig:appendix_tmin} are the same as those for cases 3C and 3 but plotting over a wider range towards low metallicity. We find differences between cases 3 and 3C only for [Fe/H] $< -4$, in which no star with measured Eu exists.
    
\begin{figure*}
	\includegraphics[width=0.86\columnwidth]{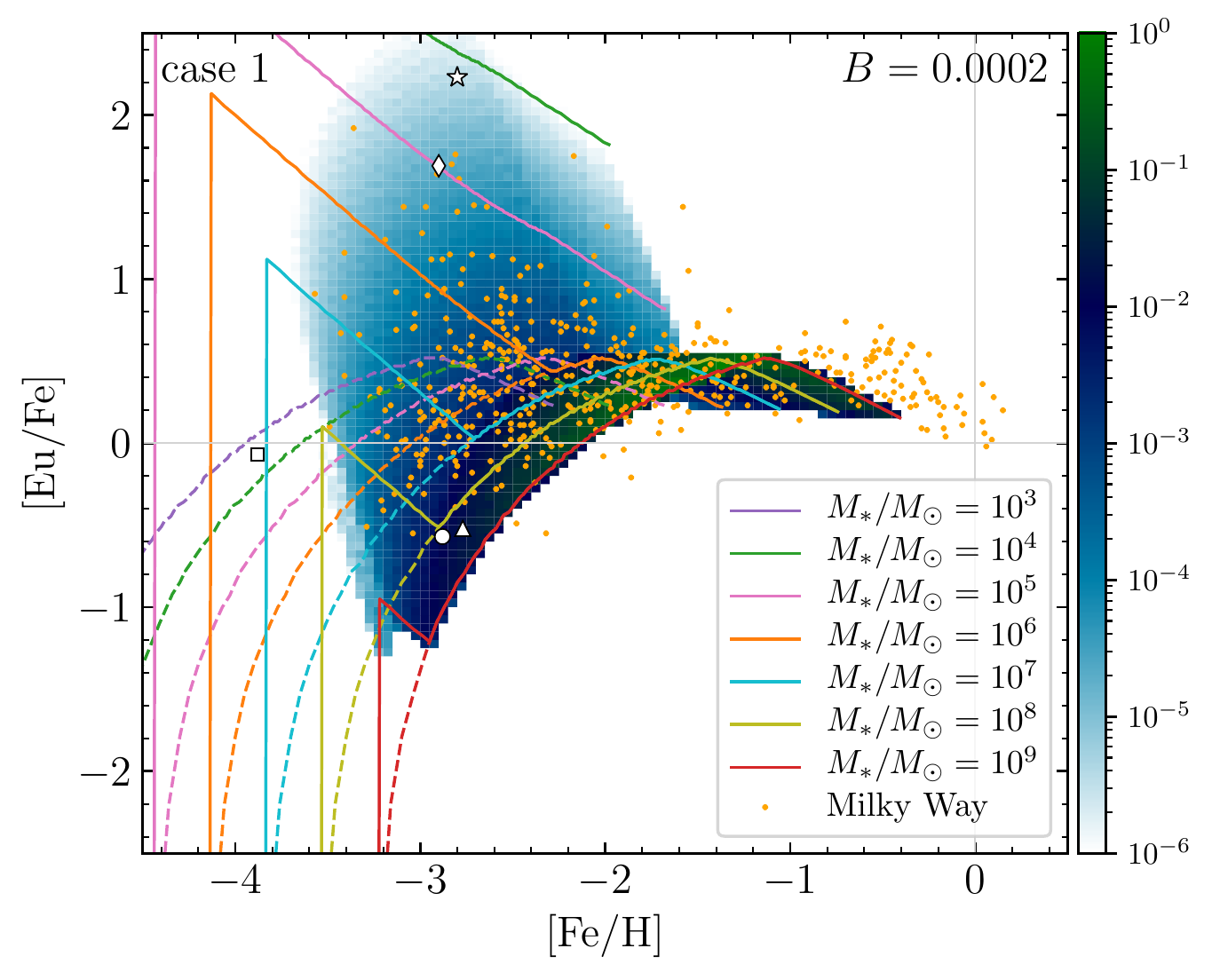}
	\includegraphics[width=0.86\columnwidth]{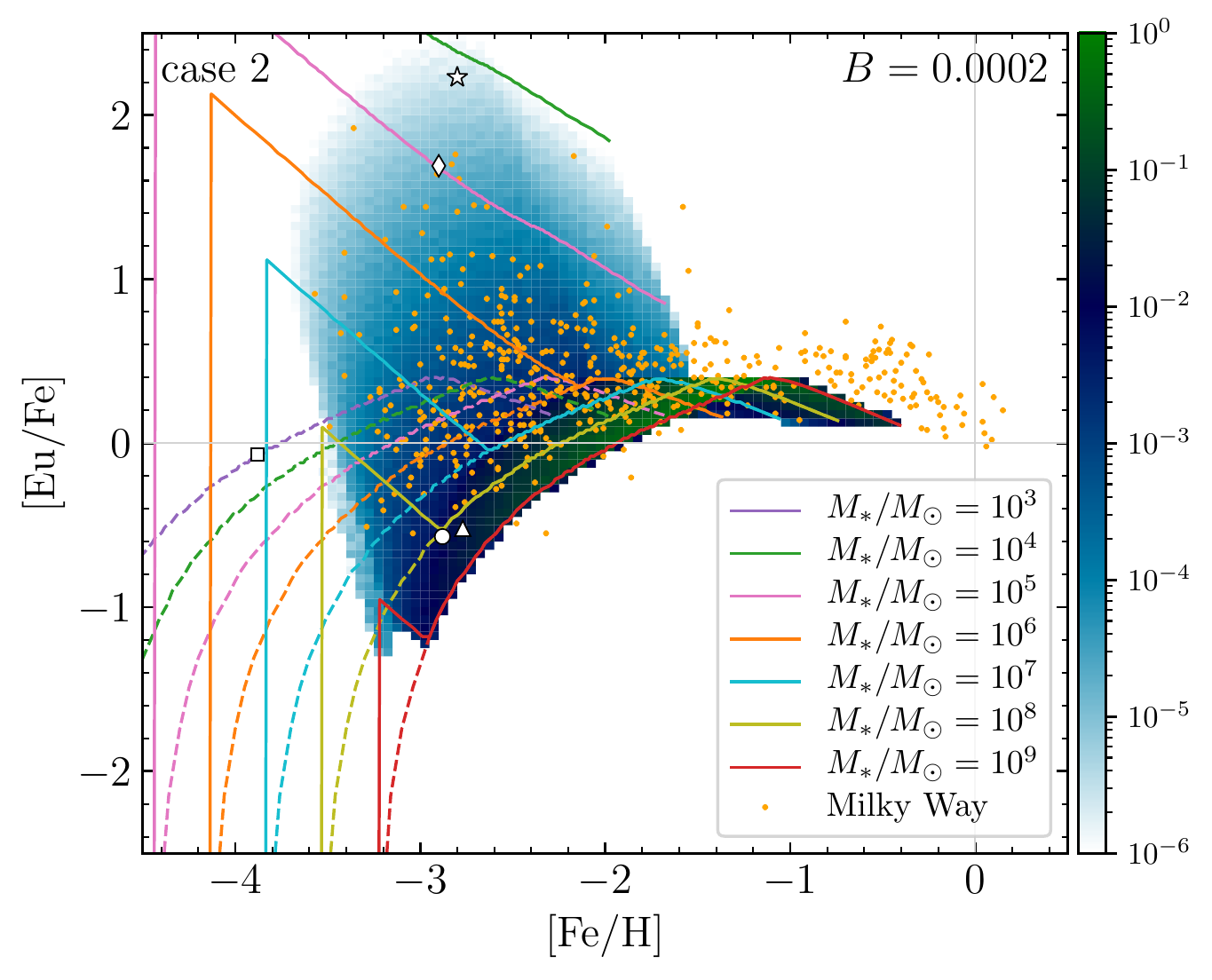}
	\includegraphics[width=0.86\columnwidth]{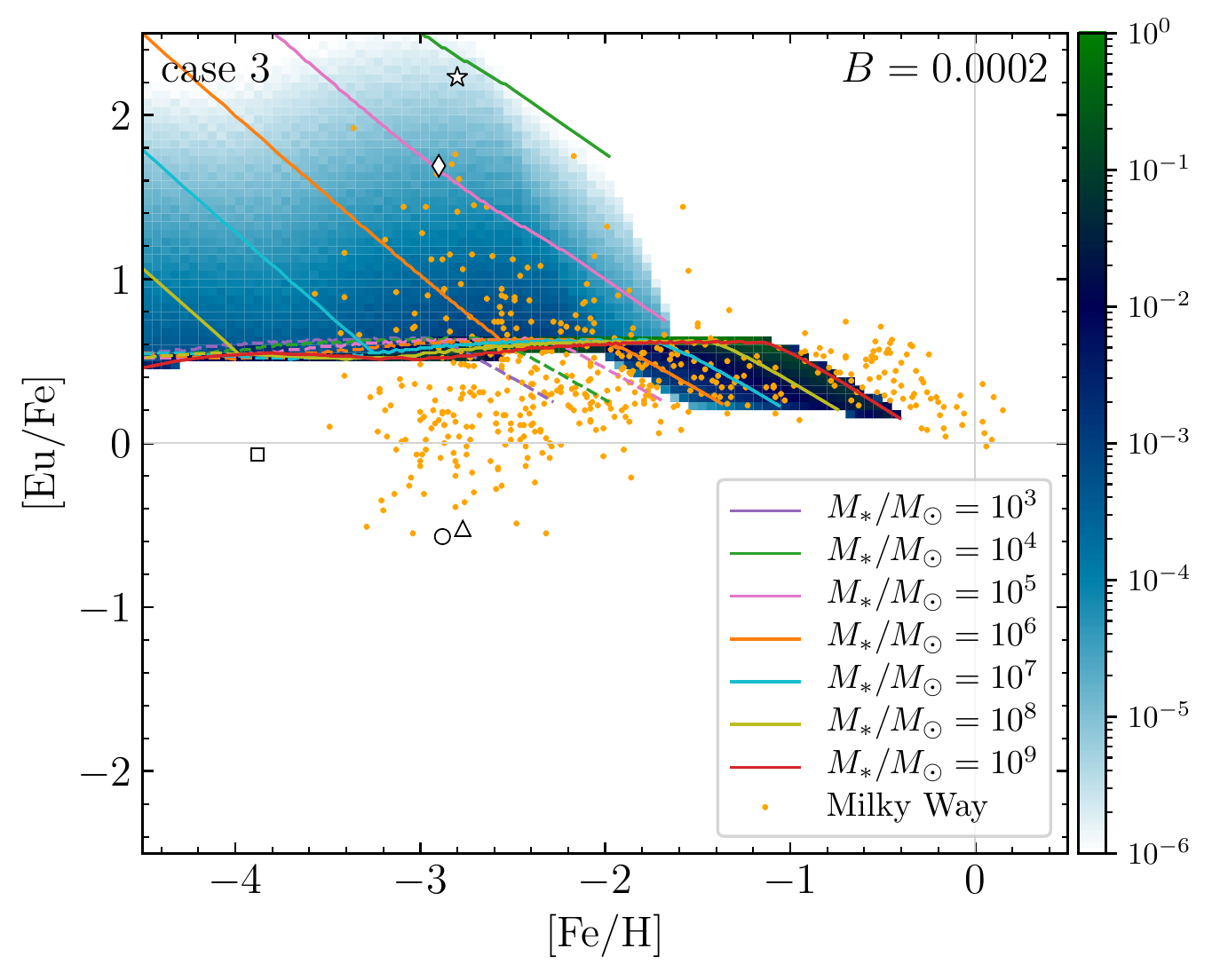}
	\includegraphics[width=0.86\columnwidth]{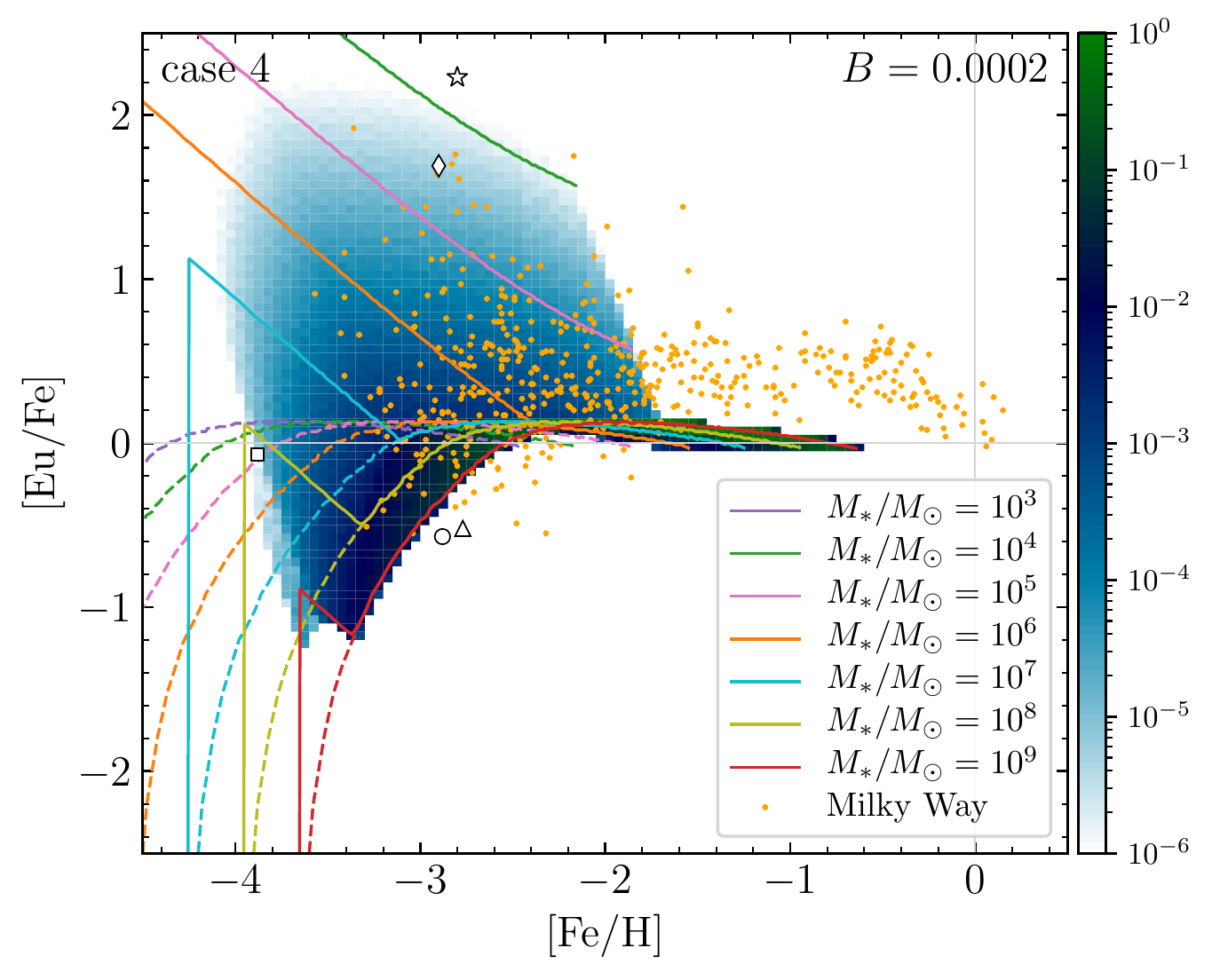}
    \caption{Same as Fig.~\ref{fig:eufe_halo}, but with the frequency of events decreased by a factor of 10 ($B = 0.0002$) and the Eu yield per event increased by a factor of 10. }
    \label{fig:appendix_freq}
\end{figure*}

Finally, Fig.~\ref{fig:appendix_freq} illustrates the results for cases 1--4 (same as Fig.~\ref{fig:eufe_halo}) but with the frequency of events decreased by a factor of 10 (or with the binary fraction in \S~\ref{subsec:yield} replaced by $B = 0.0002$) and the Eu yield per event increased by a factor of 10. We find that the values of [Eu/Fe] for $\langle N_\mathrm{NSM} \rangle < 1$ and $\langle N_\mathrm{NSM} \rangle > 1$ increase by a factor of 10 and unchanged, respectively, for each building block. However, the overall probability distributions of stars (colour scale) are similar to those in the original cases (Fig.~\ref{fig:eufe_halo}) because of the resulting smaller number of building blocks  experiencing NSMs (or subsets of CCSNe) for $\langle N_\mathrm{NSM} \rangle < 1$.

\bsp	
\label{lastpage}
\end{document}